\documentclass[a4paper,11pt]{article}
\pdfoutput=1 

\usepackage{jheppub} 
\usepackage[utf8x]{inputenc}
\usepackage[T1]{fontenc} 
\newcommand{\be}{\begin{equation}}
\newcommand{\ee}{\end{equation}}
\newcommand{\bea}{\begin{array}}
	\newcommand{\ea}{\end{array}}
\newcommand{\beqa}{\begin{eqnarray}}
\newcommand{\eeqa}{\end{eqnarray}}
\newcommand{\nn}{\nonumber}
\usepackage{physics}
\newcommand{\del}{\partial}
\usepackage{bm}
\usepackage{bbm}
\usepackage{multirow}
\usepackage[table,xcdraw]{xcolor}
\title{\boldmath Chaos from Equivariant Fields on Fuzzy $S^4$}

\author[a,b]{Ü. H. Coşkun,}
\author[a]{S.Kürkçüoğlu,}
\author[a,c]{G.C. Toga,}
\author[a]{G. Ünal}

\affiliation[a]{ Middle East Technical University, Department of Physics,\\Dumlupinar Boulevard, 06800, Ankara, Turkey}
\affiliation[b]{University of Kentucky, Department of Physics and Astronomy,\\Lexington, KY 40506, U.S.A.}
\affiliation[c]{Gazi University, Department of Physics,\\Teknikokullar, 06500, Ankara, Turkey}

\emailAdd{u.coskun@uky.edu}
\emailAdd{kseckin@metu.edu.tr}
\emailAdd{toga.can@metu.edu.tr}
\emailAdd{gonulunal23@gmail.com}

\abstract{We examine the $5d$ Yang-Mills matrix model in $0+1$-dimensions with $U(4N)$ gauge symmetry and a mass deformation term. We determine the explicit $SU(4)\approx SO(6)$ equivariant parametrizations of the gauge field and the fluctuations about the classical four concentric fuzzy four sphere configuration and obtain the low energy reduced actions(LEAs) by tracing over the $S_F^4$s for the first five lowest matrix levels. The LEAs so obtained have potentials bounded from below indicating that the equivariant fluctuations about the $S_F^4$ do not lead to any instabilities. These reduced systems exhibit chaos, which we reveal by computing their Lyapunov exponents. Using our numerical results, we explore various aspects of chaotic dynamics emerging from the LEAs. In particular, we model how the largest Lyapunov exponents change as a function of the energy. We also show that, in the Euclidean signature, the LEAs support the usual kink type soliton solutions, i.e. instantons in $1+0$-dimensions, which may be seen as the imprints of the topological fluxes penetrating the concentric $S_F^4$s due to the equivariance conditions, and preventing them to shrink to zero radius. Relaxing the Gauss law constraint in the LEAs in the manner recently discussed by Maldacena and Milekhin leads to Goldstone bosons.}

\begin{document} 
	\maketitle
	\flushbottom
	
	\section{Introduction}
	
	Matrix models associated to M-theory and string theories  \cite{Banks:1996vh, Berenstein:2002jq, Dasgupta:2002hx} have been under investigation from various perspectives ever since their discovery over twenty years ago. The broad span of interest on the subject is reflected in the literature (for a recent review, see \cite{Ydri:2017ncg} and references therein.)
	Among these, the BFSS model \cite{Banks:1996vh} is a supersymmetric $U(N)$ gauge theory consisting of nine $N \times N$ matrices in its bosonic part, whose entries depend on time only and it also goes by the common name of matrix quantum mechanics in the literature. It is associated to type II-A string theory \cite{kiritsis2011string,Ydri:2017ncg,Ydri:2016dmy} and appears as the DLCQ (discrete light-cone quantization) of M-theory on flat backgrounds \cite{kiritsis2011string, Ydri:2017ncg, Ydri:2016dmy}. The massive deformation of the BFSS theory, preserving the supersymmetry, is known as the BMN model and describes the DLCQ of $M$-theory on pp-wave backgrounds \cite{Berenstein:2002jq,Dasgupta:2002hx}. These matrix models describe systems of $N$ coincident $D0$-branes, respectively in flat and spherical backgrounds. The latter is due to the fact that fuzzy 2-spheres appear as vacuum configurations in the BMN model. At large $N$ and strong coupling/low temperature limit, $D0$-branes form a black brane, i.e. a string theoretic black hole \cite{kiritsis2011string,Ydri:2017ncg}; the structure of this gravity dual is discussed in varying amount of detail in several references \cite{kiritsis2011string, Ydri:2017ncg}. In \cite{Iizuka:2013kha}, a model on how the fuzzy spheres in the BMN model collapse to form a black hole is discussed.  
	
	The perspectives gained from the matrix models have recently started to motivate numerous investigations oriented to acquire new information on the properties of black holes, such as their thermalization, evaporation processes as well as their microscopic constituents via the study of BFSS and BMN models at large $N$, using both analytical insights and ever increasingly numerical techniques \cite{Anagnostopoulos:2007fw, Catterall:2008yz, Hanada:2008ez, Asplund:2011qj, Asplund:2012tg, Shenker:2013pqa, Gur-Ari:2015rcq, Berenstein:2016zgj, Maldacena:2015waa, Berkowitz:2016jlq}. In some of these works, for the BMN and BFSS models in the large $N$ and high temperature limit numerical evidence for the fast thermalization is obtained \cite{Asplund:2011qj, Asplund:2012tg, Shenker:2013pqa, Gur-Ari:2015rcq}. From a more general perspective, the latter is an example of the fast scrambling conjecture, which has been proposed by Sekino and Susskind \cite{Sekino:2008he} and which may be stated as the fact that in black holes, chaotic dynamics set in faster than in any other physical system and the rate at which this occurs is logarithmic in the number degrees of freedom of the black hole, that is, it is proportional to the logarithm of its Bekenstein-Hawking entropy. Even though the aforementioned limit is distinct from the one in which the gravity dual is obtained, it is the natural limit in which the classical mechanics provides a good approximation to the quantum mechanical matrix model\footnote{It may be emphasized that, this picture is only valid for quantum mechanics but not for quantum field theory, since in the latter high temperature theory does not a have good classical limit due to the UV catastrophe, as already noted in \cite{Gur-Ari:2015rcq}.}. This limit is free from fermions as the latter contributes to the dynamics of the bosonic matrices only at low temperatures. It has been also noted that \cite{Gur-Ari:2015rcq}, since, numerical studies performed so far do not show a phase transition occurring between the low and the high temperature limits of these matrix models \cite{Anagnostopoulos:2007fw, Catterall:2008yz, Hanada:2008ez,Filev:2015hia,Filev:2015cmz}, it is reasonable to expect that features like fast scrambling of blackholes in the gravity dual could survive at the high temperature limit too. In fact, chaotic dynamics in the BFSS model is studied in \cite{Gur-Ari:2015rcq} in this classical limit by calculating the Lyapunov spectrum, where it was also demonstrated that a classical analogue of fast scrambling is valid for this system as the scrambling time is found to be proportional to $\log N^2$. In \cite{Rinaldi:2017mjl} simulations of the BFSS model is performed at intermediate temperatures to numerically study the black hole horizon. 
	
	Even the matrix models at small values of $N$ appear to be highly non-trivial many-body system whose complete solution evades us to date. Recently, there also has been some interest in examining the chaotic dynamics emerging from such models \cite{Asano:2015eha, Berenstein:2016zgj} (See, also \cite{Arefeva:1997oyf}, for earlier attempts). These studies also provided some qualitative implications on the black hole phases in such models; for instance, in \cite{Berenstein:2016zgj} it has been argued that the edge of the chaotic region found in a $SU(2)$ YM matrix model with only two matrices, i.e. the smallest matrix model with non-trivial dynamics, corresponds to the end of the black hole phase. In \cite{Hubener:2014pfa} a novel approach has been developed to estimate the ground state energy of this smallest matrix model. Authors of\cite{Asano:2015eha}, considered simple ansatzes for the BMN model at $N=2,3$ satisfying the Gauss law constraint to probe the chaotic dynamics. 
	
	Fuzzy two sphere and its direct sums are not the only compact spherical geometries appearing in M-theory. In fact, it has been known for a quite long time that fuzzy four spheres make their appearance in matrix models as longitudinal five branes \cite{Castelino:1997rv}. For the purposes of  this paper however, presence of fuzzy four sphere solutions in Yang Mills 5-matrix model with massive deformation term plays the central role. Pure YM $5$-matrix odel has also been known in the literature for quite a while \cite{Kimura:2002nq} and may be obtained as the reduction of the YM theory in $5+1$-dimensions to $0+1$-dimensions keeping only the time dependence of the matrix elements. Together with the mass term it can also be conceived as a deformation of a subsector of the bosonic part of the BFSS model, as we will explain in more detail in the next section. Contrary to the fuzzy two sphere solutions in the BMN model, fuzzy four spheres in the mass deformed YM-matrix model are classical solutions for negative mass squared ($\mu^2=-8$), which may be an indication of tachyonic instabilities. Nevertheless, it was recently shown by Steinacker \cite{Steinacker:2015dra} that in pure YM 5-matrix model, one-loop quantum corrections stabilizes the radius of the fuzzy four sphere and prevents its collapse by shrinking to zero radius. In this paper, we  mainly focus on a mass deformed $U(4N)$ YM 5-matrix model and consider the exact parametrization of $SU(4)$ equivariant fluctuation modes about the four concentric fuzzy four sphere configurations. Using the equivariant parameterization of the gauge field and the fluctuations, we perform the traces over the fuzzy four spheres at first five lowest lying levels and obtain the corresponding low energy reduced actions (LEAs). We demonstrate that the potentials of all of these reduced effective actions are bounded from below, from which we infer that the negativity of $\mu^2$ does not actually cause any instabilities under equivariant fluctuations. As we will briefly discuss in section 3, this feature of our treatment may also be viewed as a consequence of the fact that the equivariant parametrization of the fluctuations introduces topological fluxes through the fuzzy four sphere, preventing it to shrink to zero radius. 
	
	Equivariant parametrizations breaks the $U(4)$ symmetry of the concentric $S_F^4$ configuration down to $U(1) \times U(1) \times U(1)$ and this is further reduced to only $U(1) \times U(1)$ in LEAs as one of the gauge fields completely decouple after tracing over $S_F^4$. The gauge fields in the reduced actions are not dynamical, and their equations of motion lead to the constraints, which may, in fact, be seen as the residue of the Gauss law constraint on the matrix model enforcing the physical states to be the singlets of the gauge symmetry group. In the LEAs the latter condition simply translates to the requirement that the two complex fields appearing in the LEAs be real, that is, uncharged under the abelian gauge fields. This breaks the $U(1) \times U(1)$ symmetry further down to ${\mathbb Z}_2 \times {\mathbb Z}_2$.\\
	\indent We utilize the LEAs to explore the chaotic structure emerging from the matrix model with the fuzzy four sphere background. For the reduced action obtained after tracing at the matrix levels $4N = 16,40,80,120$ and $224$ corresponding to the first five $S_F^4$ levels, $n=1,\cdots,5$, we numerically solve the Hamilton's equations of motion and compute the Lyapunov spectrum at several different energies, revealing the chaotic dynamics. We explore various features of the chaotic dynamics using our data. We show that the Largest Lyapunov Exponents(LLE) have a dependence on energy, which fits very well with the functional relation $\lambda_n(E) = \alpha_n + \beta_n \frac{1}{\sqrt E}$. The data on LLEs also enables us to probe the onset of chaos in the LEAs' dynamics. In fact, we are able to estimate the energies at which appreciable amount of chaotic dynamics is observed at each matrix level ($n=1,\cdots,5$) and also compute  the rate at which LLEs change to be proportional to $E^{-\frac{3}{2}}$.
	
	Except at $n=1$ the phase space of the LEAs are all ten dimensional, meaning that there are ten Lyapunov exponents associated to the LEAs at the matrix levels $n \geq 2$. At $n=1$, however, out of five of the generalized coordinates and corresponding velocities in the LEA, three of them combine to appear only in a particular form thereby reducing the dimension of the phase space to six. At low energies $n=1$ model exhibit different features compared to those for $n \geq 2$ and this is discussed in through detail section $4.2$. Plots of the time development of all the Lyapunov exponents at several different values of energy at all of the matrix levels ($n=1,\cdots,5$) is given in the Appendix B.4 and exhibit, in particular, that all the Lyapunov exponents at a given energy sum up to zero, as is expected to happen in all Hamiltonian systems. Some technical features of the computation of the Lyapunov spectrum is outlined in section 4.2 for completeness.  
	
	\indent The paper is organized as follows. In section 2, for completeness, we give a brief review of $S_F^4$ and how it appears in Yang-Mills matrix models. In section 3, we first determine the exact parametrization of the gauge field and the fluctuations, which are restricted to transform as a scalar and vectors, respectively, under the combined adjoint action of $SO(5)$ isometry of $S_F^4$ and the $SU(4)$ gauge symmetry generators in $SO(5)$. In this section we also obtain the LEAs by tracing over the $S_F^4$ at several different matrix levels, and elaborate on their structure. In section $4$ we focus on the dynamical structure of the LEAs. After discussing the implications of the Gauss law constraint, we present our results exhibiting the chaotic dynamics emerging from the LEAs. In section 5, we examine the properties of the LEAs in Euclidean signature, and make evident through a number of examples that, they possess ${\mathbb Z}_2$ kink solutions, i.e. instantons in $1+0$-dimensions. These may be seen as the imprints of the non-trivial topological fluxes piercing the $S_F^4$, which were mentioned above. Motivated by the recent work of Maldacena and Milehkin \cite{Maldacena:2018vsr}  on relaxing the Gauss law constraint in BFSS and BMN models (see \cite{Berkowitz:2018qhn} for supporting numerical work), in Section 6, we revisit the gauge symmetry of the LEAs and present a concise treatment on the consequences of not imposing the Gauss law constraint, which leads us to conclude the presence of massless excitations (Goldstone bosons) associated to these LEAs. We close the paper by summarizing our findings. Appendices collect reference formulas, intermediate steps of some of the analytic calculations, explicit form of the LEAs for $n \geq 2 $, the whole sets of the corresponding minima of the associated potentials, as well as all the time series plots of the Lyapunov spectrum at all matrix levels ($n=1,\cdots,5$) and at several different values of the energy.
	
	\section{Fuzzy $S^4$ in Yang Mills Matrix Models}
	
	\subsection{Basics}
	
	We launch the developments in this section by considering the Yang-Mills $5$-matrix model in Minkowski signature and with $U(4N)$\footnote{The reason for taking the gauge symmetry group $U(4N)$ will be come clear in the next section.} gauge symmetry, whose action may be given as \cite{Ydri:2017ncg,Kimura:2002nq,Steinacker:2015dra}
	\be
	{\cal S}_{YM}= \int dt  \, {\cal L}_{YM} = \frac{1}{g^2}\int d t  \, Tr \left ( \frac{1}{2}({\cal D}_t {\cal X}_a)^2 + \frac{1}{4} [{\cal X}_a, {\cal X}_b ]^2 \right ) \,, 
	\label{YMAction1}
	\ee 
	where ${\cal X}_a$ $(a:1,\dots 5)$ are $4N \times 4N$ Hermitian matrices transforming under the adjoint representation of $U(4N)$ as
	\be
	{\cal X}_a \rightarrow U^{\dagger} {\cal X}_a U \,, \quad U \in U(4N) \,,
	\label{gt1}
	\ee
	${\cal D}_t {\cal X}_a=\del_t {\cal X}_a - i [{\cal A}, {\cal X}_a]$ are the covariant derivatives, ${\cal A}$ is a $U(4N)$ gauge field transforming as 
	\be
	{\cal A} \rightarrow U^{\dagger}{\cal A} U- i U^{\dagger}\partial_t U \,,
	\label{gt2}
	\ee
	and $Tr$ stands for the normalized trace $Tr \mathbbm{1}_{4N} = 1$. For future reference we write out the potential part of ${\cal L}_{YM}$ separately as
	\be
	{\cal V}_{YM} = - \frac{1}{4 g^2} Tr [{\cal X}_a ,{\cal X}_b ]^2 \,.
	\ee
	Clearly, ${\cal S}_{YM}$ is invariant under the $U(4N)$ gauge transformations given by (\ref{gt1}) and (\ref{gt2}). ${\cal S}_{YM}$ is also invariant under the global $SO(5)$ rotations of ${\cal X}_a$. It can be obtained from the dimensional reduction of the $U(4N)$ gauge theory in $5+1$-dimensions to $0+1$-dimensions, where the $SO(5,1)$ Lorentz symmetry of the latter yields to the global $SO(5)$ of the reduced theory. 
	
	There are two distinct deformations of ${\cal S}_{YM}$ preserving its $U(4N)$ gauge and the $SO(5)$ global symmetries. One of these is obtained by adding a fifth order Chern-Simons term to ${\cal S}_{YM}$ (i.e. a Myers like term) which is given as \cite{Kimura:2002nq}
	\be
	{\cal S}_{CS}= \frac{1}{g^2} \int d t \, Tr  \, \frac{\lambda}{5}\epsilon^{abcde} {\cal X}_a {\cal X}_b {\cal X}_c {\cal X}_d {\cal X}_e \,,
	\label{csaction}
	\ee
	while the other is a massive deformation term of the form
	\be
	{\cal S}_{Mass}= - \frac{1}{g^2} \int \, d t \, Tr \, \mu^2 {\cal X}_a^2 \,.
	\ee
	It is useful to write out the potential terms for ${\cal S}_{1} = {\cal S}_{YM}+{\cal S}_{Mass}$ and ${\cal S}_{2} = {\cal S}_{YM}+{\cal S}_{CS} $ explicitly:
	\be
	{\cal V}_1 = \frac{1}{g^2} Tr  \left(-\frac{1}{4} [{\cal X}_a ,{\cal X}_b ]^2 + \mu^2 {\cal X}_a^2 \right) \,, \quad {\cal V}_2 = -\frac{1}{g^2} Tr \left( \frac{1}{4} [{\cal X}_a ,{\cal X}_b ]^2 + \frac{\lambda}{5}\epsilon^{abcde}{\cal X}_a {\cal X}_b {\cal X}_c {\cal X}_d {\cal X}_e \right) \,. 
	\ee
	In this paper, our main interest is on the massive deformations and hence we will focus on the dynamics emerging from ${\cal S}_{1}$, but will, although very briefly, also consider the consequences of some of the developments presented in the paper for ${\cal S}_{3} = {\cal S}_{YM}+ {\cal S}_{Mass}+ {\cal S}_{CS}$.
	
	We may as well note that ${\cal S}_{1}$, ${\cal S}_{2}$ and ${\cal S}_{3}$ may be thought as deformations of a subsector of the bosonic part of the BFSS \cite{Banks:1996vh} matrix quantum mechanics. As is already well-known, BFSS model can be conceived to emerge from the dimensional reduction of the YM theory in $9+1$ dimensions to $0+1$- dimensions \cite{Ydri:2017ncg} with the $SO(9,1)$ of the former yielding to the global $SO(9)$ of BFSS on the nine matrices ${\cal X}_I$ $(I:1\dots9)$, while the latter may be further broken to $SO(5) \times SO(4)$, via the addition of ${\cal S}_{Mass}$ and/or ${\cal S}_{CS}$ terms. To be more precise, these deformations terms, involving only ${\cal X}_a$ $(a:1,\dots 5)$, spontaneously break $SO(9)$ down to $SO(5) \times SO(4)$ and naturally split the ${\cal X}_I$ to an $SO(5)$ vector ${\cal X}_a$ and a $SO(4)$ vector ${\cal X}_\alpha$ $(\alpha:1\dots 4)$. Then, ${\cal S}_{1}$, ${\cal S}_{2}$ and ${\cal S}_{3}$ emerges by focusing on the sector of ${\cal X}_a$'s only.
	
	${\cal V}_2$ is extremized by the matrices fulfilling
	\be
	\lbrack {\cal X}_b , \lbrack {\cal X}_a , {\cal X}_b \rbrack \rbrack + \lambda \epsilon_{abcde} {\cal X}_b {\cal X}_c {\cal X}_d {\cal X}_e = 0 \,.
	\label{extrema1}
	\ee
	This equation has two immediate solutions \cite{Kimura:2002nq}, one of which is the diagonal matrices
	\be
	{\cal X}_a=\text{diag}({\cal X}_a^{(1)},{\cal X}_a^{(2)},\dots,{\cal X}_a^{(4N)}) \,,
	\label{diagextrema}
	\ee
	while the other is given by a fuzzy four sphere $S_F^4$, where $\lambda$ is forced to take on a fixed value $\lambda = \frac{2}{n+2}$, which depends on the matrix level of $S_F^4$. This latter fact requires attention, since it implies that the direct sum of fuzzy $S^4$ solve (\ref{extrema1}) if and only if they are fuzzy spheres of the same matrix level. For the configuration (\ref{diagextrema}) the potential take the value zero as is readily observed from ${\cal S}_{CS}$, whereas a simple calculation shows that  \cite{Kimura:2002nq} it, in fact, has a lower (negative) value for the fuzzy $S^4$ solutions. Thus at the classical level, fuzzy $S^4$ appears to be a more stable solution than the diagonal commuting matrices. Further numerical studies have, however, revealed that the fuzzy $S^4$ is not a minimum of ${\cal V}_2$, but instead a saddle point \cite{Azuma:2004yg}.\footnote{This is in contrast with the third order CS term appearing in the BMN deformation of the BFSS, which is an order less than the quadric YM potential and leads to stable fuzzy two sphere solutions \cite{Berenstein:2002jq}, while the fifth order CS term in the present model is an order higher than the YM potential and this may be seen as the underlying reason for fuzzy four spheres not being a minimum of ${\cal V}_2$ \cite{Azuma:2004yg}.}
	
	As for the massive deformation the potential part of ${\cal S}_{1}$ is extremized by the matrices fulfilling 
	\be
	\lbrack {\cal X}_b , \lbrack {\cal X}_a , {\cal X}_b \rbrack \rbrack - 2 \mu^2 {\cal X}_a=0 \,.
	\label{mass extrema}
	\ee
	Fuzzy four spheres $S_F^4$ and their direct sums (even from different matrix levels) are solutions of this equation for $\mu^2 = -8$. In a recent article Steinacker \cite{Steinacker:2015dra} showed that quantum corrections in the pure YM $5$-matrix model stabilizes the radius of the fuzzy four sphere. We will see that superficial instability implied by negativity of $\mu^2$ does not actually lead to a problem when we consider the equivariant fluctuations in ${\cal S}_{1}$ about the $S_F^4$ backgrounds. The reason for this is essentially that the potential of the emergent equivariantly reduced action is bounded from below at any finite matrix level. We may as well interpret this outcome as being due to the fact that the equivariant parametrization of the fluctuations introduces topological fluxes through the $S_F^4$ stabilizing its radius. We will further elucidate on these points in the Sections $3$ and $4$. Non-trivial topological fluxes leaves its imprints as kink type solutions of the reduced action in Euclidean signature, as we will exhibit in Section $5$.
	
	For the action ${\cal S}_3 = {\cal S}_{YM}+{\cal S}_{CS}+{\cal S}_{Mass}$ the potential is extremized by the matrices fulfilling \cite{Kimura:2002nq}
	\be
	\lbrack {\cal X}_b , \lbrack {\cal X}_a , {\cal X}_b \rbrack \rbrack - 2 \mu^2 {\cal X}_a + \lambda\epsilon_{abcde} {\cal X}_b {\cal X}_c {\cal X}_d {\cal X}_e = 0 \,,
	\label{mixedextrema}
	\ee
	whose fuzzy four sphere solutions need to satisfy the relation $\lambda = \frac{8 +\mu^2}{4 (n+2)}$, which includes the previous two cases of  ${\cal S}_1$ and ${\cal S}_2$ as $\mu^2 = - 8$ and $\mu^2 =0$, respectively. The equation of motion for ${\cal S}_3$ is 
	\be
	\ddot{{\cal X}}_a + \lbrack {\cal X}_b , \lbrack {\cal X}_a , {\cal X}_b \rbrack \rbrack - 2 \mu^2 {\cal X}_a + \lambda\epsilon_{abcde} {\cal X}_b {\cal X}_c {\cal X}_d {\cal X}_e = 0 \,,
	\label{meom}
	\ee
	in the gauge ${\cal A}=0$ and subject to the Gauss law constraint $\sum_a \lbrack {\cal X}_a \,, \dot{{\cal X}}_a \rbrack = 0$. Equations of motion for ${\cal S}_1$ and ${\cal S}_2$ are readily inferred from (\ref{meom}) by the remark following (\ref{mixedextrema}). Being static, $S_F^4$ configurations satisfying any one of the equations (\ref{extrema1}), (\ref{mass extrema}), (\ref{mixedextrema}) also satisfy the corresponding equation of motions as well as the Gauss law constraint.
	
	
	\subsection{Models in the Euclidean Signature}
	
	Wick rotating to the Euclidean signature we make the changes, $t \rightarrow - i \tau$, $\partial_t \rightarrow  i \partial_\tau$, ${\cal A} \rightarrow i {\cal A}$, ${\cal D}_t \rightarrow i {\cal D}_\tau$ and $L \rightarrow - L$. Euclidean action (in $1+0$-dimensions) is then
	\be
	{\cal S}_1^E = \frac{1}{g^2} \int d \tau \, Tr \left ( \frac{1}{2}({\cal D}_\tau {\cal X}_a)^2 - \frac{1}{4} [{\cal X}_a, {\cal X}_b ]^2  + \mu^2 {\cal X}_a^2 \right ) \,.
	\label{Euclideanaction}
	\ee
	In section 5, we will consider the kink-type solutions, that is to say, the instantons of the low energy effective actions in $1+0$-dimensions, which we obtain from the equivariant reduction of ${\cal S}_1^E $.
	
	\subsection{Brief Review of Fuzzy $S^4$}
	
	In this subsection we collect some of the main features of the fuzzy $S^4$ construction \cite{Grosse:1996mz,Castelino:1997rv, Kimura:2002nq, Ramgoolam:2001zx, Abe:2004sa, Ydri:2017ncg}. To start with we note that $S^4$ is embedded in $\mathbb{R}^5$ as 
	\be
	S^{4} \equiv \big \langle  \vec{X} = (X_1, X_2, \cdots, X_{5})  \in \mathbb{R}^{5} \big | {\vec X} \cdot {\vec X} = R^2 \big \rangle \,.
	\ee 
	
	Construction of fuzzy $S^4$ proceeds as follows. Let us denote by $\Gamma_a$, $(a:1,\cdots ,5)$ the Hermitian $4 \times 4 $  gamma matrices associated to $SO(5)$ fulfilling the defining anticommutation relations  
	\be
	\lbrace \Gamma_a \,, \Gamma_b \rbrace = 2 \delta_{ab} \,.
	\ee  
	For concreteness we take them to be given in the form
	\begin{align}
	\Gamma^i = \left(%
	\begin{matrix}
	0 & -i\sigma^i \\
	i \sigma^i & 0 \\
	\end{matrix}
	\right),\quad \Gamma^{4} = \left(%
	\begin{matrix}
	0 & \mathbbm{1}_{2} \\
	\mathbbm{1}_{2} & 0 \\
	\end{matrix}
	\right) \,, \quad \Gamma^{5} 
	= \left(%
	\begin{matrix}
	\mathbbm{1}_{2} & 0 \\
	0 & -\mathbbm{1}_{2} \\
	\end{matrix}
	\right) \,, \quad  i : 1,2,3 \,.
	\end{align}
	We introduce the $n$-fold tensor product 
	\be
	X_{a} := (\Gamma_{a} \otimes \mathbbm{1}_4\otimes \dots \otimes \mathbbm{1}_4 + \dots + \mathbbm{1}_4 \otimes \mathbbm{1}_4 \dots \otimes \Gamma_{a})
	\ee	
	acting on the $n$-fold completely symmetrized tensor product space
	\be
	\mathcal{H}_n := \bigotimes_n^{Sym} \mathbb{C}^4 = (\mathbb{C}^4\otimes\dots\otimes\mathbb{C}^4)_{Sym} \,,
	\ee
	which is the carrier space of the $(0,n)$ IRR\footnote{Throughout the text we label the IRRs by their Dynkin Indices. In Appendix A, we provide a short dictionary between Dynkin and highest weight labelling schemes and also give a short account of the branching rules used in the text.} of $SO(5)$. Obviously the latter is equivalent to the completely symmetric tensor product $\bigotimes_n^{Sym} (0,1)$ of the fundamental $4$-dimensional spinor representation of $SO(5)$ acting on $\mathbb{C}^4$. Dimension of this representation and hence that of $\mathcal{H}_n$ is given as 
	\be
	N := \mbox{dim} (0,n) = \frac{1}{6}(n+1)(n+2)(n+3) \,.
	\label{Dimension}
	\ee
	$X_a$ are then $N \times N$ Hermitian matrices satisfying the relations
	\begin{subequations}
		\beqa
		X_a X_a &=& n(n+4) \mathbbm{1}_N \label{radius} \,, \\
		\epsilon^{abcde} X_a X_b X_c X_d &=& 8(n+2) X_e \,,
		\label{epsilon1}
		\eeqa
	\end{subequations}
	which are the defining relations for the fuzzy four sphere, $S_F^4$.  In fact (\ref{radius}) may be seen as the $SO(5)$ invariant condition giving the radius of $S_F^4$ as $r_n := \sqrt{n(n+4)}$. This construction appears to be quite analogous to that of the fuzzy two sphere \cite{Balachandran:2005ew} with $SO(3)$ of the latter replaced with $SO(5)$, nevertheless only to the extent until one recognizes that the commutation relations of $X_a$ do not close but instead they are given by
	\be
	[X_a, X_b] =:  2 G_{ab} \,,
	\label{Gab}
	\ee
	where $G_{ab}$ are the ten generators of $SO(5)$ in its $(0,n)$ IRR satisfying the commutation relations,
	\be
	\lbrack G_{ab} , G_{cd} \rbrack = 2 ( \delta_{bc}G_{ad} + \delta_{ad}G_{bc} - \delta_{ac} G_{bd}-\delta_{bd}G_{ac}) \,.
	\label{Gabcd}
	\ee
	$G_{ab}$ are anti-hermitian by the definition (\ref{Gab}). Under the $SO(5)$ transformations generated by $G_{ab}$, $X_a$ transform as vectors (i.e. in the $(1,0)$ IRR of $SO(5)$) of $SO(5)$ since
	\be
	\comm{X_a} {G_{bc}} = 2 (\delta_{ab}X_c - \delta_{ac} X_b) \,.
	\label{Gabc}
	\ee
	We have that
	\begin{subequations}
		\beqa
		G_{ab}G_{ba} &=& 4 n (n+4) \mathbbm{1}_N \,, \label{GabCasimir} \\
		G_{ab}G_{bc} &=& n (n+4)  \delta_{ac} + G_a G_c - 2 G_c G_c  \,,
		\eeqa 
	\end{subequations}
	where (\ref{GabCasimir}) is twice the quadratic Casimir, $C_2^{SO(5)} = \sum_{a < b}^5 G_{ab} G_{ab}^\dagger = - \sum_{a < b}^5 G_{ab} G_{ab}$, of $SO(5)$ in the IRR $(0,n)$.
	
	It is also useful note that, $G_{ab}$ and $X_a$ together generate the $SO(6)$ in its $(n,0,0)$ IRR. This structure may be compactly expressed by writing
	the $15$ generators of $SO(6)$ as
	\be
	G_{AB} \equiv (G_{ab}, G_{a6}) = (G_{ab}, -i X_a) \,, \quad A, B : 1\,, \cdots ,6 \,,
	\ee
	with the commutation relations taking the usual form
	\be
	\lbrack G_{AB} , G_{CD} \rbrack = 2 (\delta_{BC}G_{AD} + \delta_{AD} G_{BC} - \delta_{AC} G_{BD} - \delta_{BD} G_{AC}) \,.
	\label{GABCD}
	\ee
	We now observe that the $SO(5)$ invariant condition (\ref{radius}) can be expressed as the difference of the quadratic Casimir operators:
	\be
	X_a X_a = C_2^{SO(6)}((n,0,0)) - C_2^{SO(5)}((0,n)) = n(n+4) \mathbbm{1}_N \,,
	\ee
	as $(n,0,0)$ of $SO(6)$ branches solely to $(0,n)$ of $SO(5)$ and hence they are of the same dimension $N$.
	
	The relation (\ref{epsilon1}) can also be expressed equivalently in the form 
	\be
	G_{ab} =- \frac{1}{2(n+2)}\epsilon^{abcde}G_{cd}X_{e}=- \frac{1}{2(n+2)}\epsilon^{abcde}X_{c}X_{d}X_{e} \,.
	\label{epsilon2}
	\ee  
	
	Another noteworthy feature of $S^4_F$ is that there is a $S^2_F$ attaching to each point of  $S^4_F$. In other words, there is a $S^2_F$ bundle over $S^4_F$ with the total space being ${\mathbb C}P^3_F$. This fact is reflected in the fields defined on $S^4_F$ carrying an intrinsic spin, whose rank can be up to $n$ \cite{Kimura:2002nq, Ramgoolam:2001zx}. 
	
	We may also record that the commutative limit is achieved by taking $n \rightarrow \infty$. The intricate structure of $S^4_F$ with $S^2_F$ fibers leads to
	\be
	\Omega_{AB} \equiv (\omega_{ab}, x_a) := \lim \limits_{n\rightarrow \infty}\frac{G_{AB}}{n} \,,
	\label{comcord}
	\ee
	where $x_a$, with $x_a x_a =1$, are the coordinates of $S^4 \subset \mathbb{R}^5$ and $\omega_{ab}$ is antisymmetric in its indices, satisfy $x_a \omega_{ab} = 0 $, as seen by taking the commutative limit of the first equality in (\ref{epsilon2}) and generate, in fact, the $S^2$ in the fibration $S^2 \rightarrow S^4 \rightarrow {\mathbb C}P^3$. Detailed discussion on this may be found in \cite{Ydri:2016osu}. 
	
	In the commutative limit, adjoint action of $X_a$ and $G_{ab}$ become the differential operators \cite{Kimura:2002nq} 
	\begin{subequations}
		\beqa
		\mbox{ad} \, X_a &\rightarrow& \nabla_a := 2i \left(\omega_{ab}\del_{x_b} - x_{b} \del_{\omega_{ab}} \right), \\
		\mbox{ad} \, G _{ab} &\rightarrow& \nabla_{ab} := 2 \left(x_a \del_{x_b}-x_b \del_{a} - \omega_{ac}\del_{\omega_{cb}} + \omega_{bc}\del_{\omega_{ca}}\right),
		\eeqa
	\end{subequations}
	where the derivatives with respect to $\omega_{ab}$'s are given via 
	\be
	\frac{\del\omega_{cd}}{\del\omega_{ab}}=\delta_{ac}\delta_{bd}-\delta_{ad} \delta_{bc} \,.
	\label{omegader}
	\ee
	We may note for future use that $x_a {\nabla}_a= 2i \left(x_a \omega_{ab}\del_{x_b} - x_b x_{b} \del_{\omega_{ab}} \right) =0$, since the first term in the r.h.s is already noted to vanish and the second term vanishes due to the antisymmetry of $\omega_{ab}$.
	
	\section{Equivariant Fields on $S_F^4$}
	
	\subsection{Symmetry Constraints and Parametrization of the Fields}
	
	We now focus on the YM model with mass term, whose action is given by ${\cal S}_{1}$
	\be
	{\cal S}_{1} = {\cal S}_{YM}+{\cal S}_{Mass} \,.
	\ee
	The potential ${\cal V}_1$ has an extremum given by four concentric $S_F^4$, 
	\be
	{\cal X}_a = X_a \otimes  \mathbbm{1}_{4} \,, \quad  X_a \in \mbox{Mat} (N) \,,
	\label{ext1}
	\ee	
	satisfying the equation (\ref{mass extrema}).
	
	We observe that $U(4N)$ gauge symmetry of the action ${\cal S}_{1}$ is broken down to $U(N) \times U(4)$, and the commutant of (\ref{ext1}) is just $U(4)$.
	Let us denote with $F_a$ the fluctuations about  (\ref{mass extrema}). We may write 
	\be
	{\cal X}_a = X_a \otimes  \mathbbm{1}_{4}  + F_a \equiv X_a  + F_a \,,
	\label{fluc}
	\ee
	where the r.h.s is introduced as a self-evident short hand notation, which will be used in the ensuing developments. We are interested in finding the equivariant fluctuations about the configuration (\ref{ext1}) \footnote{Similar analysis was previously performed in \cite{Harland:2009kg, Kurkcuoglu:2010sn, Kurkcuoglu:2012vf, Kurkcuoglu:2015qha, Kurkcuoglu:2015eta, Kurkcuoglu:2016gcp} for massive deformations of the ${\cal N} =4$ SUSY YM theory with vacua consisting of products of fuzzy two spheres, as well as $SU(N)$ gauge theories coupled to adjoint scalar matter multiplets with fuzzy sphere vacua emerging after dynamical breaking of the gauge symmetry, complementing the Kaluza Klein mode expansion approach given in \cite{Aschieri:2006uw, Chatzistavrakidis:2009ix }.}. To be more precise, we would like to concentrate on those $F_a$, which are invariant under the $SO(5)$ rotations of $S_F^4$ up to $SU(4) \subset U(4)$ gauge transformations. For this purpose we proceed as follows. We introduce the equivariant symmetry generators as
	\be
	W_{ab} = G_{ab}\otimes \mathbbm{1}_{4} + \mathbbm{1}_{N} \otimes \Sigma_{ab} \,,
	\ee 
	where $\Sigma_{ab} = \frac{1}{2} \lbrack \Gamma_a,\Gamma_b \rbrack$ are the generators of $SO(5)$ in the $4$-dimensional fundamental spinor representation $(0,1)$. They constitute the subset of generators $\Sigma_{AB} :=(\Sigma_{ab}, \Sigma_{a6} = - i \Gamma_a )$ of $SO(6) \equiv \frac{SU(4)}{\mathbb{Z}_2}$ in the fundamental $(1,0,0)$ spinor representation of $SO(6)$. Evidently, $W_{ab}$ satisfies the $SO(5)$ commutation relations and its $SO(5)$ representation content is given by the decomposition of the tensor product $(0,n) \otimes (0,1)$ into a direct sum of IRRs as 
	\be
	(0,n) \otimes (0,1) \equiv (0,n+1) \oplus(1,n-1) \oplus(0,n-1) \,.
	\label{so5irrs}
	\ee
	We digress a moment to note that, this structure can be lifted to $SO(6)$ group by writing $W_{AB} = (W_{ab}, W_{a6})$ with the representation content
	\be
	(n,0,0) \otimes (1,0,0) \equiv (n+1,0,0) \oplus (n-1,1,0) \,,
	\label{so6irrs}
	\ee
	whose branching under the $SO(5)$ simply yields (\ref{so5irrs}). Same branching under $SO(5)$ also holds for the complex conjugate representation $(0,0,n) \otimes (0,0,1) \equiv (0,0,n+1) \oplus (0,1,n-1)$.
	
	Let us examine the adjoint action of $W_{ab}$. It consists of two terms, first of which generates the infinitesimal $SO(5)$ rotations of the $S_F^4$, while the second term is responsible for generating the $SO(6)$ gauge transformations in $SO(5)$. Adjoint action of $W_{ab}$ carries the tensor product of the reducible representation (\ref{so5irrs}) with itself and using general formulas on tensor product of $SO(5)$ IRRs, it has the following decomposition in terms of $SO(5)$ IRRs
	\be
	{\bm 3} (0,0) \oplus {\bm 7} (1,0) \oplus \text{Higher dimensional IRRs} \,, \quad n \geq 2 \,,
	\label{adjointdecop1}
	\ee
	with the coefficients in bold indicating the multiplicities of the respective IRRs.  For the case of $n=1$, the decomposition takes the form 
	\be
	{\bm 3} (0,0) \oplus {\bm 5} (1,0) \oplus \text{Higher dimensional IRRs} \,, \quad n =1 .
	\label{adjointdecop2}
	\ee
	
	To study the equivariant fluctuations about the configuration (\ref{ext1}), we impose the symmetry constraints, that is the equivariance conditions,
	\begin{subequations}
		\beqa
		\lbrack W_{ab}, {\cal A} \rbrack &=& 0 \,, \label{eqcon1} \\
		\lbrack W_{ab} , F_{c} \rbrack &=& -2 (\delta_{ac} F_b -\delta_{bc} F_a)\,,
		\label{eqcon2}
		\eeqa
	\end{subequations}		
	first of which means that the gauge field ${\cal A}$ is simply required to transform as a scalar of $SO(5)$ under the adjoint action of $W_{ab}$, which is quite natural as it does not carry any $SO(5)$ index, while the second requires that the fluctuations $F_a$ introduced in (\ref{fluc}) transform as a vector of $SO(5)$ to comply with the equivariance condition. 
	
	We infer from the decomposition (\ref{adjointdecop1}) that the space of rotational invariants that may be constructed form the intertwiners of $(0,n)$ and $(0,1)$ IRRs of $SO(5)$ is three-dimensional, while the space of vectors that may be constructed in term of the intertwiners and $X_a$ is of dimension seven. The aforementioned intertwiners may be introduced via the projection operators to the IRRs appearing in the r.h.s. of (\ref{so5irrs}). We may express these three projections as 
	\be
	P_I = \prod_{J\neq I} \frac{-(G_{ab}+\Sigma_{ab})^2 - 2 \, C_2(\lambda_J)}{ 2 \, C_2(\lambda_I) - 2 \, C_2(\lambda_J)} \,, \quad P_I^2 =P_I \,, \quad P_I^\dagger = P_I \,, \quad I:1,2,3 \,, 
	\label{projectorformula}
	\ee
	where the factors of two in front of Casimirs are due to the unrestricted sum over $a$'s and $b$'s. $P_I$ are projections to the IRRs of $SO(5)$ in the order given in the r.h.s of (\ref{so5irrs}) and $C_2(\lambda_I)= -\sum_{a < b}^5 M_{ab} M_{ab}$ stand for the quadratic Casimirs of $SO(5)$ in the IRRs labeled by $\lambda_I \equiv ( (0,n+1), (1,n-1), (0,n-1) )$. Using (\ref{projectorformula}) and the fact that idempotents can be given as $Q_I = \mathbbm{1}_{4N} - 2 P_I$ we may list the intertwiners as the idempotents
	\begin{subequations}
		\beqa
		Q_1 &=& \frac{(G\cdot\Sigma-4)(G\cdot\Sigma -4n-16)-16(n+1)(n+2)}{16(n+1)(n+2)} \,, \\
		Q_2 &=&\frac{(G\cdot\Sigma+4n)(G\cdot\Sigma -4n-16)+2(2n+2)(2n+6)}{-2(2n+2)(2n+6)} \,, \\
		Q_3 &=& \frac{(G\cdot\Sigma-4)(G\cdot\Sigma + 4n)-16(n+3)(n+2)}{16(n+3)(n+2)} \,.
		\eeqa
	\end{subequations}	
	By construction we do have $Q_I^2 = \mathbbm{1}_{4N}$ and $Q_I^\dagger = Q_I$. Let us also note that $Q_I$ are not all independent from each other as we have $\sum_I Q_I = -  \mathbbm{1}_{4N}$. A straightforward, but a long calculation, whose details are provided in Appendix A, gives
	\be
	(G \cdot \Sigma)^2 = 12 \, \Gamma_a \Gamma_b G_{ab}  + 8n(n+2) X_a \Gamma_a + 8n(n+4) \mathbbm{1}_{4N} \,,
	\label{Gabcomlim}
	\ee
	and will be useful in what follows.
	
	Adjoint representation of $SO(6)$ branches under $SO(5)$ as ${\bm {15}} \rightarrow {\bm 5} \oplus {\bm {10}}$, or in the Dynkin notation:
	\be
	(1,0,1) \equiv (1,0) \oplus (0, 2) \,.
	\ee  
	Thus, further insight on how $SU(4) \approx SO(6)$ generators sits in these intertwiners is gained by observing that $Q_I$ contain, ten of these generators as $\Sigma_{ab}$, and the remaining five as $\Gamma_a$, as seen from (\ref{Gabcomlim}).
	
	Using (\ref{comcord}), commutative limit of (\ref{Gabcomlim}) takes the form
	\be
	\lim\limits_{n\rightarrow \infty} \frac{(G\cdot \Sigma)^2}{n^2} = 8(x_a \Gamma_a + \mathbbm{1}_{4} ) \,.
	\ee
	Consequently, we find for $q_I := \lim\limits_{n\rightarrow \infty} Q_I$: 
	\begin{subequations}\label{qcoml}
		\beqa
		q_1 &=& \frac{1}{2} \left ( x_a \Gamma_a - \sum_{a<b} \omega_{ab}\Sigma_{ab} - \mathbbm{1}_{4} \right ) \,, \\
		q_2 &=&- x_a \Gamma_a \,,
		\label{q2} \\
		q_3 &=&\frac{1}{2} \left ( x_a \Gamma_a + \sum_{a<b} \omega_{ab}\Sigma_{ab} - \mathbbm{1}_{4} \right ) \,.
		\eeqa
	\end{subequations}	
	
	We may argue that, equivariant parametrization of the fluctuations introduces topological fluxes through the concentric $S_F^4$s, preventing the latter to shrink to zero radius. Without going into any technicalities regarding the $S^2$ fibre coordinates $\omega_{ab}$ over $S^4$, using (\ref{q2}), this reasoning is supported by the fact that in the commutative limit the topological flux piercing the $S_F^4$ may be linked to the second Chern number on $S^4$: 
	\be
	c_2(S^4) =  \frac{1}{8 \pi^2} \int_{S^4} p_2 \, d \, p_2 \, d \, p_2 = 1 \,, 
	\ee
	where $p_2 = \frac{1}{2} (1- q_2)$ stand for the rank four projectors generating the projective module over the algebra of functions over $S^4$.
	
	We may solve the constraints given in (\ref{eqcon1}) and (\ref{eqcon2}) as follows. To satisfy (\ref{eqcon1}), we may choose to parameterize the gauge field ${\cal A}$ as 
	\be
	{\cal A} = \frac{1}{2} \alpha_1 Q_1 + \frac{1}{2} \alpha_2 \mathbbm{1}_{4N} + \frac{1}{2} \alpha_3 Q_3 \,,
	\label{A0par}
	\ee
	where we have introduced $\alpha_\mu \equiv \alpha_\mu(t)$ are real functions of time only, and eliminated $Q_2$ in favor of $\mathbbm{1}_{4N}$ using $\sum_I Q_I = -  \mathbbm{1}_{4N}$. From this form of the gauge field, it is readily observed that the $U(4)$ gauge symmetry is broken down to $U(1) \times U(1) \times U(1)$. However, later on we will see that term proportional to identity matrix in (\ref{A0par}) decouples after the dimensional reduction and the gauge symmetry of the reduced actions will eventually be $U(1) \times U(1)$.
	
	For the fluctuations satisfying (\ref{eqcon2}), a convenient parameterization that befits the ensuing developments turns out be  
	\begin{multline}
	F_a= i\frac{\phi_1}{2}[X_a,Q_1]+i\frac{\chi_1}{2}[X_a,Q_3]+\frac{\phi_2 +1}{2}Q_1[X_a,Q_1]+\frac{\chi_2+1}{2}Q_3[X_a,Q_3]  \\
	+ \phi_3 \left (\acomm{{\hat X}_a}{Q_1}-Q_3[{\hat X}_a,Q_3] \right) + \chi_3 \left (\acomm{{\hat X}_a}{Q_3} - Q_1[{\hat X}_a,Q_1] \right) + \phi_4 \left ({\hat X}_a + {\hat \Gamma}_a + Q_3[{\hat X}_a,Q_3] \right) \,.
	\label{vectorinv2}
	\end{multline}
	where we have introduced $\phi_\mu =\phi_\mu(t)$ and $\chi_\nu =\chi_\nu(t)$ as real functions of time only and used the notation
	\be
	{\hat X}_a = \frac{X_a \otimes \mathbbm{1}_{4}}{n} \equiv \frac{X_a}{n} \,, \quad {\hat \Gamma}_a = \frac{\mathbbm{1}_{N} \otimes \Gamma_a}{n} \equiv
	\frac{\Gamma_a}{n} \,.
	\ee
	The $\frac{1}{n}$ factors appearing in the last three terms of $F_a$ via, ${\hat X}_a$ and ${\hat \Gamma}_a$ are naturally expected for the convergence of $F_a$. Similar analysis performed in \cite{Harland:2009kg, Kurkcuoglu:2010sn, Kurkcuoglu:2012vf, Kurkcuoglu:2015qha} on equivariant parametrization of fluctuations over $S_F^2$ and $S_F^2 \times S_F^2$, shares the same features. Indeed, as $n \rightarrow \infty$, we find
	\begin{multline}
	F_a \rightarrow f_a := i \frac{\phi_1}{2} \nabla_a q_1 + i\frac{\chi_1}{2} \nabla_a q_3 + \frac{\phi_2 +1}{2} q_1 \nabla_a q_1 +\frac{\chi_2+1}{2} q_3 \nabla_a q_3 \\ + \phi_3 2 x_a q_1 + \chi_3 2 x_a q_3 + \phi_4 x_a \,.
	\label{FA}
	\end{multline}
	Demanding the fluctuations $f_a$ to be tangential to $S^4$ means that $x_a f_a =0$ has to be fulfilled. Since $x_a \nabla_a = 0$, as noted after (\ref{omegader}), the latter condition is satisfied if and only if $\phi_3(t)$, $\chi_3(t)$ and $\phi_4(t)$ all vanish in this limit.
	
	\subsection{Reduction over $S_F^4$ and the Low Energy Effective Action}
	
	We are now in a position to exploit the explicit parameterizations given (\ref{A0par}) and (\ref{vectorinv2}) to determine the low energy effective action that emerges from the action ${\cal S}_{1}$ by tracing over the concentric $S_F^4$'s.  
	
	After a straightforward but a long calculation, with some intermediate steps relegated to the Appendix B, we may write down the covariant derivatives in the form	
	\be
	\label{D03}
	\begin{split}
		{\cal D}_t {\cal X}_a=&\frac{i}{2}(D_t\phi_1-iQ_1D_t\phi_2)\comm{X_a}{Q_1}+\frac{i}{2}(D_t\chi_1-iQ_3D_t\chi_2)\comm{X_a}{Q_3}\\
		&\del_t\phi_3 \left ( \acomm{{\hat X}_a}{Q_1}-Q_3\comm{{\hat X}_a}{Q_3} \right) + \del_t \chi_3 \left( \acomm{{\hat X}_a}{Q_3} - Q_1\comm{{\hat X}_a}{Q_1} \right) \\&+\del_t \phi_4 \left ( {\hat X}_a + {\hat \Gamma}_a+Q_3\comm{{\hat X}_a}{Q_3} \right) \,,
	\end{split}
	\ee
	where we have used
	\be
	D_t\phi_i=\del_t\phi_i + \epsilon_{ji}\alpha_1\phi_j \,, \quad 
	D_t\chi_i=\del_t\chi_i +  \epsilon_{ji}\alpha_3 \chi_j \,.
	\ee
	for the fields $\phi_i(t)$ and $\chi_i(t)$ for $(i:1,2)$ only.
	
	Trace of the kinetic term is evaluated to be
	\be
	\begin{split}
		\frac{1}{2} Tr ({\cal D}_t {\cal X}_a)^2& = \frac{n(n+4)}{(n+1)^2} |D_t \phi|^2 + \frac{n(n+4)}{(n+3)^2} |D_t \chi|^2 \\
		& +\frac{2(n+4)(n^5+8n^4+18n^3+8n^2-11n)}{n^2(n+1)^2(n+3)^2} (\del_t\phi_3)^2 - \frac{12n(n+4)}{n^2(n+1)(n+3)}(\del_t\phi_3\del_t\chi_3) \\
		& +\frac{n(n+4)(-n^3-3n^2+17n+35)}{n^2(n+1)(n+3)^2}  (\del_t\phi_3\del_t\phi_4) - \frac{n(n+4)(n+5)}{n^2(n+3)} (\del_t\phi_4\del_t\chi_3) \\
		&+\frac{(n^4+10n^3+30n^2+34n+45)}{2 n^2(n+3)^2} (\del_t\phi_4)^2 \\
		&+\frac{2n(n+4)(n^4+8n^3+18n^2+8n-11)}{n^2 (n+3)^2(n+1)^2} (\del_t\chi_3)^2 \,,
	\end{split}
	\label{kinetic1}
	\ee
	where we have introduced the complex fields $\phi(t) = \phi_1(t) + i \phi_2(t)$ and $\chi(t) = \chi_1(t) + i \chi_2(t)$, with the covariant derivatives taking the usual form $D_t \phi = \partial_t \phi + i \alpha_1 \phi$, $D_t \chi = \partial_t \chi + i \alpha_3 \chi$. 
	
	Although (\ref{kinetic1}) does not appear to be manifestly positive definite, a simple analysis using Mathematica confirms that it is so, as it should be by construction. Thus, it is possible and useful to make a linear field redefinition in the sector spanned by $\phi_3$, $\phi_4$ and $\chi_3$, i.e. convert to a basis, in which the kinetic term (\ref{kinetic1}) is diagonalized. In the next section, we will naturally work with such linearly redefined fields, which diagonalize the kinetic term for specific values of $n$ from $n=1$ to $n=5$.   
	
	We can now proceed to evaluate the trace of the mass term in the action ${\cal S}_{1}$. This can be written out as
	\be
	\begin{split}
		&- \mu^2 Tr ({\cal X}_a {\cal X}_a) = - \mu^2 Tr (X_a X_a + 2 X_a F_a +F_a F_a) \\
		&= - \mu^2 \Bigg ( \frac{2n(n+4)}{(n+1)^2} |\phi|^2 + \frac{2n(n+4)}{(n+3)^2} |\chi|^2 +\frac{4 (n+4)(n^4+8n^3+18n^2+8n-11))}{n (n+1)^2 (n+3)^2} (\phi_3^2+\chi_3^2) \\ 
		& \quad \quad + \frac{n^4+10n^3+30n^2+34n+45}{n^2 (n+3)^2} \phi_4^2 + \frac{2 (n+4)(-n^3-3n^2-17n+35)}{n (n+1)(n+3)^2} \phi_3\phi_4 \\
		& \quad \quad - \frac{24 (n+4)}{n (n+1)(n+3)} \phi_3\chi_3 -\frac{2(n+4)(n+5)}{n(n+3)} \chi_3\phi_4 + \frac{(n+4)(-n^3-4n^2+7n+22)}{ (n+3)^2(n+1)} \phi_3 \\
		& \quad \quad + \frac{(n+4)(-n^3-8n^2-9n+6)}{(n+1)^2(n+3)} \chi_3 + \frac{(n+4)(n^2+6n+5)}{(n+3)^2} \phi_4 + C(n) \Bigg ) \,.
	\end{split}
	\label{massterm}
	\ee
	A few remarks are in order. Firstly, we see that there are terms which are linear in the fields $\phi_3$, $\phi_4$ and $\chi_3$. For finite values of $n$, which is essentially going to be our main focus in the next section, these terms cause no harm. In the $n \rightarrow \infty$ the coefficients of these terms diverge linearly with $n$. Thus, alluding to our previous remark following (\ref{FA}), for the finiteness of the large $n$ limit, it will suffice to assume that $\phi_3$, $\phi_4$ and $\chi_3$ vanish faster than $\frac{1}{n}$. The mass terms then converges to $- 2\mu^2 ( |\phi|^2 +  |\chi|^2)$, while the kinetic term is given by $\abs{D_t\phi}^2 + \abs{D_t\chi}^2$ in this limit.  Let us also note that, we have $\mu^2 = -8$, since we are inspecting this term about the $S_F^4$ configuration satisfying (\ref{mass extrema}). The exact form of the constant term $C(n)$ is immaterial; in the next section we will adjust the overall constant term in the reduced Lagrangians so as to set the minimum value of the potential to zero.
	
	Analytic calculation of the trace of the interaction term $\frac{1}{4} Tr [{\cal X}_a ,{\cal X}_b ]^2$ in the action ${\cal S}_{1}$ for the equivariant parametrizations (\ref{A0par}) and (\ref{vectorinv2}) turns out to be quite a formidable task as it involves large number of rather complicated traces. As an alternative approach, we have evaluated the traces for this term for the values of $n=1,2,3,4,5$ using Mathematica. These already correspond to reasonably large span of matrix sizes $4N = 40, 80, 140, 224$, respectively and gives us ample information for exploring the dynamics of the low energy reduced action. This is what we take up in the next section.
	\section{Dynamics of the Low Energy Reduced Action}
	
	\subsection{Gauge Symmetry \& the Gauss Law Constraint}
	
	The equivariantly reduced action obtained from ${\cal S}_{1}$ is invariant under the $U(1)\times U(1)$ gauge transformations
	\beqa
	\phi^\prime &=& e^{-i \Lambda_1(t)} \phi \,, \quad \alpha_1^\prime (t) = \alpha_1(t) + \del_t \Lambda_1(t) \,, \nn \\
	\chi^\prime &=& e^{-i \Lambda_3(t)} \chi \,, \quad  \alpha_3^\prime (t)= \alpha_3(t) + \del_t \Lambda_3 (t) \,
	\eeqa
	with the remaining fields $\phi_3$, $\phi_4$ and $\chi_3$ being real and thus uncharged under $U(1)\times U(1)$. We observe this manifestly from (\ref{kinetic1}) and the mass term (\ref{massterm}). The interaction term has the same gauge symmetry too by construction and can be manifestly seen from (\ref{LC1}) at the level $n=1$. 
	
	As the time derivatives of $U(1)$ gauge fields $\alpha_1(t)$ and $\alpha_3(t)$ appear nowhere in the action, these fields have no dynamics of their own. Their equations of motion give us the Gauss law constraints:
	\beqa
	\frac{1}{2i} \frac{1}{|\phi|^2} (\phi (\del_t \phi)^* - (\del_t\phi) \phi^*)  &=&  \alpha_1(t) \,,\nn \\
	\frac{1}{2i} \frac{1}{|\chi|^2} (\chi (\del_t \chi)^* - (\del_t\chi) \chi^*)  &=&  \alpha_3 (t) \,.
	\label{glc}
	\eeqa
	We make the gauge choice $\alpha_1(t) = 0 $ and $\alpha_3(t) = 0$, which essentially amounts to the reality conditions $\phi^*= \phi$ and $\chi^*= \chi$. We can be more precise by first noting that the Gauss law constraints does not break the $U(1) \times U(1)$ gauge symmetry completely, but a residual ${\mathbb Z}_2 \times {\mathbb Z}_2$ remains. Indeed, writing $\phi \equiv (\phi_1, \phi_2) = |\phi| (\cos \theta , \sin \theta)$ and $\chi \equiv (\chi_1, \chi_2) = |\chi| (\cos \sigma , \sin \sigma)$, we may express the constraints in the form
	\be
	\partial_t \theta = \frac{1}{|\phi|^2} \varepsilon_{ij} \phi_i \partial_t \phi_j = \partial_t \Lambda_1 =0  \,, \quad   \partial_t \sigma = \frac{1}{|\chi|^2} \varepsilon_{ij} \chi_i \partial_t \chi_j = \partial_t \Lambda_3 = 0 \,.
	\label{gauss1}
	\ee
	Therefore, the remaining ${\mathbb Z}_2 \times {\mathbb Z}_2$ symmetry is encoded in the gauge functions as $\Lambda_1(t) = \Lambda_1^0 + \pi k$ and $\Lambda_3(t) = \Lambda_3^0 + \pi k$, where $\Lambda_1^0$ and $\Lambda_3^0$ are constants and $k \in {\mathbb Z}_2$. This indicates that, for both of the gauge functions, $\Lambda_1$ and $\Lambda_3$, we have more generally
	\be
	\int_{- \infty}^\infty dt \, \partial_t \Lambda = \Lambda (\infty) - \Lambda (- \infty) = \pi k
	\label{gaugefreedom}
	\ee
	Due to (\ref{gauss1}), we have $\theta(t) = \theta^0 + \pi k$ and $\sigma(t) = \sigma^0 + \pi k$, and (\ref{gaugefreedom}) holds for both $\theta(t)$ and $\sigma(t)$, as well. Having noted these points, in what follows we set $\phi_2(t)$ and $\chi(t)$ to zero (i.e., we have both $\theta^0$ and $\sigma^0$ set to zero). Then, the ${\mathbb Z}_2 \times {\mathbb Z}_2$ symmetry is implemented by $(\phi_1 \,, \chi_1) \rightarrow (\pm \phi_1 \,, \pm \chi_1)$ \& $(\phi_1 \,, \chi_1) \rightarrow (\pm \phi_1 \,, \mp \chi_1)$. 
	
	In Section $5$ we consider the structure of the LEAs in the Euclidean time $\tau$. Due to the ${\mathbb Z}_2 \times {\mathbb Z}_2$ symmetry, we will be able to explore kink type solutions of the LEAs by imposing appropriate boundary conditions on $\phi_1(\tau)$ and $\chi_1(\tau)$ as $\tau \rightarrow \pm \infty$. Availability to impose topologically non-trivial boundary conditions on the latter can be attributed to the property (\ref{gaugefreedom}) of the restricted gauge functions, which holds the same in the Euclidean signature.
	
	There is also the possibility of not imposing the Gauss law constraint as recently discussed for BFSS and BMN matrix models in \cite{Maldacena:2018vsr}. This leads to presence of Goldstone bosons for the LEAs that we have obtained, as we will briefly discuss and demonstrate in Section 6.
	
	\subsection{Dynamics of the Reduced Action and Chaos} \label{sssec:ChaosSec}
	
	The explicit form of the equivariantly reduced Lagrangian for $n=2$ is given below
	\be
	\begin{split}
		L_{(n=2)} = &\frac{1}{2}\left(0.96 \dot{\chi}_1^2+2.7 \dot{\phi}_1^2+12.94 \dot{\phi}_3^2+6.32 \dot{\phi}_4^2+0.88 \dot{\chi}_3^2\right)-1.09 \chi _1^4-0.252 \chi _3^4\\
		&-2.03 \chi _3^3+6.99 \chi _1^2-0.26 \chi _1^2 \chi _3^2-4.80 \chi _3^2+2.69 \chi _1^2 \chi _3\\
		&+0.11 \chi _3-4.8 \chi _1^2 \phi _1^2-0.10 \chi _3^2 \phi _1^2+3.77 \chi _3 \phi _3 \phi _1^2-0.77 \chi _3 \phi _4 \phi _1^2\\
		&-2.79 \chi _3 \phi _1^2-1.46 \chi _3^3 \phi _3+0.44 \chi _3^3 \phi _4-1.62 \chi _1^2 \phi _3^2-2.71 \chi _1^2 \phi _4^2\\
		&+5.02 \chi _1^2 \phi _3-5.11 \chi _1^2 \phi _3 \phi _4+3.81 \chi _1^2 \phi _4-3.36\chi _3^2 \phi _3^2\\
		&-0.33 \chi _3^2 \phi _4^2-8.51 \chi _3^2 \phi _3+1.92 \chi _3^2 \phi _3 \phi _4+2.75 \chi _3^2 \phi _4-0.64\chi _3 \phi _3^3\\
		&-0.67 \chi _3 \phi _4^3-19.2 \chi _3 \phi _3^2-1.45\chi _3 \phi _3 \phi _4^2+1.80 \chi _3 \phi _4^2\\
		&-1.36 \chi _1^2 \chi _3 \phi _3-2.51 \chi _1^2 \chi _3 \phi _4-13.05 \chi _3 \phi _3+2.16 \chi _3 \phi _3^2 \phi _4\\
		&+10.25 \chi _3 \phi _3 \phi _4+1.07 \chi _3 \phi _4-3.70 \phi _1^4-32.51 \phi _3^2 \phi _1^2\\
		&+0.90 \phi _4^2 \phi _1^2+41.66 \phi _3 \phi _1^2+19.59\phi _3 \phi _4 \phi _1^2-20.62\phi _4 \phi _1^2\\
		&+12.20 \phi _1^2-14.33 \phi _3^4-5.46 \phi _4^4+41.31 \phi _3^3-5.89 \phi _3 \phi _4^3\\
		&+28.77 \phi _4^3-28.88 \phi _3^2-3.423 \phi _3^2 \phi _4^2+22.60 \phi _3 \phi _4^2-43.37 \phi _4^2\\
		&-46.70 \phi _3+4.18 \phi _3^3 \phi _4+3.42 \phi _3^2 \phi _4-15.50 \phi _3 \phi _4+16.80 \phi _4 - 29.6 \,.
	\end{split}
	\label{L2}
	\ee
	while the explicit form of $L_{(n)}$ for $n=3,4,5$ are given in Appendix B.
	
	Let us note that in $L_{(n)}$ for  $n=2,3,4,5$, we have  {\it i)} performed the linear transformation among the fields $\phi_3 \rightarrow \phi_3^\prime$, $\phi_4 \rightarrow \phi_4^\prime$, $\chi_3 \rightarrow \chi_3^\prime$ which diagonalizes the kinetic term (\ref{kinetic1}), and dropped the $^\prime$'s in the final form, {\it ii)} have set $\mu^2 =-8$ as we have already remarked to do so in the paragraph after (\ref{massterm}), {\it iii)} have imposed the Gauss law constraints as we just discussed, {\it iv)} adjusted the constant in the final form of each $L_{(n)}$, so that the associated potentials, $V_{(n)}$, take the value zero at their minima, {\it v)} introduced an over-dot, ${\dot {(...)}}$, to denote the time derivatives and {\it vi)} set the coupling constant $g$ to one, as it has no effect on the classical physics save for determining a global normalization in the energy unit.
	
	For $n=1$, the reduced action takes a simpler form as compared to the cases for $n \geq 2$, which is given as 
	\be
	\begin{split}
		L_{(n=1)} = \frac{5}{16} \dot{\chi}_1^2 + \frac{5}{4} \dot{\phi}_1^2 + \frac{15}{4} \dot{\Phi}^2 - \frac{5}{8} \left(\phi _1^2+\chi _1^2-4\right){}^2 - & \frac{15}{8} \left(\phi _1^2+4 \Phi (1+ \Phi)-3\right){}^2 \\
		& - \frac{5}{2} \phi _1^2 \chi _1^2 - \frac{15}{2} (1+2 \Phi)^2 \phi _1^2 \,.
	\end{split}
	\label{L1}
	\ee	
	where we have introduced $\Phi =  \phi_3 +\phi_4 - \chi_3$, which is the only combination of the constituent dynamical variables $\phi_3$, $\phi_4$ and $\chi_3$ upon which $L_{(n=1)}$ depends. This is to be expected in view of (\ref{adjointdecop2}). For $L_{(n=1)}$ too, all items following (\ref{L2}) are performed as well, except the item {\it i)}, which is already taken care of with the introduction of $\Phi(t)$. 
	
	An important feature of the reduced Lagrangians is that their potentials are all bounded from below. Therefore, we infer that at any level $n$ the equivariant fluctuations about the concentric $S_F^4$ solutions, with $\mu^2 =- 8$  do not cause any instability. The absolute minima of the potentials are given in (\ref{n2vacua}), (\ref{minima3}), (\ref{minima4}) and (\ref{minima5}).
	
	We find that the reduced Lagrangians, $L_{(n)}$, have chaotic dynamics. One of the basic tools to probe the presence of chaos in a dynamical system is to compute the Lyapunov exponents, which measures the exponential growth in perturbations \cite{ott2002chaos}. If, say, $x(t)$ is a phase space coordinate, in a chaotic system the perturbation in $x(t)$, denoted by $\delta x(t)$, deviates exponentially from its initial value at $t=0$; $|\delta x(t)| = |\delta x(0)| e^{\lambda_L t}$, $\lambda_L$ being the Lyapunov exponent corresponding to the phase space variable $x(t)$. 
	
	The phase space corresponding to the LEA are $10$-dimensional, except for the $n=1$ case, and spanned by 
	\be
	(\phi_1, \chi_1,\phi_3, \chi_3, \phi_4, p_{\phi_1}, p_{\chi_1}, p_{\phi_3},p_{\chi_3}, p_{\phi_4})
	\ee
	where $p_i$ are the corresponding conjugate momenta and the Hamiltonians, $H_{(n)}$, are obtained from $L_{(n)}$  in the usual manner using $H = p_i {\dot q}_i - L$. Using numerical solutions for the Hamilton's equations of motion, we have performed calculations of all of the Lyapunov spectrum for the models at the levels $n = 2,3,4,5$ at the energies 
\be
E = 15,20,25,30,40,50,100,250,500,1000,2000 \,,
\ee
and obtained their time series, which are given in the appendix (\ref{timeseries}) in figures (\ref{fig:n1E15}-\ref{fig:n5E2000}). To be more precise, we ran a Matlab code, which calculated the mean of the time series for 20 runs with randomly selected initial conditions for all of the Lyapunov exponents at each $n$ and for the energies given above. The initial conditions are randomly selected by the code from a sector of the $10$-dimensional phase space for $(n =2,3,4,5)$. The latter is specified by giving the initial values of the eight of the phase space coordinates, while the code randomly selects an initial value for one phase space coordinate and calculates the last one to satisfy the given value of the energy. We have checked that, increasing the number of the randomly selected coordinates somewhat increases the computation time, but does not have any significant impact on our results. For $n=1$, dimension of the phase space reduces to $6$ as easily observed from $L_{(1)}$ in (\ref{L1}). A similar analysis to the one described above is also performed at this level, whose marked differences and similarities with the rest will also be pointed out in what ensues. Table 1 summarizes our findings for the largest Lyapunov exponents $\lambda_{max}$ at $(n = 2,3,4,5)$ at the listed values of the energy. Table 2 and 3 give the LLE data at a larger set of energies for $n=1$ to probe especially the low energy region, which appears to have different features.
	
	
	\begin{table}[]
		\centering
		\label{LLEtable1}
		\begin{tabular}{|l|l|l|l|l|}
			\hline
			Energy 	  & n=2      & n=3      & n=4      & n=5      \\ \hline
			E=15  	     & 0.094  & 0.035  & 0.016  & 0.039  \\ \hline
			E=20  	     & 0.2788 & 0.086  & 0.055  & 0.056  \\ \hline
			E=25  	    & 0.4204 & 0.2159 & 0.1563 & 0.1180  \\ \hline
			E=30  	   & 0.4893 & 0.3515 & 0.2623 & 0.2372 \\ \hline
			E=40  	   & 0.6370 & 0.5371 & 0.4453 & 0.4393 \\ \hline
			E=50       & 0.7265 & 0.6450 & 0.5710 & 0.5276 \\ \hline
			E=100      & 0.9645 & 0.8430 & 0.7578 & 0.6972 \\ \hline
			E=250  	   & 1.099  & 1.0699 & 0.9922 & 0.9054 \\ \hline
			E=500  		& 1.1574 & 1.1439 & 1.1064 & 1.0405 \\ \hline
			E=1000		& 1.2138 & 1.1983 & 1.1610  & 1.0982 \\ \hline
			E=2000 		& 1.3087 & 1.2949 & 1.2412 & 1.1566 \\ \hline
		\end{tabular}
		\caption{LLE  Values for $n\geq2$}
	\end{table}
	
		\begin{table}[]
			\label{LLEtable2}
		\centering
				\begin{minipage}[t]{0.4\textwidth}
			\begin{tabular}{|l|l|}
				\hline
				Energy  & n=1      \\ \hline
				10 & 0.03361 \\ \hline
				15 & 0.1056  \\ \hline
				20 & 0.1831  \\ \hline
				25 & 0.4218  \\ \hline
				30 & 0.5416  \\ \hline
				35 & 0.5744  \\ \hline
				38 & 0.3329  \\ \hline
				40 & 0.1284   \\ \hline
				45 & 0.1152  \\ \hline
			\end{tabular}
			\caption{LLE  Values for $n=1$, $E\leq45$}
				\end{minipage}
					\begin{minipage}[t]{0.4\textwidth}
						\begin{tabular}{|l|l|}
							\hline
							Energy    & n=1     \\ \hline
							50   & 0.1386 \\ \hline
							75   & 0.1879 \\ \hline
							100  & 0.3356 \\ \hline
							150  & 0.5786 \\ \hline
							200  & 0.9476 \\ \hline
							300  & 1.1487 \\ \hline
							500  & 1.2307  \\ \hline
							1000 & 1.3303  \\ \hline
							2000 & 1.5061  \\ \hline
						\end{tabular}
						\caption{LLE  Values for $n=1$,$E>45$}
					\end{minipage}
		\end{table}

\begin{figure}[!htb]\centering
		\begin{minipage}[t]{0.5\textwidth}
			\centering
			\includegraphics[width=1\textwidth]{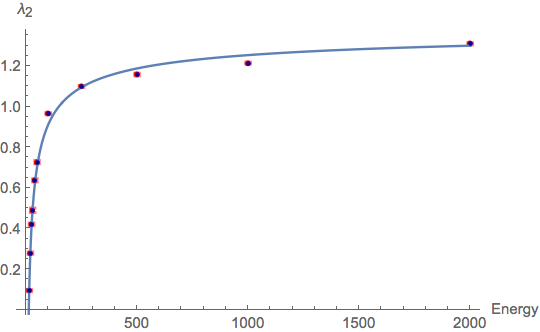}
			\caption{$\lambda_2(E) = 1.41- 5.00 \frac{1}{\sqrt E} $} 
			\label{fig:Plot_LLE_En2}
		\end{minipage}%
		\begin{minipage}[t]{0.5\textwidth}
			\centering	
			\includegraphics[width=1\textwidth]{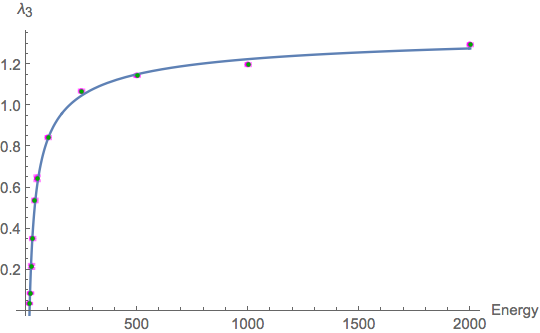}
			\caption{$\lambda_3(E) = 1.40- 5.61 \frac{1}{\sqrt E} $} 
			\label{fig:Plot_LLE_En3}
		\end{minipage}
	\end{figure}

\begin{figure}[!htb]\centering
		\begin{minipage}[t]{0.5\textwidth}
			\centering
			\includegraphics[width=1\textwidth]{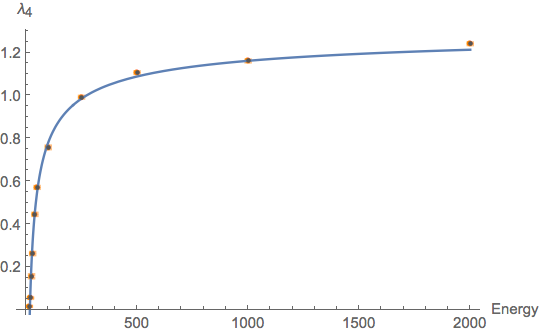}
			\caption{$\lambda_4(E) = 1.34 - 5.58 \frac{1}{\sqrt E} $} 
			\label{fig:Plot_LLE_En4}
		\end{minipage}%
		\begin{minipage}[t]{0.5\textwidth}
			\centering	
			\includegraphics[width=1\textwidth]{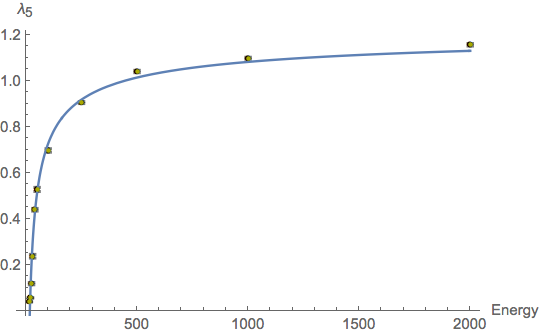}
			\caption{$\lambda_5(E) = 1.25- 5.19 \frac{1}{\sqrt E} $} 
			\label{fig:Plot_LLE_En5}
		\end{minipage}
	\end{figure}

	It is especially interesting from our data to explore the dependence of the LLEs at a given level $n$ with respect to the energy. We find that our data for LLEs fit very well with the functional relation
	\begin{equation}
	\lambda_n(E) = \alpha_n + \beta_n \frac{1}{\sqrt E}
	\label{lvE}
	\end{equation}
	where $\lambda_n(E)$ denotes the LLE as a function of energy at fixed $n$.
The plots for the data and corresponding fits are given in the figures (Fig.\ref{fig:Plot_LLE_En2}, Fig.\ref{fig:Plot_LLE_En3}, Fig.\ref{fig:Plot_LLE_En4}, Fig.\ref{fig:Plot_LLE_En5}). We have also calculated the standard error for the largest Lyapunov exponents from the standard deviation of the final mean value of $\lambda_n(E)$ using the Largest Lyapunov exponents from each of the 20 runs at each level $n$ and at the energies listed above. The errors are quite small, typically remain around $0.0050$ with the span being from $\pm 0.0018$ to $\pm 0.011$ and thus appear very small in the figures.
		
	\begin{table}[!htb]
		\label{LvEfit}
		\centering
			\begin{tabular}{|l|l|l|}
				\hline
				$\lambda_i(E)$  & $\alpha_n$ & $\beta_n$      \\ \hline
				$\lambda_2(E)$ & 1.41 &-5.00 \\ \hline
				$\lambda_3(E)$ & 1.40 &-5.61  \\ \hline
				$\lambda_4(E)$ & 1.34 &-5.58 \\ \hline
				$\lambda_5(E)$ & 1.25 &-5.19 \\ \hline
			\end{tabular}
			\caption{ $\alpha_n$ \& $\beta_n$ values for the fit in (\ref{lvE})}
			\end{table}
Our findings appear to be quite novel in the sense that, to the best of our knowledge, they constitute the only result in the literature within the context of matrix Yang-Mills theories at zero temperature, in which the dependence of the LLEs on the energy is predicted from numerical data at several of the lowest lying matrix levels. The data and the corresponding fits indicate that after $E=500$, increase in LLEs becomes very slow and from the last row of the values of LLE at $E=2000$ given in table 1, we see that the $\alpha_n$ values of the fits provide a reasonably good estimates of the values of $\lambda_n(E=2000)$, which is within a margin of $\leq 0.1$ only.

At the level $n=1$ we find a markedly different behaviour of LLEs with increasing energy for $E \leq 50$ as can be seen from the data and corresponding plots given in the figures  Fig.(\ref{fig:n1higherfit}) and  Fig.(\ref{fig:n1lowerfit}). Sudden decrease in the value of the Lyapunov exponents around $E=40$ can be attributed to the fact that the potential $V_{(1)}$ takes the value $40$ at its local maximum. Although the value of LLE decreases around this nonstable equilibrium point it resumes to attain the growing profile for $E \geq 45$ and the same functional form which befit those of $n \geq 2$ also fits well with the numerical results as can be seen from Fig.(\ref{fig:n1higherfit}). We have found the suitable fit to be {$\lambda_1(E) = 1.78 - 12.9 \frac{1}{\sqrt E} $ for $E \geq 45$.
	
	\begin{figure}[!htb]
	\centering
	\begin{minipage}[t]{0.5\textwidth}	
		\centering
		\includegraphics[width=\linewidth,height=0.25\textheight]{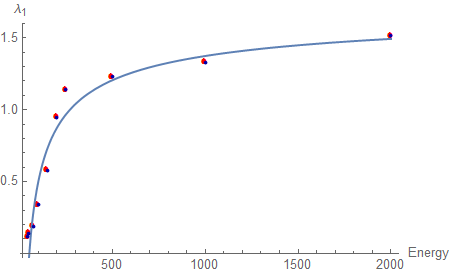}
		\caption{$\lambda_1(E) = 1.78 - 12.9 \frac{1}{\sqrt E} $ for $E \geq 45$}
		\label{fig:n1higherfit}
	\end{minipage}%
	\begin{minipage}[t]{0.5\textwidth}	
		\centering
		\includegraphics[width=\linewidth,height=0.25\textheight]{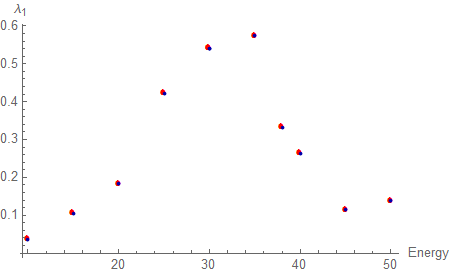}
		\caption{Sample LLE values for $10 \leq E \leq 50 $}
		\label{fig:n1lowerfit}
	\end{minipage}
\end{figure}
	
	These results also enable us to probe the onset of chaos in our models. To do so, let us first remark that the numerical calculations are performed for a finite computation time only, therefore even at low energies numerical values of the LLEs are small numbers compared to the characteristic scale of LLE values within a given model, but may not be seen to vanish in finite time. Secondly, it is well-known that in Hamiltonian systems periodic dynamics and chaotic dynamics can coexist and that there is in general no sharp passage from one to another \cite{hilborn2000chaos}. Keeping these two facts in mind, it thus appears reasonable to set a critical lower bound on the LLE value at and above which the models have appreciable amount of chaotic dynamics, (i.e. there are comparable number of chaotic and periodic trajectories.) Inspecting the data from table 1 and the time series plots of the Lyapunov spectrum  given in the appendix (\ref{timeseries}), we see that a reasonable choice would be to take this bound to be about $0.1$. Then, we find that for $n=2,3,4,5$, critical energies for the onset of chaos turns out to be $E \approx 16, 22, 23, 24$, respectively, with the corresponding LLE values being $\lambda_2(E=16) =0.1167 $, $\lambda_3(E=22) =0.1039$, $\lambda_4(E=23) = 0.0996$, $\lambda_5(E=24) = 0.1178$. Lyapunov exponents become smaller below these energies as can be inspected from table 1. In fact, using equation (\ref{lvE}), we predict that they get smaller at an increasing rate of $\frac{d \lambda_n(E)}{d E} = -\frac{1}{2} \beta_n E^{-\frac{3}{2}}$, ($\beta_n < 0)$, as $E$ decreases. From the fits we find that for $n=2,3,4,5$, LLEs vanish at the energies $\approx 12.6, 16, 17.4, 17.3$, respectively, which is consistent with the small values obtained for LLE  at $E=15$. Below $E=15$, part of the initial conditions become too small for the numerical integrator built into Matlab to handle and we can not obtain any healthy data on the LLE in this region. However, there appears no reason to expect that any significant amount of chaos remains below $E=15$.
	
	It is also worthwhile to explore the change in LLE values as $n$ takes on the values $n=2,3,4,5$ at fixed value of the energy. Figure \ref{fig:Plot_LLE_N} depicts this at several different values of the energy. It turns out that logaritmic functions of the form
	\be
	\lambda_E(n) = \gamma_E - \delta_E \log n \,, \quad n = 2,3,4,5 \,,
	\ee
	provide a good fit to the data. Here $\lambda_E(n)$ stand for LLE as a function of $n$ at fixed energy. The dashed lines in figure (\ref{fig:Plot_LLE_N}) are provided for visual guidance as $n$ takes on only the integer values. Values of $\gamma_E$, $\delta_E$ are also provided below for convenience. Only important feature we infer from these fits is that at a given energy $E$, decrease in $\lambda_E(n)$ appears to be logarithmically slow, suggesting that, even for $n>5$, we may expect to have chaotic dynamics at moderate values of the energy.  
	\begin{figure}[!htb]\centering
		\centering
		\includegraphics[width=1\textwidth]{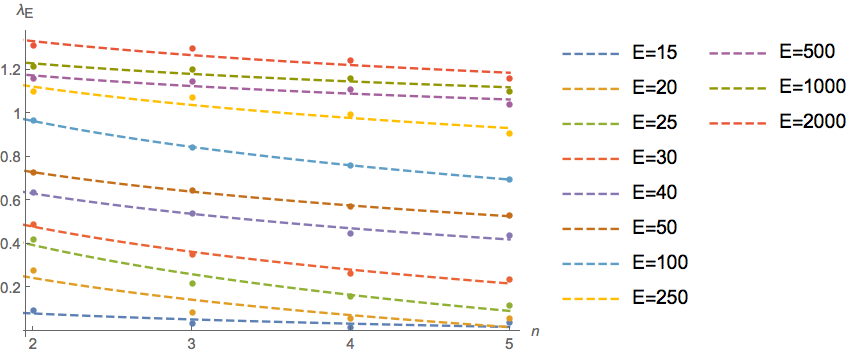}
		\caption{$\lambda_E$ versus $n$ (Dashed lines are provide only for visual guidance)} 
		\label{fig:Plot_LLE_N}
	\end{figure}
	
	\begin{table}[!htb]
		\centering
		\label{fitvalues}
		\begin{tabular}{|l|l|l|}
			\hline
			Energy   & $\gamma_E$    & $\delta_E$    \\ \hline
			E=15 & 0.13  &  0.07 \\ \hline
			E=20 & 0.41  &  0.25 \\ \hline
			E=25 & 0.62  &  0.32 \\ \hline
			E=30 &  0.67 &  0.28 \\ \hline
			E=40 & 0.79  &  0.23 \\ \hline
			E=50 &  0.88 & 0.22 \\ \hline
			E=100 & 1.17   & 0.29  \\ \hline
			E=250 & 1.26  & 0.20  \\ \hline
			E= 500 & 1.25  & 0.12  \\ \hline
			E= 1000 & 1.31  & 0.16   \\ \hline
			E= 2000 & 1.44  & 0.16   \\ \hline
		\end{tabular}
		\caption{$\gamma_E$ and $\delta_E$ fit values}
	\end{table}
	
Before closing this section we want to note that the LEAs treated in this paper have only ${\mathbb Z}_2 \times {\mathbb Z}_2$ symmetry after gauge fixing as already explained in section 4.1. This residual gauge symmetry is spontaneously broken by the degenerate vacuum configurations of zero energy, which are given section 5.2 and in appendix B.5 (the potential $V_{(n)}$ and hence the corresponding vacua are the same both in real time and Euclidean time), as well as by the instanton solutions given in the next section. We see that the configurations that respect the ${\mathbb Z}_2 \times {\mathbb Z}_2$ symmetry are only those for which both $\phi_1$ and $\chi_1$ are vanishing, as these are the only two fields changing sign under the ${\mathbb Z}_2 \times {\mathbb Z}_2$ action. We thus easily infer that those configurations respecting this ${\mathbb Z}_2 \times {\mathbb Z}_2$ symmetry and possessing the smallest energy are the static solutions of the equations of motion minimizing $V_{(n)}(\phi_1 = 0, \chi_1=0, \phi_3, \phi_3, \chi_3)$. Using Mathematica we have calculated that these configurations have the energies $ E = 10, 15.1, 18.0, 19.7, 20.8$, respectively for $n=1,2,3,4,5$. At each of $n=2,3,4,5$ there are eight distinct configurations at the given energy, while at $n=1$ there are only two.
	
	\section{Kink Solutions}
	
	We now consider the matrix model in the Euclidean signature ${\cal S}_1^E$ given in (\ref{Euclideanaction}) and the corresponding LEAs. The latter have multiple degenerate vacua supporting kink solutions, i.e. instantons in $1+0$-dimensions, whose features are sketched out in what follows. 
	
	\subsection{Kinks at level $n=1$} 
	
	The Lagrangian is given as
	\be
	\begin{split}
		L_{(n=1)} =\frac{1}{8} {\chi}_1^{\prime 2} + \frac{1}{2} \phi_1^{\prime 2} + \frac{3}{2} \Phi^{\prime 2} + \frac{1}{4} \left(\phi _1^2+\chi _1^2-4\right){}^2 + & \frac{3}{4} \left(\phi _1^2+4 \Phi (1+ \Phi)-3\right){}^2 \\
		&+ \phi _1^2 \chi _1^2 + 3 (1+2 \Phi)^2 \phi _1^2 \,.
	\end{split}
	\label{LE1}
	\ee	
	where $^\prime$ stands for derivatives with respect to the Euclidean time $\tau$. We have also scaled out an unimportant factor of $\frac{5}{2}$ compared to (\ref{L1}).
	We can easily see from (\ref{LE1}) that there are three different pairs of vacua, which are given by the configurations
	\beqa
	\phi_1 = \pm 2 \,, \quad \chi_1 = 0 \,, \quad \Phi &=& - \frac{1}{2} \,, \nn \\
	\phi_1 = 0 \,, \quad \chi_1 = \pm 2 \,, \quad \Phi &=& \frac{1}{2} \, \mbox{or} \, - \frac{3}{2} \,,
	\label{z2vacua}
	\eeqa
	Since either $\phi_1$ or $\chi_1$ vanish in these vacua, we infer that, the kink solutions could be of the type with topological indices $(\pm 1, 0)$ or $(0, \pm1) \in {\mathbb Z}_2 \oplus {\mathbb Z}_2$. These are the familiar kink solutions \cite{coleman1988aspects,Rajaraman:1982is}. Indeed, we find that the equations of motion are of the form
	\beqa
	\phi_1^{\prime \prime} - \left ( 4 \phi _1^3 +3 \chi _1^2 \phi _1+ \phi _1 (7+6 \Phi) (6 \Phi -1) \right) &=& 0 \,, \nn \\
	\chi_1^{\prime \prime} - 4 \left  ( \chi _1^3+3 \phi _1^2 \chi _1-4 \chi _1 \right ) &=& 0 \,, \\
	\Phi^{\prime \prime} - \left ( 2 (1+2 \Phi) \left(3 \phi _1^2+4 \Phi (1+\Phi)-3\right) \right ) &=& 0 \nn \,.
	\eeqa
	which have the following solutions
	\beqa
	&&\phi_1(\tau) =  2 \tanh (2 \sqrt2 \, \tau) \,, \quad \chi_1(\tau) = 0 \,, \quad \Phi(\tau) = -\frac{1}{2} \,, \quad \phi_1(\pm \infty) = \pm 2 \,,\\
	&&\phi_1(\tau) = 0 \,, \quad \chi_1(\tau) = 2 \tanh (2 \sqrt2 \, \tau)  \,, \quad \Phi(\tau) = \frac{1}{2} \, \mbox {or} \, - \frac{3}{2} \,,
	\quad \chi_1(\pm \infty) = \pm 2 \,.
	\eeqa
	
	\subsection{Kinks at levels $n \geq 2$}
	
	For $n=2,3,4,5$, the number of degenerate vacua increases. This may be expected due to the larger number degrees of freedom in the LEAs. Nevertheless, a similar structure in vacuum configurations to that of $n=1$ is observed, and allow for the kink solutions. At $n=3$, for instance, we have eight pairs of degenerate vacua, which are given as
	\be
	\begin{split}
		&\left\{\phi _1\to 0.,\phi _3\to 2.56,\phi _4\to 3.42,\chi _1\to \pm2.,\chi _3\to -11.5\right\},\\
		&\left\{\phi _1\to 0.,\phi _3\to -0.28,\phi _4\to 0.55,\chi _1\to \pm2.,\chi _3\to 1.46\right\},\\
		&\left\{\phi _1\to 0.,\phi _3\to 2.55,\phi _4\to -0.60,\chi _1\to \pm2.,\chi _3\to -5.26\right\},\\
		&\left\{\phi _1\to 0.,\phi _3\to -0.27,\phi _4\to 4.60,\chi _1\to \pm 2.,\chi _3\to -4.79\right\},\\
		&\left\{\phi _1\to \pm2.,\phi _3\to 2.30,\phi _4\to 4.13,\chi _1\to 0,\chi _3\to -3.01\right\},\\
		&\left\{\phi _1\to \pm2.,\phi _3\to -0.02,\phi _4\to -0.16,\chi _1\to 0,\chi _3\to -7.04\right\},\\
		&\left\{\phi _1\to \pm2.,\phi _3\to 0.28,\phi _4\to -0.55,\chi _1\to 0,\chi _3\to -1.46\right\},\\
		&\left\{\phi _1\to \pm2.,\phi _3\to 2.00,\phi _4\to 4.51,\chi _1\to 0,\chi _3\to -8.59\right\},\\
	\end{split}
	\label{minima3}
	\ee
	
	The equations of motion for $L_{(n=3)}$ are coupled non-linear differential equations, which are not easy to solve. We may look at the linearized system of equations around one of the minima. For notational simplicity, let us write $(\phi_1, \chi_1, \phi_3, \phi_4, \chi_3) \equiv (S_1,S_2,S_3,S_4,S_5) := {\bm S}$ and also introduce ${\bm S} = {\bm S}^0 + {\bm s}$, where ${\bm S}^0$ is one of the vacuum configurations in (\ref{minima3}) and ${\bm s}$ stands for the fluctuations. The linearized system of equations is given by
	\be
	{\cal C}_i s_i^{\prime \prime} = \frac{\partial^2 V_{(3)}}{\partial s_i \partial s_j} \Bigg |_{{\bm S}^0} s_j \,,
	\ee
	where no sum over repeated indices is implied on th l.h.s and ${\cal C}_i$ can be easily read off from (\ref{l3}). For, say, ${\bm S}^0_\pm \equiv \left\{\phi _1\to \pm2.,\phi _3\to 2.00,\phi _4\to 4.51,\chi _1\to 0,\chi _3\to -8.59\right\}$, these take the form 
	\beqa
	2.6 s_1^{\prime \prime} -125.3 s_1-30.1 s_3+51.9 s_4+5.41 s_5 &=& 0 \,, \nn \\
	0.52 s_2 ^{\prime \prime} -38.91 s_2, &=& 0 \nn \,, \\
	9.8 s_3^{\prime \prime} - 30.1 s_1- 216.3 s_3+29.7 s_4+3.7 s_5, &=& 0 \,, \\
	6.9 s_4^{\prime \prime} +51.9 s_1+29.7 s_3 -110.8 s_4 -6.4 s_5, &=& 0 \,, \nn \\
	0.92 s_5 ^{\prime \prime} +5.4 s_1+3.7 s_3-6.4 s_4-5.8 s_5 &=& 0 \,.\nn 
	\eeqa   
	The leading order profiles of the solutions of these equations  which are regular as $\tau \rightarrow \infty$ are given below, while their complete forms are given in appendix B.
	\beqa
	&& s_1(\tau) \approx \left(-0.98 c_1-0.18 c_2+4.98 c_3 +1.11 c_4\right) e^{-2.38 \tau} \,, \quad s_2(\tau) = c_{5} e^{-8.65068 \tau} \,, \nn \\ 
	&& s_3(\tau) \approx \left(-0.65 c_1- 0.12 c_2+3.28 c_3 + 0.73 c_4\right) e^{-2.38 \tau} \,, \nn \\
	&& s_4(\tau) \approx \left(7.75 c_1+1.49 c_2 - 39.25 c_3 - 8.73 c_4\right) e^{-2.38 \tau} \,, \nn \\
	&& s_5(\tau) \approx \left(-73.40 c_1-13.99 c_2+404.89 c_3+89.86c_4\right) e^{-2.38 \tau} \,,
	\label{linear}
	\eeqa
	where $c_i$ $(i:1\, \cdots \, 5)$ are arbitrary constants. (\ref{linear}) provides the asymptotic profile of the kink solutions near ${\bm S}_+^0$.
	
	\subsection{Kinks in the Presence of the Chern Simons Term}
	
	Here we will confine our discussion only to the level $n=1$.  In the Euclidean signature, the reduced action obtained from ${\cal S}_3$ takes the form
	\be
	\begin{split}
		&L_{(n=1)} = \frac{5}{16} {\chi}_1^{\prime 2} + \frac{5}{4} \phi_1^{\prime 2} + \frac{15}{4} \Phi^{\prime 2} + \frac{5}{8} \left(\phi _1^2+\chi _1^2 +\frac{1}{2} \mu^2 \right)^2 +  \frac{15}{8} \left(\phi _1^2 + 4 \Phi (1+ \Phi) + (1 + \frac{1}{2} \mu^2 )\right)^2 \\
		& + \frac{5}{2} \phi _1^2 \chi _1^2 + \frac{15}{2} (1+2 \Phi)^2 \phi _1^2 -\frac{1}{8} \left(\mu ^2+8\right) (2 \Phi +1) \left(5 \phi _1^2 \left(\chi _1^2+(2 \Phi +1)^2\right) + 
		(2 \Phi +1)^4 + 5 \phi _1^4\right) \,.
	\end{split}
	\label{LECS1}
	\ee	
	Last term in (\ref{LECS1}) is the contribution coming from the reduction of ${\cal S}_{CS}$ as can be clearly observed from the fifth order terms it contains. Due to presence of this term the potential of (\ref{LECS1}) does not have an absolute minimum, in fact, naturally, it is not bounded from below. Nevertheless, for $\mu^2 < 0$, it is a matter of a simple calculation to see that the potential still have degenerate local minima. This still allows for kink solutions, which are, however, only metastable and will decay under sufficiently large perturbations. As a concrete example, we have, for instance for $\mu^2=-1$, the local minima occurring at 
	\be
	\phi_1 = 0 \,, \quad \chi_1 =  \pm \frac{1}{\sqrt2} \,, \quad \Phi \approx -0.804 \,,
	\ee 
	and a kink solution to the equations of (\ref{LECS1}) is given by
	\be
	\phi_1 = 0 \,, \quad \chi_1 = \frac{1}{\sqrt2} \tanh {\sqrt2} \tau  \,, \quad \Phi \approx -0.804 \,.
	\ee
	
	\section{Gauge Symmetry Revisited}
	
	In \cite{Maldacena:2018vsr} Maldacena and Milekhin considered the BFSS model without imposing the Gauss law constraint, i.e. without the $SU(N)$-singlet condition on the physical states. This is based on the fact that the BFSS model with ${\cal A} = 0$ is still well-defined even in the absence of the Gauss law constraint, at the expense that the $SU(N)$ is no longer a local but a global gauge symmetry group in this situation. 
	
	We observe that such a possibility is also valid and applicable to the low energy reduced actions that we have obtained in this paper. The latter have $U(1) \times U(1)$ gauge symmetry and the Gauss law constraint was given in (\ref{glc}), imposing the $U(1) \times U(1)$ singlet condition meaning that the complex fields $\phi(t)$ and $\chi(t)$ are uncharged, i.e. real, under the gauge fields $\alpha_1(t)$ and $\alpha_3(t)$, respectively. If we do not impose the constraint we can still set the gauge fields $\alpha_1(t)$ and $\alpha_3(t)$ to zero in the LEAs at any level $n$. In this case, the LEAs have a global $U(1) \times U(1)$ symmetry, which is spontaneously broken by several different vacuum configurations, and hence imply the existence of Goldstone bosons in these LEAs. For instance, at the level $n=1$, we have the action
	\be
	\begin{split}
		L_{(n=1)} =\frac{1}{8} |\dot{\chi}|^2 + \frac{1}{2} |\dot{\phi}|^2 + \frac{3}{2} \dot{\Phi}^2 - \frac{1}{4} \left(|\phi|^2 + |\chi|^2 - 4 \right)^2 - & \frac{3}{4} \left(|\phi|^2 + 4 \Phi (1+ \Phi) - 3 \right)^2 \\ &- |\phi|^2 |\chi|^2 - 3 (1+2 \Phi)^2 |\phi|^2 \,.
	\end{split}
	\label{LC1}
	\ee	
	with three different vacuum configurations, as easily recognized from (\ref{z2vacua})
	\beqa
	|\phi| = 2 \,, \quad \chi = 0 \,, \quad \Phi &=& - \frac{1}{2} \,, \nn \\
	\phi = 0 \,, \quad |\chi| = 2 \,, \quad \Phi &=& \frac{1}{2} \, \mbox{or} \, - \frac{3}{2} \,,
	\label{degvacua}
	\eeqa
	spontaneously breaking the $U(1) \times U(1)$ symmetry. Thus, we immediately infer that the fluctuations around each of these vacuum configurations give rise to one Goldstone boson in the usual manner that it arises in an abelian gauge theory with degenerate vacua.  As a concrete example, we have, with $\phi = 2 + \sigma_1 + i \sigma_2$, $\Phi = - \frac{1}{2} + \rho$ and denoting the fluctuation around $\chi = 0$ still with $\chi$, potential part of $L_{(n=1)}$ takes the form
	\be
	\begin{split}
		V_{(n=1)}(\sigma_1, \sigma_2, \chi, \rho) = \frac{1}{4} (\sigma_1^2 +\sigma_2^2 + 2 \sigma_1 + |\chi|^2)^2 + & \frac{3}{4} (\sigma_1^2 +\sigma_2^2 + 2 \sigma_1 + 4 \rho^2)^2 \\ 
		& + \left((2 +\sigma_1)^2 + \sigma_2 \right) ( |\chi|^2 + 6 \rho ) \,,
	\end{split}
	\ee
	showing that $\sigma_2$ is massless, i.e. it is the Goldstone boson associated with this particular vacuum configuration. Similar analysis show the existence of Goldstone bosons for the other two vacuum configurations in (\ref{degvacua}). Finally, we note that for $n =2,3,4,5$ we infer from (\ref{n2vacua}), (\ref{minima3}), (\ref{minima4}) and (\ref{minima5}) that, at each value of $n$, there are eight distinct vacuum configurations determined by the values of $\phi$, $\chi$ and the real fields $\phi_3$, $\chi_3$ and $\phi_4$, each of which comes with a Goldstone boson. 
	
	\section{Conclusions}
	
	In this paper we have studied the $5d$ mass-deformed Yang-Mills matrix model with $U(4N)$ gauge symmetry. We have found the exact form of the $SU(4)\approx SO(6)$ equivariant parametrizations of the gauge field and the fluctuations about the classical four concentric fuzzy four sphere configuration and used them to calculate the LEAs by performing traces over the $S_F^4$s for the first five lowest matrix levels. The LEA's obtained in this manner have potentials bounded from below, which indicates that the equivariant fluctuations about the $S_F^4$ configurations with a tachyonic mass term ($\mu^2= -8$)do not lead to any instabilities. We have showed through detailed numerical computations that  these reduced systems have chaotic dynamics and exhibited its various features. In particular, based on the numerical calculations of the Lyapunov spectrum we deterined how the LLE behaves as a function of energy, and also were able to comment on the aspects of the onset of chaos in these models. In the Euclidean signature, we have demonstrated that the LEAs support the usual kink type solutions, i.e. instantons in $1+0$-dimensions. The latter may be viewed as the residual topologically non-trivial configurations, linked to the topological fluxes   penetrating the concentric $S_F^4$s due to the equivariance conditions, and preventing them to shrink to zero radius .     
	
	\vskip 2em
	
	{\bf \large Acknowledgements}
	
	\vskip 1em
	
	We thank O. Oktay for discussions and sharing his computer code with us for the calculation of the Lyapunov exponents. We also thank K. Baskan for his help on our computer code at various instances. S.K. thanks H. Steinacker for discussions during the annual workshop of COST Action MP1405 held in Sofia, Bulgaria, February 2018. The work of S.K. and G.C.Toga is supported by the Middle East Technical University under Project GAP-105-2018-2809.
	
	\vskip 1em
	
	\bibliography{articleref}

\providecommand{\href}[2]{#2}\begingroup\raggedright\begin{thebibliography}{10}

\bibitem{Banks:1996vh}
T.~Banks, W.~Fischler, S.~H. Shenker and L.~Susskind, \emph{{M theory as a
  matrix model: A Conjecture}},
  \href{https://doi.org/10.1103/PhysRevD.55.5112}{\emph{Phys. Rev.} {\bfseries
  D55} (1997) 5112--5128},
  [\href{https://arxiv.org/abs/hep-th/9610043}{{\ttfamily hep-th/9610043}}].

\bibitem{Berenstein:2002jq}
D.~E. Berenstein, J.~M. Maldacena and H.~S. Nastase, \emph{{Strings in flat
  space and pp waves from N=4 superYang-Mills}},
  \href{https://doi.org/10.1088/1126-6708/2002/04/013}{\emph{JHEP} {\bfseries
  04} (2002) 013}, [\href{https://arxiv.org/abs/hep-th/0202021}{{\ttfamily
  hep-th/0202021}}].

\bibitem{Dasgupta:2002hx}
K.~Dasgupta, M.~M. Sheikh-Jabbari and M.~Van~Raamsdonk, \emph{{Matrix
  perturbation theory for M theory on a PP wave}},
  \href{https://doi.org/10.1088/1126-6708/2002/05/056}{\emph{JHEP} {\bfseries
  05} (2002) 056}, [\href{https://arxiv.org/abs/hep-th/0205185}{{\ttfamily
  hep-th/0205185}}].

\bibitem{Ydri:2017ncg}
B.~Ydri, \emph{{Review of M(atrix)-Theory, Type IIB Matrix Model and Matrix
  String Theory}},  \href{https://arxiv.org/abs/1708.00734}{{\ttfamily
  1708.00734}}.

\bibitem{kiritsis2011string}
E.~Kiritsis, \emph{String Theory in a Nutshell}.
\newblock In a Nutshell. Princeton University Press, 2011.

\bibitem{Ydri:2016dmy}
B.~Ydri, \emph{{Lectures on Matrix Field Theory}},
  \href{https://doi.org/10.1007/978-3-319-46003-1}{\emph{Lect. Notes Phys.}
  {\bfseries 929} (2017) pp.1--352},
  [\href{https://arxiv.org/abs/1603.00924}{{\ttfamily 1603.00924}}].

\bibitem{Iizuka:2013kha}
N.~Iizuka, D.~Kabat, S.~Roy and D.~Sarkar, \emph{{Black Hole Formation in Fuzzy
  Sphere Collapse}},
  \href{https://doi.org/10.1103/PhysRevD.88.044019}{\emph{Phys. Rev.}
  {\bfseries D88} (2013) 044019},
  [\href{https://arxiv.org/abs/1306.3256}{{\ttfamily 1306.3256}}].

\bibitem{Anagnostopoulos:2007fw}
K.~N. Anagnostopoulos, M.~Hanada, J.~Nishimura and S.~Takeuchi, \emph{{Monte
  Carlo studies of supersymmetric matrix quantum mechanics with sixteen
  supercharges at finite temperature}},
  \href{https://doi.org/10.1103/PhysRevLett.100.021601}{\emph{Phys. Rev. Lett.}
  {\bfseries 100} (2008) 021601},
  [\href{https://arxiv.org/abs/0707.4454}{{\ttfamily 0707.4454}}].

\bibitem{Catterall:2008yz}
S.~Catterall and T.~Wiseman, \emph{{Black hole thermodynamics from simulations
  of lattice Yang-Mills theory}},
  \href{https://doi.org/10.1103/PhysRevD.78.041502}{\emph{Phys. Rev.}
  {\bfseries D78} (2008) 041502},
  [\href{https://arxiv.org/abs/0803.4273}{{\ttfamily 0803.4273}}].

\bibitem{Hanada:2008ez}
M.~Hanada, Y.~Hyakutake, J.~Nishimura and S.~Takeuchi, \emph{{Higher derivative
  corrections to black hole thermodynamics from supersymmetric matrix quantum
  mechanics}},
  \href{https://doi.org/10.1103/PhysRevLett.102.191602}{\emph{Phys. Rev. Lett.}
  {\bfseries 102} (2009) 191602},
  [\href{https://arxiv.org/abs/0811.3102}{{\ttfamily 0811.3102}}].

\bibitem{Asplund:2011qj}
C.~Asplund, D.~Berenstein and D.~Trancanelli, \emph{{Evidence for fast
  thermalization in the plane-wave matrix model}},
  \href{https://doi.org/10.1103/PhysRevLett.107.171602}{\emph{Phys. Rev. Lett.}
  {\bfseries 107} (2011) 171602},
  [\href{https://arxiv.org/abs/1104.5469}{{\ttfamily 1104.5469}}].

\bibitem{Asplund:2012tg}
C.~T. Asplund, D.~Berenstein and E.~Dzienkowski, \emph{{Large N classical
  dynamics of holographic matrix models}},
  \href{https://doi.org/10.1103/PhysRevD.87.084044}{\emph{Phys. Rev.}
  {\bfseries D87} (2013) 084044},
  [\href{https://arxiv.org/abs/1211.3425}{{\ttfamily 1211.3425}}].

\bibitem{Shenker:2013pqa}
S.~H. Shenker and D.~Stanford, \emph{{Black holes and the butterfly effect}},
  \href{https://doi.org/10.1007/JHEP03(2014)067}{\emph{JHEP} {\bfseries 03}
  (2014) 067}, [\href{https://arxiv.org/abs/1306.0622}{{\ttfamily 1306.0622}}].

\bibitem{Gur-Ari:2015rcq}
G.~Gur-Ari, M.~Hanada and S.~H. Shenker, \emph{{Chaos in Classical D0-Brane
  Mechanics}}, \href{https://doi.org/10.1007/JHEP02(2016)091}{\emph{JHEP}
  {\bfseries 02} (2016) 091},
  [\href{https://arxiv.org/abs/1512.00019}{{\ttfamily 1512.00019}}].

\bibitem{Berenstein:2016zgj}
D.~Berenstein and D.~Kawai, \emph{{Smallest matrix black hole model in the
  classical limit}},
  \href{https://doi.org/10.1103/PhysRevD.95.106004}{\emph{Phys. Rev.}
  {\bfseries D95} (2017) 106004},
  [\href{https://arxiv.org/abs/1608.08972}{{\ttfamily 1608.08972}}].

\bibitem{Maldacena:2015waa}
J.~Maldacena, S.~H. Shenker and D.~Stanford, \emph{{A bound on chaos}},
  \href{https://doi.org/10.1007/JHEP08(2016)106}{\emph{JHEP} {\bfseries 08}
  (2016) 106}, [\href{https://arxiv.org/abs/1503.01409}{{\ttfamily
  1503.01409}}].

\bibitem{Berkowitz:2016jlq}
E.~Berkowitz, E.~Rinaldi, M.~Hanada, G.~Ishiki, S.~Shimasaki and P.~Vranas,
  \emph{{Precision lattice test of the gauge/gravity duality at large-$N$}},
  \href{https://doi.org/10.1103/PhysRevD.94.094501}{\emph{Phys. Rev.}
  {\bfseries D94} (2016) 094501},
  [\href{https://arxiv.org/abs/1606.04951}{{\ttfamily 1606.04951}}].

\bibitem{Sekino:2008he}
Y.~Sekino and L.~Susskind, \emph{{Fast Scramblers}},
  \href{https://doi.org/10.1088/1126-6708/2008/10/065}{\emph{JHEP} {\bfseries
  10} (2008) 065}, [\href{https://arxiv.org/abs/0808.2096}{{\ttfamily
  0808.2096}}].

\bibitem{Filev:2015hia}
V.~G. Filev and D.~O'Connor, \emph{{The BFSS model on the lattice}},
  \href{https://doi.org/10.1007/JHEP05(2016)167}{\emph{JHEP} {\bfseries 05}
  (2016) 167}, [\href{https://arxiv.org/abs/1506.01366}{{\ttfamily
  1506.01366}}].

\bibitem{Filev:2015cmz}
V.~G. Filev and D.~O'Connor, \emph{{A Computer Test of Holographic Flavour
  Dynamics}}, \href{https://doi.org/10.1007/JHEP05(2016)122}{\emph{JHEP}
  {\bfseries 05} (2016) 122},
  [\href{https://arxiv.org/abs/1512.02536}{{\ttfamily 1512.02536}}].

\bibitem{Rinaldi:2017mjl}
E.~Rinaldi, E.~Berkowitz, M.~Hanada, J.~Maltz and P.~Vranas, \emph{{Toward
  Holographic Reconstruction of Bulk Geometry from Lattice Simulations}},
  \href{https://doi.org/10.1007/JHEP02(2018)042}{\emph{JHEP} {\bfseries 02}
  (2018) 042}, [\href{https://arxiv.org/abs/1709.01932}{{\ttfamily
  1709.01932}}].

\bibitem{Asano:2015eha}
Y.~Asano, D.~Kawai and K.~Yoshida, \emph{{Chaos in the BMN matrix model}},
  \href{https://doi.org/10.1007/JHEP06(2015)191}{\emph{JHEP} {\bfseries 06}
  (2015) 191}, [\href{https://arxiv.org/abs/1503.04594}{{\ttfamily
  1503.04594}}].

\bibitem{Arefeva:1997oyf}
I.~{\relax Ya}. Aref'eva, P.~B. Medvedev, O.~A. Rytchkov and I.~V. Volovich,
  \emph{{Chaos in M(atrix) theory}},
  \href{https://doi.org/10.1016/S0960-0779(98)00159-3}{\emph{Chaos Solitons
  Fractals} {\bfseries 10} (1999) 213--223},
  [\href{https://arxiv.org/abs/hep-th/9710032}{{\ttfamily hep-th/9710032}}].

\bibitem{Hubener:2014pfa}
R.~Hübener, Y.~Sekino and J.~Eisert, \emph{{Equilibration in low-dimensional
  quantum matrix models}},
  \href{https://doi.org/10.1007/JHEP04(2015)166}{\emph{JHEP} {\bfseries 04}
  (2015) 166}, [\href{https://arxiv.org/abs/1403.1392}{{\ttfamily 1403.1392}}].

\bibitem{Castelino:1997rv}
J.~Castelino, S.~Lee and W.~Taylor, \emph{{Longitudinal five-branes as four
  spheres in matrix theory}},
  \href{https://doi.org/10.1016/S0550-3213(98)00291-0}{\emph{Nucl. Phys.}
  {\bfseries B526} (1998) 334--350},
  [\href{https://arxiv.org/abs/hep-th/9712105}{{\ttfamily hep-th/9712105}}].

\bibitem{Kimura:2002nq}
Y.~Kimura, \emph{{Noncommutative gauge theory on fuzzy four sphere and matrix
  model}}, \href{https://doi.org/10.1016/S0550-3213(02)00469-8}{\emph{Nucl.
  Phys.} {\bfseries B637} (2002) 177--198},
  [\href{https://arxiv.org/abs/hep-th/0204256}{{\ttfamily hep-th/0204256}}].

\bibitem{Steinacker:2015dra}
H.~C. Steinacker, \emph{{One-loop stabilization of the fuzzy four-sphere via
  softly broken SUSY}},
  \href{https://doi.org/10.1007/JHEP12(2015)115}{\emph{JHEP} {\bfseries 12}
  (2015) 115}, [\href{https://arxiv.org/abs/1510.05779}{{\ttfamily
  1510.05779}}].

\bibitem{Maldacena:2018vsr}
J.~Maldacena and A.~Milekhin, \emph{{To gauge or not to gauge?}},
  \href{https://doi.org/10.1007/JHEP04(2018)084}{\emph{JHEP} {\bfseries 04}
  (2018) 084}, [\href{https://arxiv.org/abs/1802.00428}{{\ttfamily
  1802.00428}}].

\bibitem{Berkowitz:2018qhn}
E.~Berkowitz, M.~Hanada, E.~Rinaldi and P.~Vranas, \emph{{Gauged And Ungauged:
  A Nonperturbative Test}},
  \href{https://doi.org/10.1007/JHEP06(2018)124}{\emph{JHEP} {\bfseries 06}
  (2018) 124}, [\href{https://arxiv.org/abs/1802.02985}{{\ttfamily
  1802.02985}}].

\bibitem{Azuma:2004yg}
T.~Azuma, S.~Bal, K.~Nagao and J.~Nishimura, \emph{{Absence of a fuzzy S**4
  phase in the dimensionally reduced 5-D Yang-Mills-Chern-Simons model}},
  \href{https://doi.org/10.1088/1126-6708/2004/07/066}{\emph{JHEP} {\bfseries
  07} (2004) 066}, [\href{https://arxiv.org/abs/hep-th/0405096}{{\ttfamily
  hep-th/0405096}}].

\bibitem{Grosse:1996mz}
H.~Grosse, C.~Klimcik and P.~Presnajder, \emph{{On finite 4-D quantum field
  theory in noncommutative geometry}},
  \href{https://doi.org/10.1007/BF02099720}{\emph{Commun. Math. Phys.}
  {\bfseries 180} (1996) 429--438},
  [\href{https://arxiv.org/abs/hep-th/9602115}{{\ttfamily hep-th/9602115}}].

\bibitem{Ramgoolam:2001zx}
S.~Ramgoolam, \emph{{On spherical harmonics for fuzzy spheres in diverse
  dimensions}},
  \href{https://doi.org/10.1016/S0550-3213(01)00315-7}{\emph{Nucl. Phys.}
  {\bfseries B610} (2001) 461--488},
  [\href{https://arxiv.org/abs/hep-th/0105006}{{\ttfamily hep-th/0105006}}].

\bibitem{Abe:2004sa}
Y.~Abe, \emph{{Construction of fuzzy S**4}},
  \href{https://doi.org/10.1103/PhysRevD.70.126004}{\emph{Phys. Rev.}
  {\bfseries D70} (2004) 126004},
  [\href{https://arxiv.org/abs/hep-th/0406135}{{\ttfamily hep-th/0406135}}].

\bibitem{Balachandran:2005ew}
A.~P. Balachandran, S.~Kurkcuoglu and S.~Vaidya, \emph{{Lectures on fuzzy and
  fuzzy SUSY physics}},  \href{https://arxiv.org/abs/hep-th/0511114}{{\ttfamily
  hep-th/0511114}}.

\bibitem{Ydri:2016osu}
B.~Ydri, A.~Rouag and K.~Ramda, \emph{{Emergent fuzzy geometry and fuzzy
  physics in four dimensions}},
  \href{https://doi.org/10.1016/j.nuclphysb.2017.01.023}{\emph{Nucl. Phys.}
  {\bfseries B916} (2017) 567--606},
  [\href{https://arxiv.org/abs/1607.08296}{{\ttfamily 1607.08296}}].

\bibitem{Harland:2009kg}
D.~Harland and S.~Kurkcuoglu, \emph{{Equivariant reduction of Yang-Mills theory
  over the fuzzy sphere and the emergent vortices}},
  \href{https://doi.org/10.1016/j.nuclphysb.2009.06.031}{\emph{Nucl. Phys.}
  {\bfseries B821} (2009) 380--398},
  [\href{https://arxiv.org/abs/0905.2338}{{\ttfamily 0905.2338}}].

\bibitem{Kurkcuoglu:2010sn}
S.~Kurkcuoglu, \emph{{Noncommutative Vortices and Flux-Tubes from Yang-Mills
  Theories with Spontaneously Generated Fuzzy Extra Dimensions}},
  \href{https://doi.org/10.1103/PhysRevD.82.105010}{\emph{Phys. Rev.}
  {\bfseries D82} (2010) 105010},
  [\href{https://arxiv.org/abs/1009.1880}{{\ttfamily 1009.1880}}].

\bibitem{Kurkcuoglu:2012vf}
S.~Kurkcuoglu, \emph{{Equivariant Reduction of U(4) Gauge Theory over $S_F^2 x
  S_F^2$ and the Emergent Vortices}},
  \href{https://doi.org/10.1103/PhysRevD.85.105004}{\emph{Phys. Rev.}
  {\bfseries D85} (2012) 105004},
  [\href{https://arxiv.org/abs/1201.0728}{{\ttfamily 1201.0728}}].

\bibitem{Kurkcuoglu:2015qha}
S.~Kurkcuoglu, \emph{{New fuzzy extra dimensions from $SU({\cal N})$ gauge
  theories}}, \href{https://doi.org/10.1103/PhysRevD.92.025022}{\emph{Phys.
  Rev.} {\bfseries D92} (2015) 025022},
  [\href{https://arxiv.org/abs/1504.02524}{{\ttfamily 1504.02524}}].

\bibitem{Kurkcuoglu:2015eta}
S.~Kürkçüoğlu and G.~Ünal, \emph{{Equivariant Fields in an $SU({\cal N})$
  Gauge Theory with new Spontaneously Generated Fuzzy Extra Dimensions}},
  \href{https://doi.org/10.1103/PhysRevD.93.105019}{\emph{Phys. Rev.}
  {\bfseries D93} (2016) 105019},
  [\href{https://arxiv.org/abs/1506.04335}{{\ttfamily 1506.04335}}].

\bibitem{Kurkcuoglu:2016gcp}
S.~Kürkçüoğlu and G.~Ünal, \emph{{$U(3)$ gauge theory on fuzzy extra
  dimensions}}, \href{https://doi.org/10.1103/PhysRevD.94.036003}{\emph{Phys.
  Rev.} {\bfseries D94} (2016) 036003},
  [\href{https://arxiv.org/abs/1607.00075}{{\ttfamily 1607.00075}}].

\bibitem{Aschieri:2006uw}
P.~Aschieri, T.~Grammatikopoulos, H.~Steinacker and G.~Zoupanos,
  \emph{{Dynamical generation of fuzzy extra dimensions, dimensional reduction
  and symmetry breaking}},
  \href{https://doi.org/10.1088/1126-6708/2006/09/026}{\emph{JHEP} {\bfseries
  09} (2006) 026}, [\href{https://arxiv.org/abs/hep-th/0606021}{{\ttfamily
  hep-th/0606021}}].

\bibitem{ott2002chaos}
E.~Ott, \emph{Chaos in Dynamical Systems}.
\newblock Cambridge University Press, 2002.

\bibitem{hilborn2000chaos}
R.~Hilborn, \emph{Chaos and Nonlinear Dynamics: An Introduction for Scientists
  and Engineers}.
\newblock Oxford University Press, 2000.

\bibitem{coleman1988aspects}
S.~Coleman, \emph{Aspects of Symmetry: Selected Erice Lectures}.
\newblock Cambridge University Press, 1988.

\bibitem{Rajaraman:1982is}
R.~Rajaraman, \emph{{Solitons and Instantons. An Introduction to Solitons and
  Instantons in Quatum Field Theory}}.
\newblock 1982.

\end{thebibliography}\endgroup
	\bibliographystyle{JHEP}

	\appendix
	\section{Some Formulas on Representation Theory}
	
	\subsection{Branching Rules \& Relations Among Dynkin and Heights Weight Labels}
	
	Irreducible representations of $SO(2k)$ and $SO(2k-1)$can be given in terms of the highest weight labels $[\lambda]\equiv(\lambda_1 \,, \lambda_2 \,, \cdots \,, \lambda_{k-1} \,, \lambda_k)$ and $[\mu]\equiv(\mu_1 \,, \mu_2 \,, \cdots \,, \, \,\mu_{k-1})$ respectively. Branching of the IRR $[\lambda]$ of $SO(2k)$ under $SO(2k-1)$ IRRs follows from the
	rule 
	\be
	\lambda_1 \geq \mu_1 \geq \lambda_2 \geq \mu_2 \geq \cdots \geq \mu_{k-1} \geq |\lambda_k| \,,
	\ee
	
	The relationship between Dykin labels and highest weight labels  for $SO(5)$ IRRs  is
	\be
	(p,q)_{Dynkin} \equiv (\lambda_1,\lambda_2)_{HW} \,,
	\ee
	with 
	\be
	\lambda_1=\frac{p+q}{2}  \hspace{1cm}\lambda_2=\frac{q}{2} \,.
	\ee
	
	For the $SO(6)$ IRRs the correspondence is given by
	\be
	(p,q,r)_{Dynkin} \equiv (\lambda_1,\lambda_2,\lambda_3)_{HW} \,,
	\ee
	with
	\be
	\lambda_1=q + \frac{p+r}{2} \hspace{1cm}\lambda_2=\frac{p+r}{2} \hspace{1cm} \lambda_3=\frac{p-r}{2} \,.
	\ee
	
	\subsection{Computation of $(G\cdot \Sigma)^2$}
	
	Here we present the details of the calculation leading to (\ref{Gabcomlim}). We have
	\beqa
	(G\cdot \Sigma)^2 &=& G_{ij}G_{kl} \Sigma_{ij} \Sigma_{kl} \nn \\ 
	&=& \gamma_i \gamma_j \gamma_k \gamma_l G_{ij}G_{kl} \,.
	\label{GS2}
	\eeqa
	Multiplying both sides of (\ref{GS2}) with $\epsilon^{abcdm}$ and using $\epsilon^{ijklm}\gamma_i \gamma_j \gamma_k \gamma_l=24\gamma_m$ as can be inferred from (\ref{epsilon1}) one obtains 
	\be
	\epsilon^{abcdm} \epsilon^{ijklm}\gamma_i \gamma_j \gamma_k \gamma_l G_{ab} G_{cd}=24\epsilon^{abcdm}\gamma_m G_{ab} G_{cd}
	\label{GS3}
	\ee	
	To handle the left hand side of (\ref{GS3}) we make use of the determinant 
	\begin{align}
	\epsilon^{abcdm} \epsilon^{ijklm}= \left|%
	\begin{array}{cccc}
	\delta^{ai} & \delta^{aj} & \delta^{ak} & \delta^{al} \\
	\delta^{bi} & \delta^{bj} & \delta^{bk} & \delta^{bl} \\
	\delta^{ci} & \delta^{cj }& \delta^{ck} & \delta^{cl} \\
	\delta^{di} & \delta^{dj} & \delta^{dk} & \delta^{dl} 
	\end{array}%
	\right|
	\end{align}
	Left hand side of (\ref{GS3}) then reads
	\beqa
	&& 4\gamma_a \gamma_b \gamma_c \gamma_d G_{ab}G_{cd} + 4\gamma_c \gamma_d \gamma_a \gamma_b G_{ab}G_{cd} + 4\gamma_a \gamma_c \gamma_d \gamma_b G_{ab}G_{cd} \nonumber \\
	&& +4\gamma_a \gamma_d \gamma_b \gamma_c G_{ab}G_{cd} + 4\gamma_c \gamma_a \gamma_b \gamma_d G_{ab}G_{cd}+ 4\gamma_d \gamma_a \gamma_c \gamma_b G_{ab}G_{cd} \nn \\
	&&= 4(6\gamma_a \gamma_b \gamma_c \gamma_d G_{ab}G_{cd} - 12G_{ab}G_{ab}+ 24\gamma_a \gamma_c G_{ab}G_{cb}) \,,
	\eeqa
	where the second line follows after making use of $\lbrace \Gamma_a \,, \Gamma_b \rbrace = 2 \delta_{ab}$ for rearrangements. Therefore, (\ref{GS3}) can be cast into the form
	\be
	(G\cdot \Sigma)^2 = \gamma_a \gamma_b \gamma_c \gamma_d G_{ab} G_{cd} \nn = \epsilon^{abcdm}\gamma_m G_{ab} G_{cd}+2G_{ab} G_{ab}-4\gamma_a \gamma_c G_{ab} G_{cb} \,.
	\label{GS4}
	\ee	
	Employing $G_{ab}=\frac{1}{2}[X_a ,X_b]$, we can expand the first term in r.h.s of (\ref{GS4}) as 
	\beqa
	\epsilon^{abcdm}\gamma_m G_{ab} G_{cd} &=& \frac{1}{4}\epsilon^{abcdm} \gamma_m \left(X_a X_b X_c X_d- X_a X_b X_d X_c-X_b X_a X_c X_d +X_b X_a X_d X_c  \right) \nn \\
	&=&\epsilon^{abcdm}\gamma_m  X_a X_b X_d X_c \nn \\
	&=& 8(n+2)X_m \gamma_m \,.
	\eeqa	 
	after using (\ref{epsilon1}). Substituting $G_{ab}G_{ab}=- 4n(n+4) {\mathbbm 1}_{4N}$ for the second term and simplifying the third term as
	\beqa
	-4 \gamma_a \gamma_c G_{ab}G_{cb} &=& -2 \gamma_a \gamma_c G_{ab}G_{cb} - 2 \gamma_a \gamma_c G_{ab}G_{cb} \nonumber \\
	& = & - 2 \gamma_a \gamma_c G_{ab}G_{cb} -4 G_{ab}G_{ab} + 2 \gamma_a \gamma_c G_{cb}G_{ab} \nonumber \\
	& = & -2 \gamma_a \gamma_c [G_{ab}, G_{cb}] - 4 G_{ab}G_{ab} \nonumber \\
	&=& 12\gamma_a \gamma_c G_{ac} - 4G_{ab}G_{ab}
	\eeqa
	and finally, combining all the terms together we get 
	\be
	(G \cdot \Sigma)^2 = 12 \, \Gamma_a \Gamma_b G_{ab}  + 8 (n+2) X_a \Gamma_a + 8n(n+4) \mathbbm{1}_{4N} \,.
	\label{Gabcomlim1}
	\ee
	
	\section{Details on the Dimensional Reduction of ${\cal S}_{1}$}	
	
	\subsection{Useful Identities}
	
	Some useful identities among $Q_1$, $Q_3$ and $X_a$, which simplify the analytic calculations are listed below:  
	\be
	\begin{split}
		[Q_1,[X_a,Q_3]] &=0 \,, \quad  [Q_3,[X_a,Q_1]] =0 \,, \\
		[Q_1,Q_3[X_a,Q_3]] &=0 \,, \quad [Q_3,Q_1[X_a,Q_1]] = 0 \,, \\
		[Q_3,Q_1[X_a,Q_1]] &=0 \,, \quad [Q_1,\{X_a,Q_2\}] =0 \,, \\
		[Q_1,\{X_a,Q_1\}-Q_3[X_a,Q_3]] &=0 \,, \quad [Q_3,\{X_a,Q_1\}-Q_3[X_a,Q_3]] =0 \,, \\
		[Q_1,\{X_a,Q_3\}-Q_1[X_a,Q_1]]& =0 \,, \quad [Q_3,\{X_a,Q_3\}-Q_1[X_a,Q_1]] =0 \,. 
	\end{split}
	\label{identities}	
	\ee	
	
	\subsection{Intermediate Forms of ${\cal D}_t {\cal X}_a$}
	
	Two intermediate steps in obtaining (\ref{D03}) may be listed as follows. Substituting the parametrizations in (\ref{A0par}) and (\ref{vectorinv2}) in ${\cal D}_t {\cal X}_a = \del_t {\cal X}_a - i [{\cal A}, {\cal X}_a]$, we find 
	\be
	\begin{split}
		{\cal D}_t X_a=&i\frac{\partial_t\phi_1}{2}\comm{X_a}{Q_1}+i\frac{\partial_t\chi_1}{2}\comm{X_a}{Q_3}\\
		&+\frac{\partial_t\phi_2}{2}Q_1\comm{X_a}{Q_1}+\frac{\partial_t\chi_2}{2}Q_3\comm{X_a}{Q_3} 
		X_a\\ &+\partial_t\phi_3(\acomm{{\hat X}_a}{Q_1}-Q_3\comm{{\hat X}_a}{Q_3})+\partial_t\chi_3(\acomm{{\hat X}_a}{Q_3}-Q_1\comm{{\hat X}_a}{Q_1})
		\\ &+\partial_t\phi_4({\hat X}_a+{\hat \Gamma}_a+Q_3\comm{{\hat X}_a}{Q_3})+  \frac{\alpha_1\phi_1}{4}\comm{Q_1}{\comm{X_a}{Q_1}}
		\\&+ \frac{\alpha_1\chi_1}{4}\comm{Q_1}{\comm{X_a}{Q_3}} - i \frac{\alpha_1(\phi_2+1)}{4}\comm{Q_1}{Q_1\comm{X_a}{Q_1}}
		\\ &-i \frac{\alpha_1(\chi_2+1)}{4}\comm{Q_1}{Q_3\comm{X_a}{Q_3}} -i \frac{\alpha_1 \phi_3}{2}\comm{Q_1}{\acomm{{\hat X}_a}{Q_1}-Q_3\comm{{\hat X}_a}{Q_3}}
		\\ & -i \frac{\alpha_1 \chi_3}{2}\comm{Q_1}{\acomm{{\hat X}_a}{Q_3}-Q_1\comm{{\hat X}_a}{Q_1}}- i \frac{\alpha_1 \phi_4}{2}\comm{Q_1}{{\hat X}_a+{\hat \Gamma}_a+Q_3\comm{{\hat X}_a}{Q_3}}
		\\&+ \frac{\alpha_3\phi_1}{2}\comm{Q_3}{\comm{X_a}{Q_1}}+ \frac{\alpha_3 \chi_1}{4}\comm{Q_3}{\comm{X_a}{Q_3}}- i\frac{\alpha_3(\phi_2+1)}{4}\comm{Q_3}{Q_1\comm{X_a}{Q_1}}
		\\&- i\frac{\alpha_3(\chi_2+1)}{4}\comm{Q_3}{Q_3\comm{X_a}{Q_3}}- i \frac{\alpha_3\phi_3}{2}\comm{Q_3}{\acomm{{\hat X}_a}{Q_1}-Q_3\comm{{\hat X}_a}{Q_3}}
		\\&-i \frac{\alpha_3\chi_3}{2}\comm{Q_3}{\acomm{{\hat X}_a}{Q_3}-Q_1\comm{{\hat X}_a}{Q_1}}-i\frac{\alpha_3\phi_4}{2}\comm{Q_3}{{\hat X}_a+{\hat \Gamma}_a+Q_3\comm{{\hat X}_a}{Q_3}}+
		\\ &-i \frac{\alpha_1}{2}\comm{Q_1}{X_a}- i \frac{\alpha_3}{2}\comm{Q_3}{X_a} \,.
	\end{split}
	\ee
	
	With the help of the identities listed in (\ref{identities}), this simplifies to
	\be
	\label{D02}
	\begin{split}
		{\cal D}_tX_a=&\frac{i}{2}(\partial_t\phi_1- \alpha_1\phi_2)\comm{X_a}{Q_1}+\frac{i}{2}(\partial_t\chi_1 - \alpha_3\chi_2)\comm{X_a}{Q_3}
		\\&+\frac{1}{2}(\partial_t\phi_2 + \alpha_1\phi_1)Q_1\comm{X_a}{Q_1}+\frac{1}{2}(\partial_t\chi_2 + \alpha_1\chi_2)Q_3\comm{X_a}{Q_3}
		\\&+\partial_t\phi_3(\acomm{{\hat X}_a}{Q_1}-Q_3\comm{{\hat X}_a}{Q_3})+\partial_t\chi_3(\acomm{{\hat X}_a}{Q_3}-Q_1\comm{{\hat X}_a}{Q_1})
		\\&+\partial_t\phi_4({\hat X}_a+{\hat \Gamma}_a+Q_3\comm{{\hat X}_a}{Q_3}) \,.
	\end{split}
	\ee
	
	\subsection{Explicit form of the Low Energy Reduced Actions}\label{LNS}	
	
	Below we give the equivariantly reduced Lagrangians for $n=3,4,5$:
	
	\be 
	\begin{split}
		L_{(n=3)} = & \frac{1}{2} \left(0.58 \dot{\chi}_1^2+0.92 \dot{\chi}_3^2+6.903 \dot{\phi}_4^2+2.6 \dot{\phi }_1^2+9.8 \dot{\phi}_3^2\right)-1.43 \chi _1^4\\
		&-0.05 \chi _3^4-0.81 \chi _3^3+8.10 \chi _1^2-0.30 \chi _1^2 \chi _3^2-3.63 \chi _3^2+3.34 \chi _1^2 \chi _3\\
		&+0.23 \chi _3-5.25 \chi _1^2 \phi _1^2-0.17 \chi _3^2 \phi _1^2+4.04 \chi _3 \phi _3 \phi _1^2-1.34 \chi _3 \phi _4 \phi _1^2\\
		&-3.56 \chi _3 \phi _1^2-0.447 \chi _3^3 \phi _3+0.128 \chi _3^3 \phi _4-4.09 \chi _1^2 \phi _3^2-1.79 \chi _1^2 \phi _4^2\\
		&+9.83 \chi _1^2 \phi _3-6.24 \chi _1^2 \phi _3 \phi _4+ 4.20 \chi _1^2 \phi _4-1.66 \chi _3^2 \phi _3^2\\
		&-0.13 \chi _3^2 \phi _4^2-4.70 \chi _3^2 \phi _3+0.86 \chi _3^2 \phi _3 \phi _4+1.49 \chi _3^2 \phi _4\\
		&-0.23 \chi _3 \phi _3^3-0.142 \chi _3 \phi _4^3-17.80 \chi _3 \phi _3^2-0.75 \chi _3 \phi _3 \phi _4^2\\
		&+0.34 \chi _3 \phi _4^2-2.30 \chi _1^2 \chi _3 \phi _3-1.96 \chi _1^2 \chi _3 \phi _4-12.48 \chi _3 \phi _3\\
		&+1.08 \chi _3 \phi _3^2 \phi _4+9.19 \chi _3 \phi _3 \phi _4+1.64 \chi _3 \phi _4-3.94 \phi _1^4-23.22 \phi _3^2 \phi _1^2\\
		&-1.65 \phi _4^2 \phi _1^2++13.14 \phi _1^2+35.3 \phi _3 \phi _1^2+18.84 \phi _3 \phi _4 \phi _1^2-21.27 \phi _4 \phi _1^2\\
		&-8.46 \phi _3^4-1.19 \phi _4^4 + 29.74 \phi _3^3-2.33 \phi _3 \phi _4^3+11.24 \phi _4^3-51.36 \phi _3^2\\
		&-2.91 \phi _3^2 \phi _4^2+16.41 \phi _3 \phi _4^2-29.82 \phi _4^2-45.17 \phi _3+3.54 \phi _3^3 \phi _4\\
		&+4.90 \phi _3^2 \phi _4-14.55 \phi _3 \phi _4+17.07 \phi _4 - 30.61 \,.
	\end{split}
	\label{l3}
	\ee	
	
	\be
	\begin{split}
		L_{(n=4)} = &\frac{1}{2}\left(1.3 \dot{\chi}_1^2+2.6 \dot{\phi}_1^2+8.38 \dot{\phi}_3^2+6.77 \dot{\phi}_4^2+0.88 \dot{\chi}_3^2\right)-1.66 \chi _1^4\\
		&-1.66 \chi _1^4-0.02 \chi _3^4-0.39 \chi _3^3+8.81 \chi _1^2-0.29\chi _1^2 \chi _3^2-2.75 \chi _3^2\\
		&+3.47 \chi _1^2 \chi _3+0.25 \chi _3-5.48 \chi _1^2 \phi _1^2-0.18 \chi _3^2 \phi _1^2+3.55 \chi _3 \phi _3 \phi _1^2\\
		&-1.32 \chi _3 \phi _4 \phi _1^2-3.72 \chi _3 \phi _1^2-0.16 \chi _3^3 \phi _3+0.04 \chi _3^3 \phi _4-7.09 \chi _1^2 \phi _3^2\\
		&-0.80 \chi _1^2 \phi _4^2+14.2 \chi _1^2 \phi _3-5.37 \chi _1^2 \phi _3 \phi _4+3.18 \chi _1^2 \phi _4-0.85 \chi _3^2 \phi _3^2\\
		&-0.05 \chi _3^2 \phi _4^2-2.59 \chi _3^2 \phi _3+0.40 \chi _3^2 \phi _3 \phi _4+0.79 \chi _3^2 \phi _4+0.01 \chi _3 \phi _3^3\\
		&-0.02 \chi _3 \phi _4^3-14.3 \chi _3 \phi _3^2-0.33 \chi _3 \phi _3 \phi _4^2-0.18 \chi _3 \phi _4^2-2.92 \chi _1^2 \chi _3 \phi _3\\
		&-1.22 \chi _1^2 \chi _3 \phi _4-10.38 \chi _3 \phi _3+0.40 \chi _3 \phi _3^2 \phi _4+7.01 \chi _3 \phi _3 \phi _4+1.59 \chi _3 \phi _4\\
		&-3.94 \phi _1^4-16.04 \phi _3^2 \phi _1^2-2.03 \phi _4^2 \phi _1^2+13.4 \phi _1^2+28.6 \phi _3 \phi _1^2+14.2 \phi _3 \phi _4 \phi _1^2\\
		&-17.7 \phi _4 \phi _1^2-5.49 \phi _3^4-0.21 \phi _4^4+24.2 \phi _3^3-0.73 \phi _3 \phi _4^3+3.47 \phi _4^3\\
		&-66.4 \phi _3^2-1.99 \phi _3^2 \phi _4^2+9.79 \phi _3 \phi _4^2-16.53 \phi _4^2-42.75 \phi _3\\
		&+2.24 \phi _3^3 \phi _4+5.81 \phi _3^2 \phi _4-11.5 \phi _3 \phi _4+14.5 \phi _4-31 \,.
	\end{split}
	\ee
	
	\be
	\begin{split}
		L_{(n=5)} = & \frac{1}{2}\left(1.5 \dot{\chi}_1^2+2.5 \dot{\phi}_1^2+7.549 \dot{\phi}_3^2+6.537 \dot{\phi}_4^2+0.83 \dot{\chi}_3^2\right)-1.85 \chi _1^4\\
		&-1.85 \chi _1^4-0.006 \chi _3^4-0.22 \chi _3^3+9.31 \chi _1^2-0.26 \chi _1^2 \chi _3^2-2.20 \chi _3^2+3.45 \chi _1^2 \chi _3\\
		&+0.25 \chi _3-5.63 \chi _1^2 \phi _1^2-0.18 \chi _3^2 \phi _1^2+2.8  \chi _3 \phi _3 \phi _1^2-0.97 \chi _3 \phi _4 \phi _1^2
		-3.69 \chi _3 \phi _1^2\\
		&-0.07 \chi _3^3 \phi _3+0.02 \chi _3^3 \phi _4-10.4 \chi _1^2 \phi _3^2-0.19 \chi _1^2 \phi _4^2+18.01 \chi _1^2 \phi _3\\
		&-3.23 \chi _1^2 \phi _3 \phi _4+1.60 \chi _1^2 \phi _4-0.45 \chi _3^2 \phi _3^2-0.02 \chi _3^2 \phi _4^2\\
		&-1.45 \chi _3^2 \phi _3+0.17 \chi _3^2 \phi _3 \phi _4+0.37 \chi _3^2 \phi _4+0.10 \chi _3 \phi _3^3+0.0003 \chi _3 \phi _4^3\\
		&-10.40 \chi _3 \phi _3^2-0.10 \chi _3 \phi _3 \phi _4^2-0.20 \chi _3 \phi _4^2-3.37 \chi _1^2 \chi _3 \phi _3-0.58 \chi _1^2 \chi _3 \phi _4\\
		&-8.29 \chi _3 \phi _3+0.07 \chi _3 \phi _3^2 \phi _4+4.34 \chi _3 \phi _3 \phi _4+1.16 \chi _3 \phi _4-3.88 \phi _1^4\\
		&-9.66 \phi _3^2 \phi _1^2-1.20 \phi _4^2 \phi _1^2 +13.4 \phi _1^2+20.9 \phi _3 \phi _1^2+8.14 \phi _3 \phi _4 \phi _1^2\\
		&-11.95 \phi _4 \phi _1^2-3.66 \phi _3^4-0.02 \phi _4^4+22.15 \phi _3^3-0.12 \phi _3 \phi _4^3+0.61 \phi _4^3-77.93 \phi _3^2\\
		&-0.90 \phi _3^2 \phi _4^2+3.98 \phi _3 \phi _4^2-6.40 \phi _4^2-38.93 \phi _3+0.883 \phi _3^3 \phi _4+5.42 \phi _3^2 \phi _4\\
		&-7.43 \phi _3 \phi _4+10.35 \phi _4-31 \,.
	\end{split}
	\ee
	\subsection{Time Series for Lyapunov exponents for $n=1,2,3,4,5$.}\label{timeseries}
		\begin{figure}[!htb]
			\centering
			\begin{minipage}[t]{0.5\textwidth}	
				\centering
				\includegraphics[width=\linewidth,height=0.25\textheight]{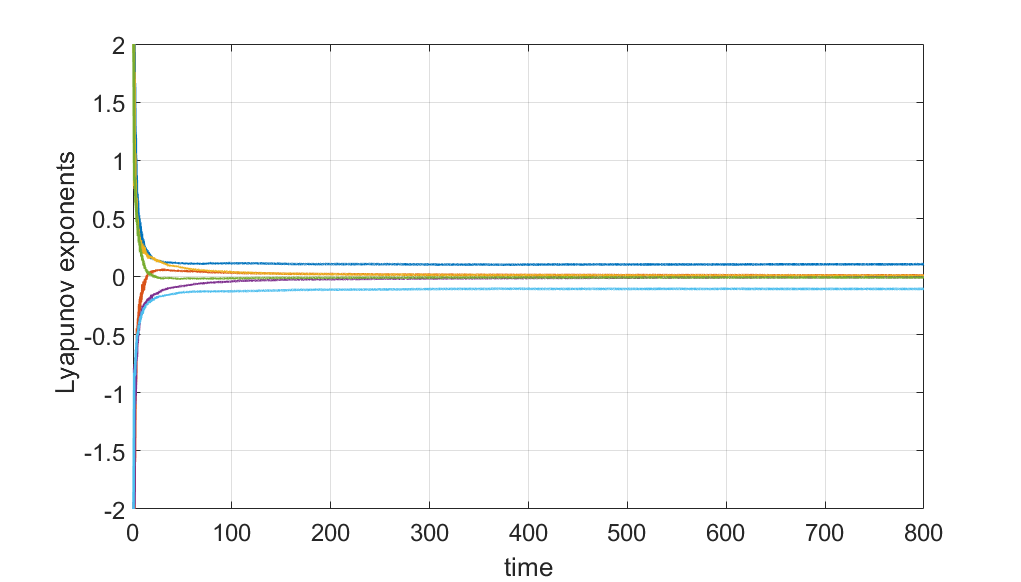}
				\caption{ $n=1$, $E=15$, $LLE=0.094$}
				\label{fig:n1E15}
			\end{minipage}%
			\begin{minipage}[t]{0.5\textwidth}	
				\centering
				\includegraphics[width=\linewidth,height=0.25\textheight]{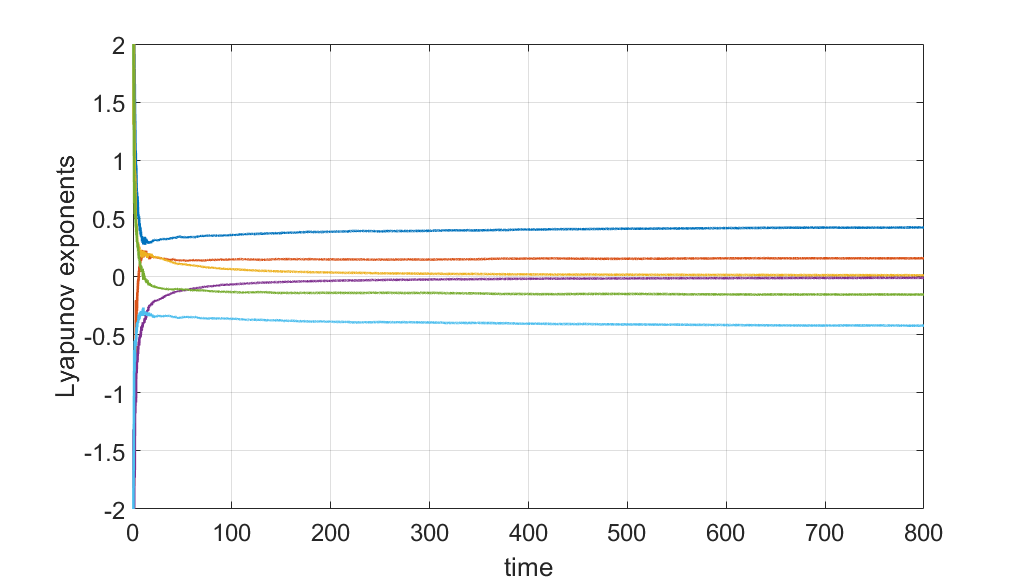}
				\caption{ $n=1$, $E=25$ ,$LLE=0.2788$}
				\label{fig:n1E25}
			\end{minipage}%
		\end{figure}
		\begin{figure}[!htb]
			\centering
			
			\begin{minipage}[t]{0.5\textwidth}	
				\centering
				\includegraphics[width=\linewidth,height=0.25\textheight]{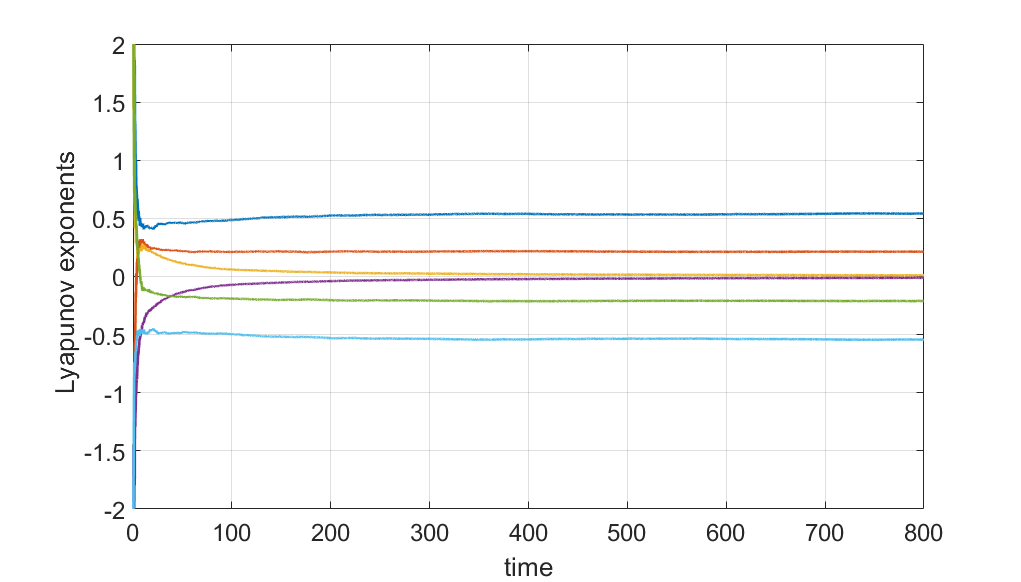}
				\caption{ $n=1$, $E=30$,$LLE = 0.4893$}
				\label{fig:n1E30}
			\end{minipage}
		\end{figure}
		\newpage
		\begin{figure}[!htb]
			\centering
			\begin{minipage}[t]{0.5\textwidth}	
				\centering
				\includegraphics[width=\linewidth,height=0.25\textheight]{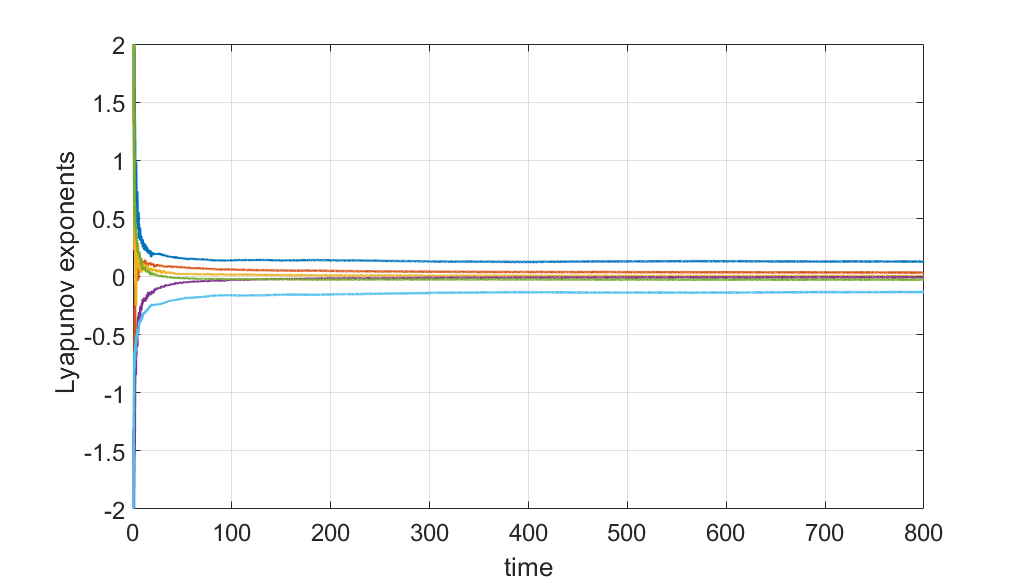}
				\caption{ $n=1$,$E=40$,$LLE=0.6370$}
				\label{fig:n1E40}
			\end{minipage}%
			\begin{minipage}[t]{0.5\textwidth}	
				\centering
				\includegraphics[width=\linewidth,height=0.25\textheight]{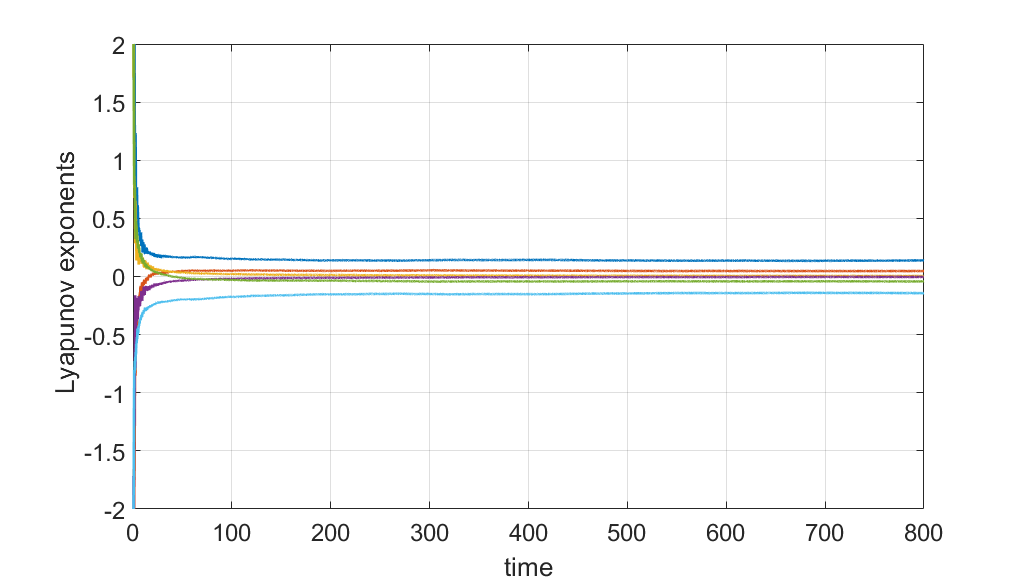}
				\caption{ $n=1$, $E=50$,$LLE=0.7265 $}
				\label{fig:n1E50}
			\end{minipage}
		\end{figure}
		\begin{figure}[!htb]
			\centering
			\includegraphics[width=0.5\linewidth,height=0.25\textheight]{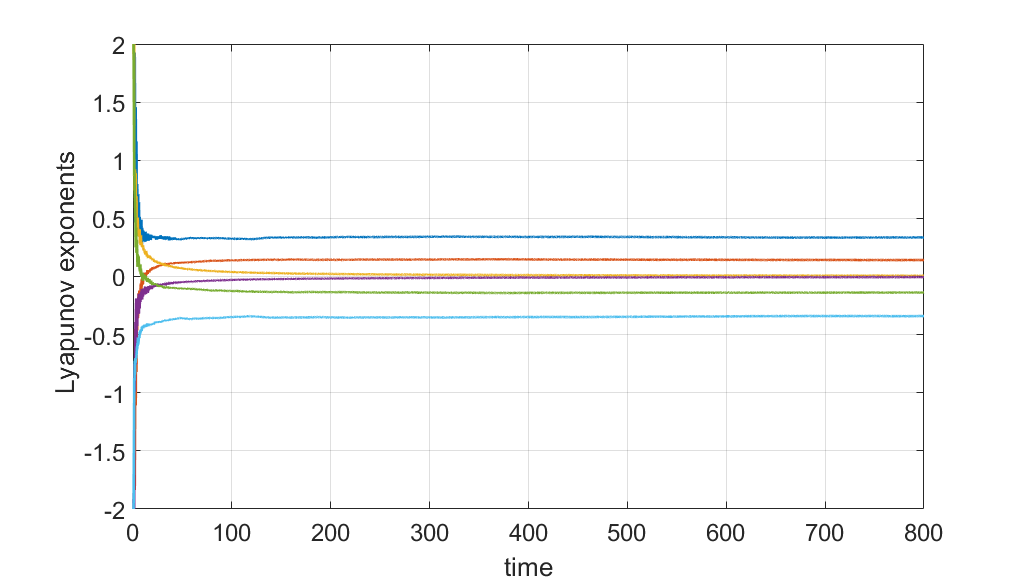}
			\caption{ $n=1$, $E=100$,$LLE=0.9645 $}
			\label{fig:n1E100}
		\end{figure}
		\begin{figure}[!htb]
			\centering
			\begin{minipage}[t]{0.5\textwidth}	
				\centering
				\includegraphics[width=\linewidth,height=0.25\textheight]{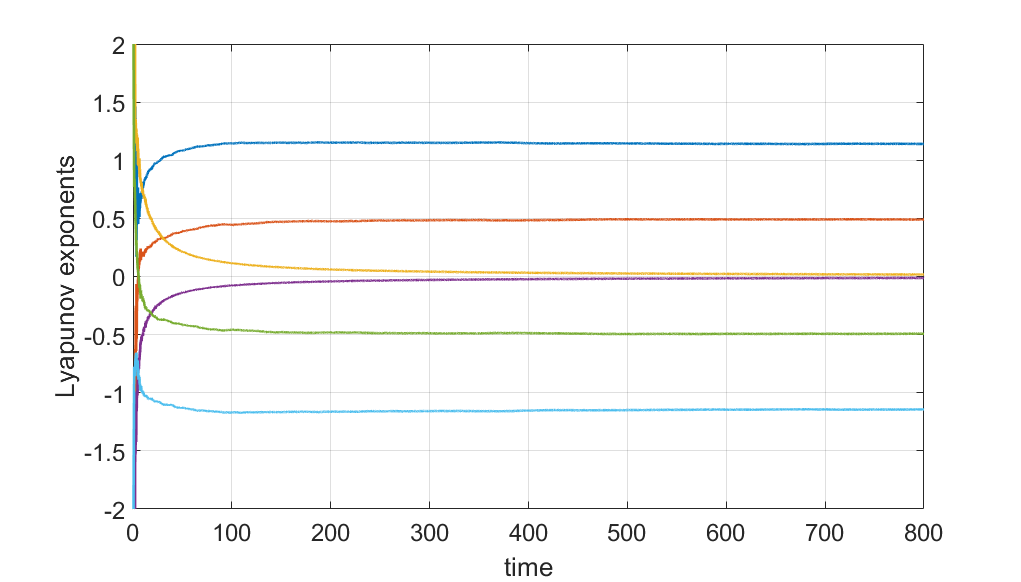}
				\caption{ $n=1$, $E=250$,$LLE=1.099 $}
				\label{fig:n1E250}
			\end{minipage}%
			\begin{minipage}[t]{0.5\textwidth}	
				\centering
				\includegraphics[width=\linewidth,height=0.25\textheight]{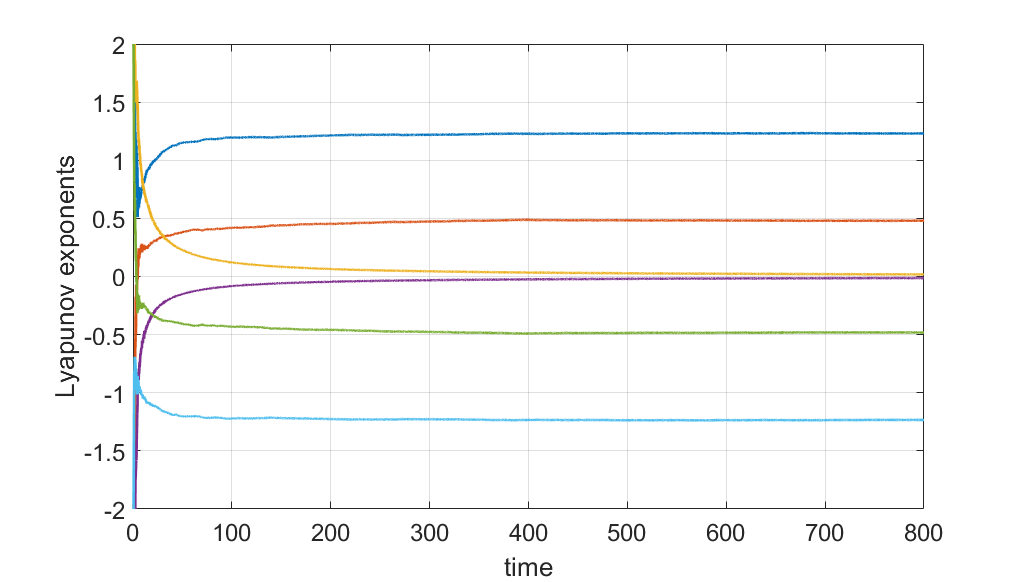}
				\caption{ $n=1$, $E=500$,$LLE= 1.1574 $}
				\label{fig:n1E500}
			\end{minipage}
		\end{figure}
		\newpage
		\begin{figure}[!htb]
			\centering
			\begin{minipage}[t]{0.5\textwidth}	
				\centering
				\includegraphics[width=\linewidth,height=0.25\textheight]{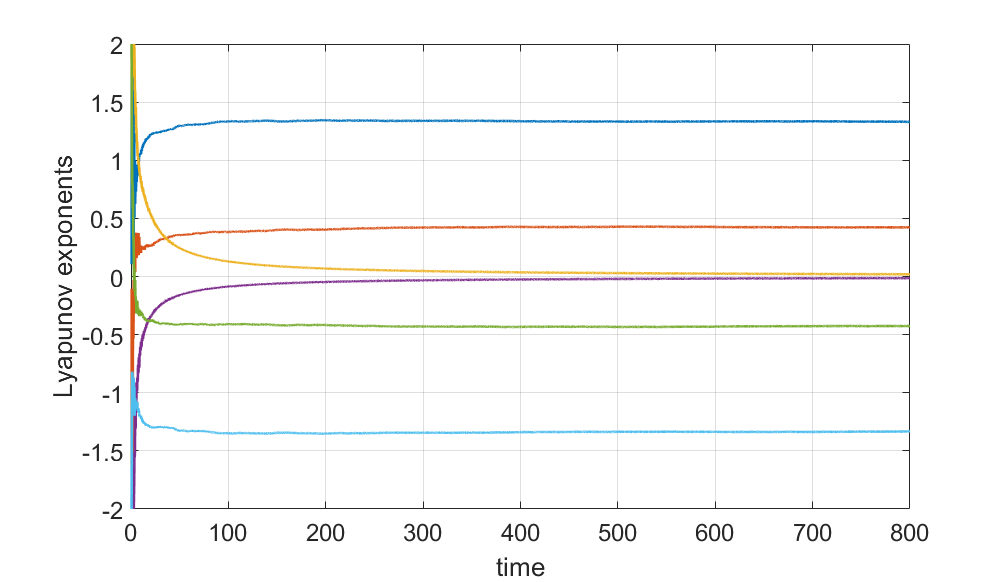}
				\caption{ $n=1$, $E=1000$,$LLE= 1.2138 $}
				\label{fig:n1E1000}
			\end{minipage}%
			\begin{minipage}[t]{0.5\textwidth}	
				\centering
				\includegraphics[width=\linewidth,height=0.25\textheight]{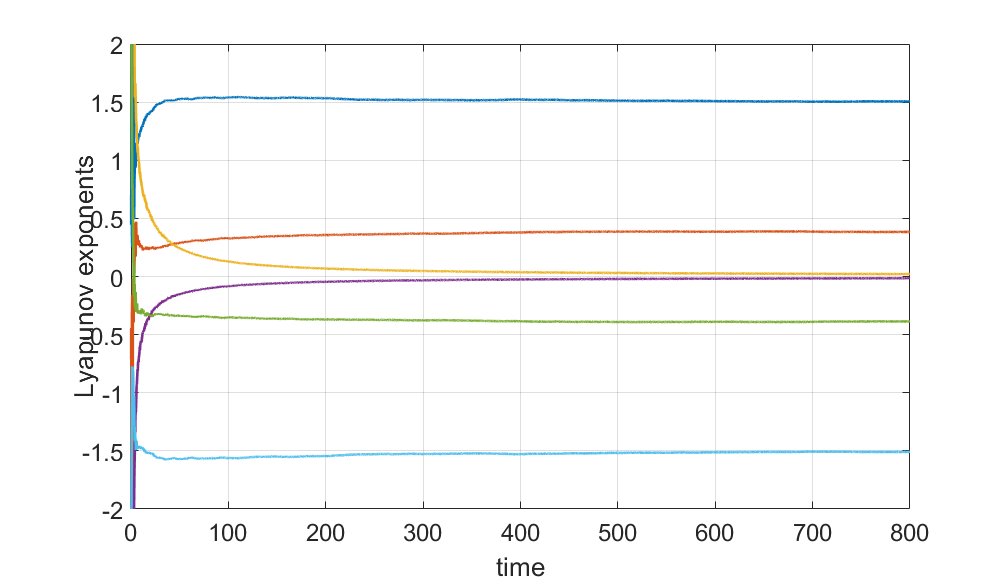}
				\caption{ $n=1$, $E=2000$,$LLE= 1.3087 $}
				\label{fig:n1E2000}
			\end{minipage}
		\end{figure}
		
		\begin{figure}[!htb]
			\centering
			\begin{minipage}[t]{0.5\textwidth}	
				\centering
				\includegraphics[width=\linewidth,height=0.25\textheight]{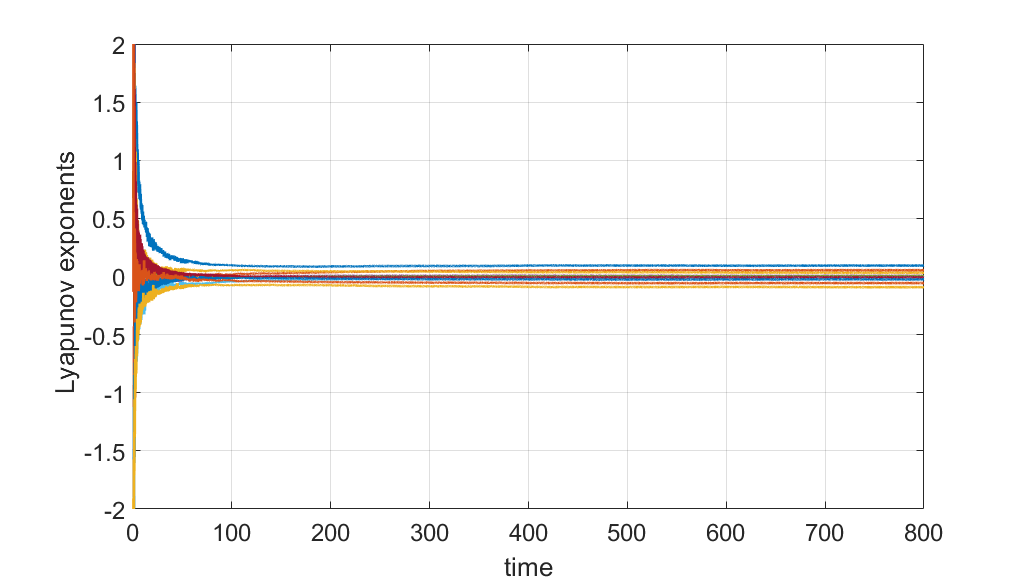}
				\caption{ $n=2$, $E=15$, $LLE=0.094$}
				\label{fig:n2E15}
			\end{minipage}%
			\begin{minipage}[t]{0.5\textwidth}	
				\centering
				\includegraphics[width=\linewidth,height=0.25\textheight]{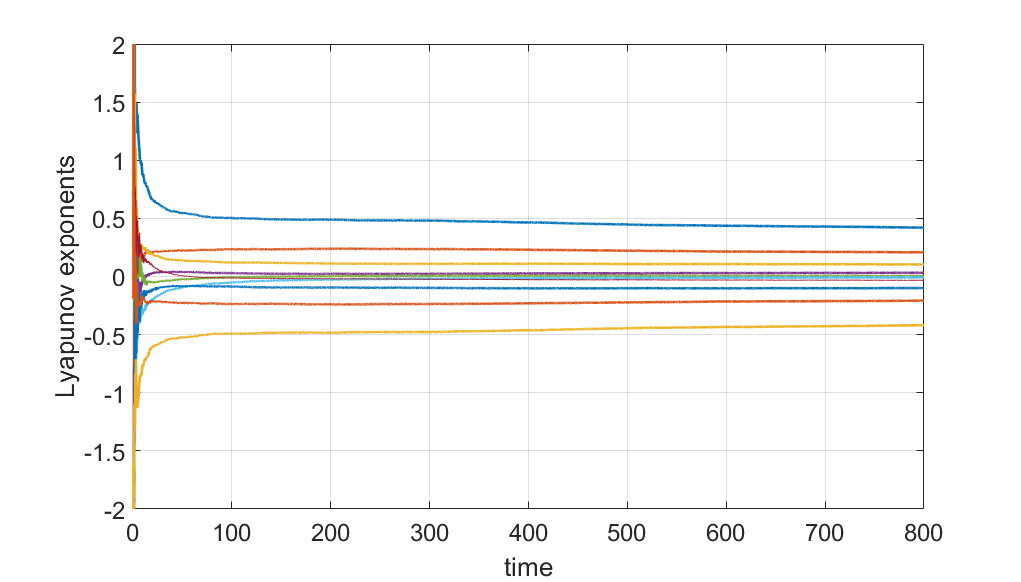}
				\caption{ $n=2$, $E=25$ ,$LLE=0.2788$}
				\label{fig:n2E25}
			\end{minipage}%
		\end{figure}
		\begin{figure}[!htb]
			\centering
			
			\begin{minipage}[t]{0.5\textwidth}	
				\centering
				\includegraphics[width=\linewidth,height=0.25\textheight]{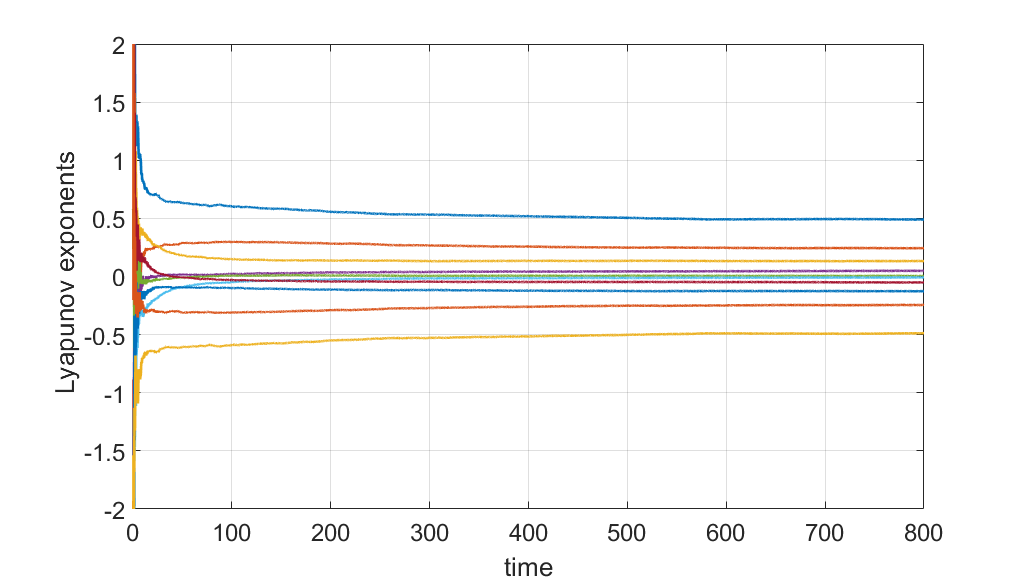}
				\caption{ $n=2$, $E=30$,$LLE = 0.4893$}
				\label{fig:n2E30}
			\end{minipage}
		\end{figure}
		\newpage
		\begin{figure}[!htb]
			\centering
			\begin{minipage}[t]{0.5\textwidth}	
				\centering
				\includegraphics[width=\linewidth,height=0.25\textheight]{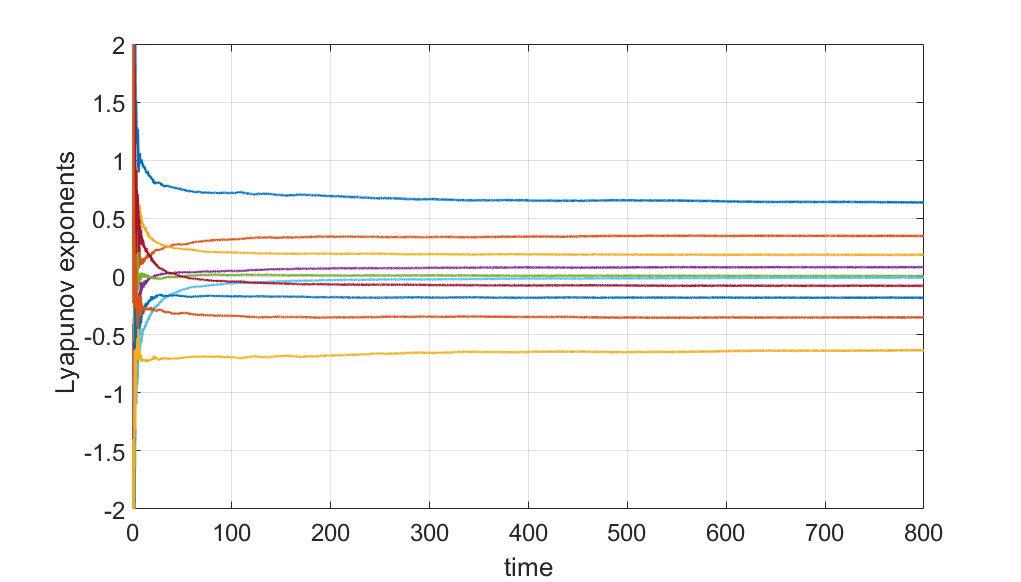}
				\caption{ $n=2$,$E=40$,$LLE=0.6370$}
				\label{fig:n2E40}
			\end{minipage}%
			\begin{minipage}[t]{0.5\textwidth}	
				\centering
				\includegraphics[width=\linewidth,height=0.25\textheight]{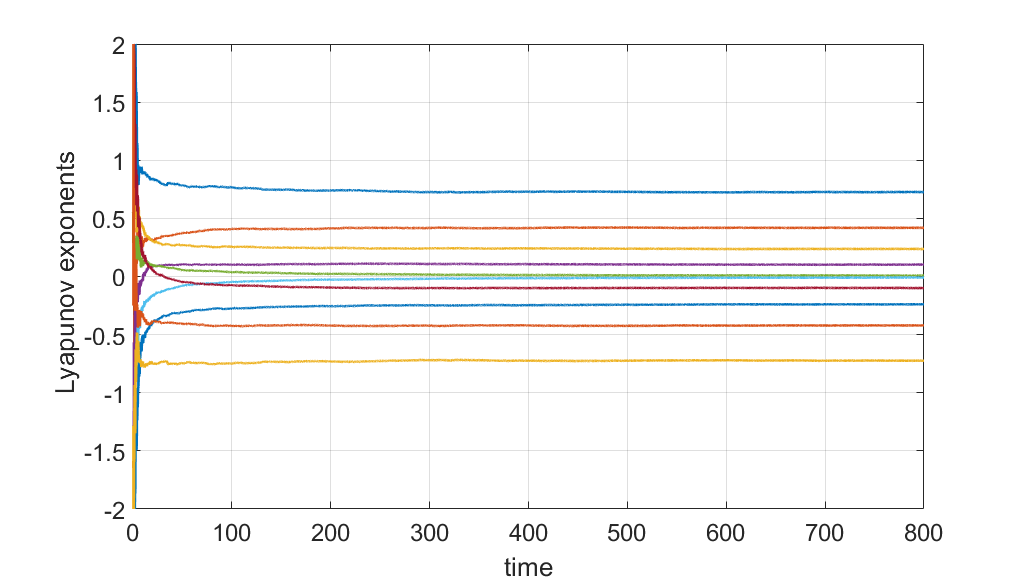}
				\caption{ $n=2$, $E=50$,$LLE=0.7265 $}
				\label{fig:n2E50}
			\end{minipage}
		\end{figure}
		\begin{figure}[!htb]
			\centering
			\includegraphics[width=0.5\linewidth,height=0.25\textheight]{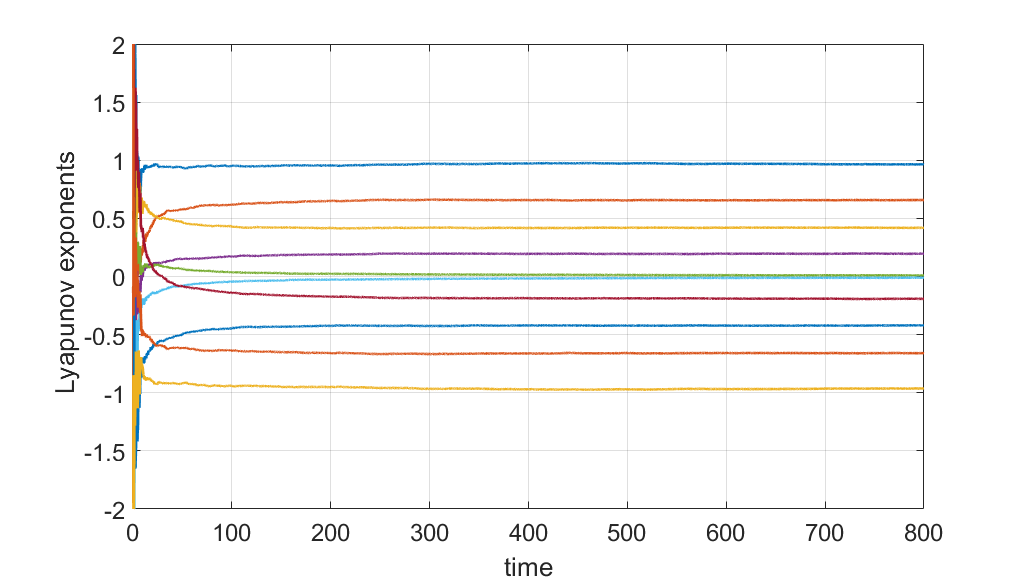}
			\caption{ $n=2$, $E=100$,$LLE=0.9645 $}
			\label{fig:n2E100}
		\end{figure}
		\begin{figure}[!htb]
			\centering
			\begin{minipage}[t]{0.5\textwidth}	
				\centering
				\includegraphics[width=\linewidth,height=0.25\textheight]{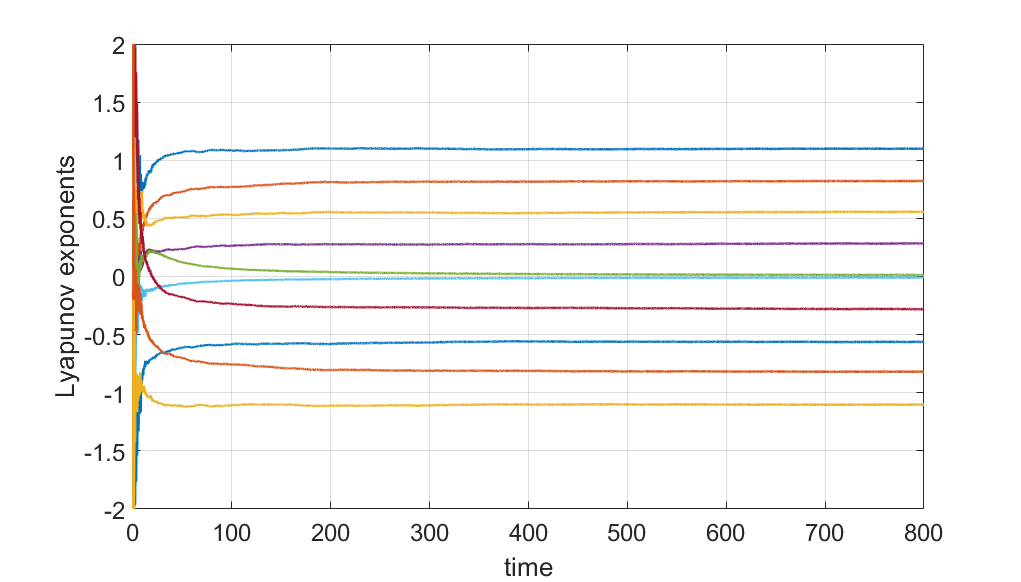}
				\caption{ $n=2$, $E=250$,$LLE=1.099 $}
				\label{fig:n2E250}
			\end{minipage}%
			\begin{minipage}[t]{0.5\textwidth}	
				\centering
				\includegraphics[width=\linewidth,height=0.25\textheight]{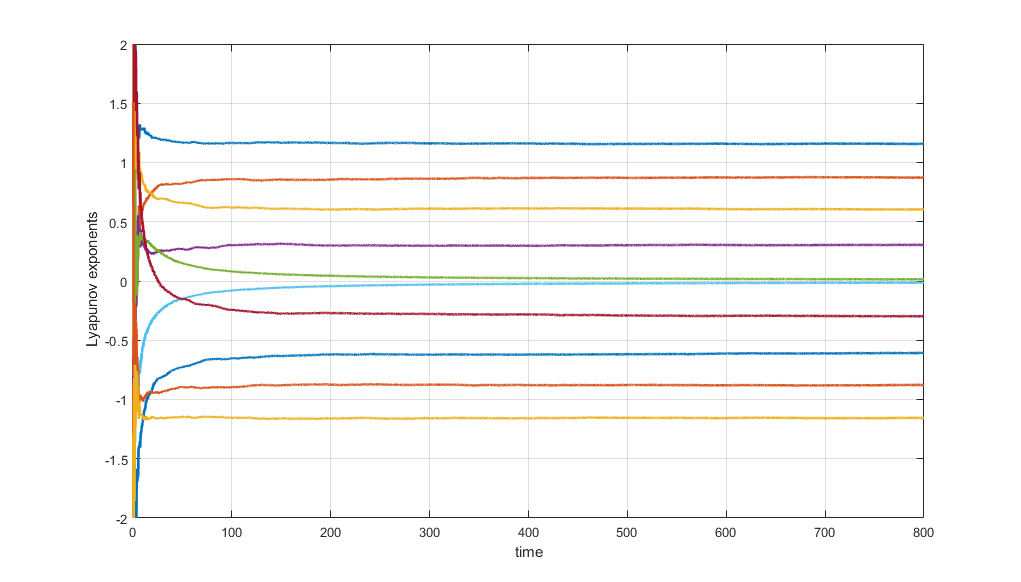}
				\caption{ $n=2$, $E=500$,$LLE= 1.1574 $}
				\label{fig:n2E500}
			\end{minipage}
		\end{figure}
		\newpage
		\begin{figure}[!htb]
			\centering
			\begin{minipage}[t]{0.5\textwidth}	
				\centering
				\includegraphics[width=\linewidth,height=0.25\textheight]{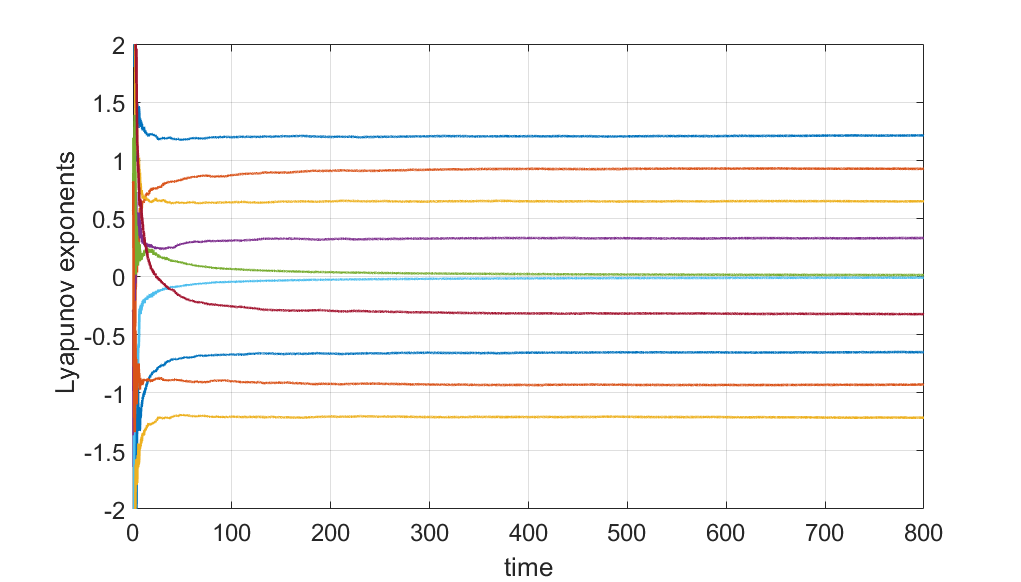}
				\caption{ $n=2$, $E=1000$,$LLE= 1.2138 $}
				\label{fig:n2E1000}
			\end{minipage}%
			\begin{minipage}[t]{0.5\textwidth}	
				\centering
				\includegraphics[width=\linewidth,height=0.25\textheight]{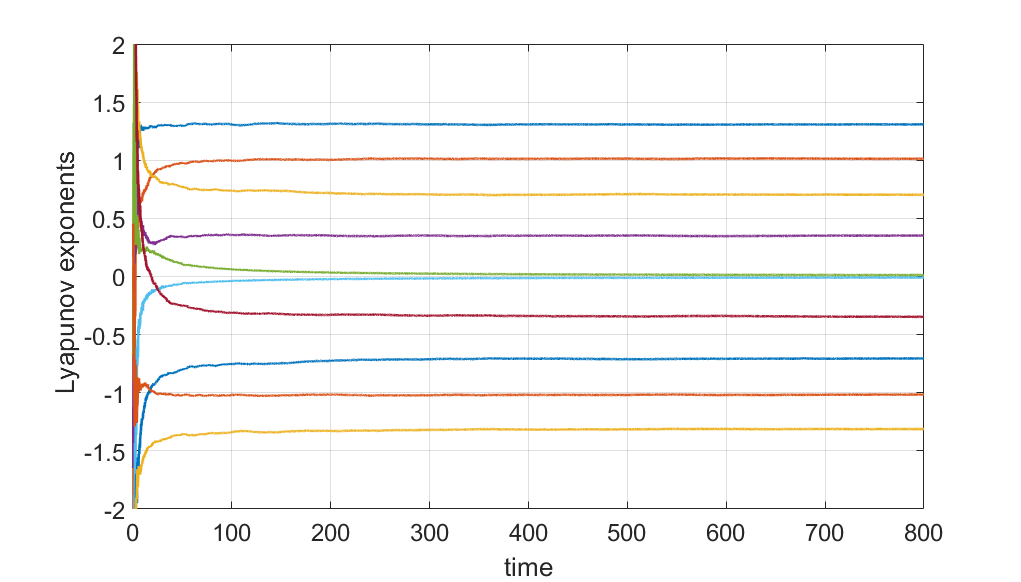}
				\caption{ $n=2$, $E=2000$,$LLE= 1.3087 $}
				\label{fig:n2E2000}
			\end{minipage}
		\end{figure}
		
	\begin{figure}[!htb]
		\centering
		\begin{minipage}[t]{0.5\textwidth}	
			\centering
			\includegraphics[width=\linewidth,height=0.25\textheight]{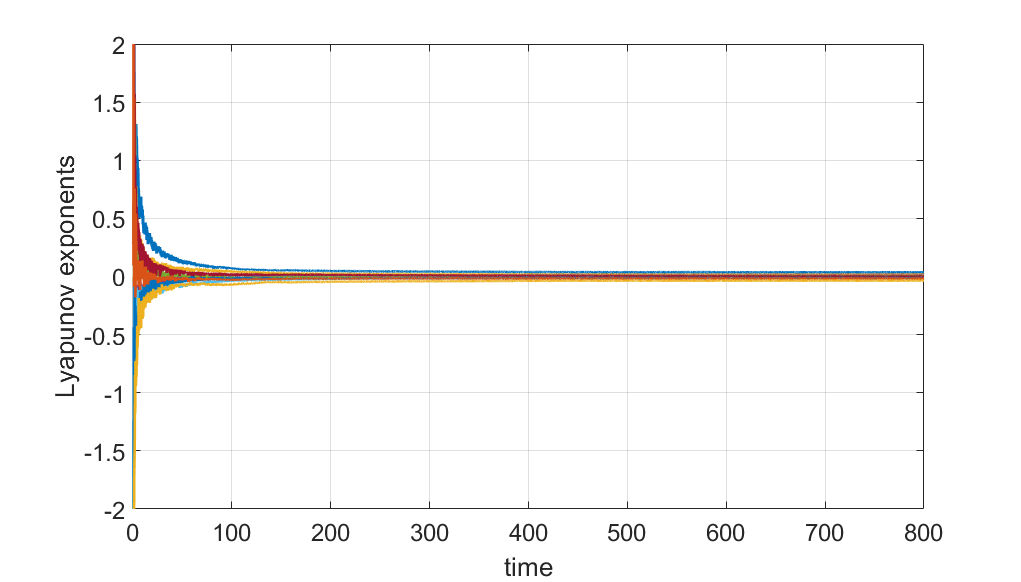}
			\caption{ $n=3$, $E=15$ ,$LLE=0.035$}
			\label{fig:n3E15}
		\end{minipage}%
		\begin{minipage}[t]{0.5\textwidth}	
			\centering
			\includegraphics[width=\linewidth,height=0.25\textheight]{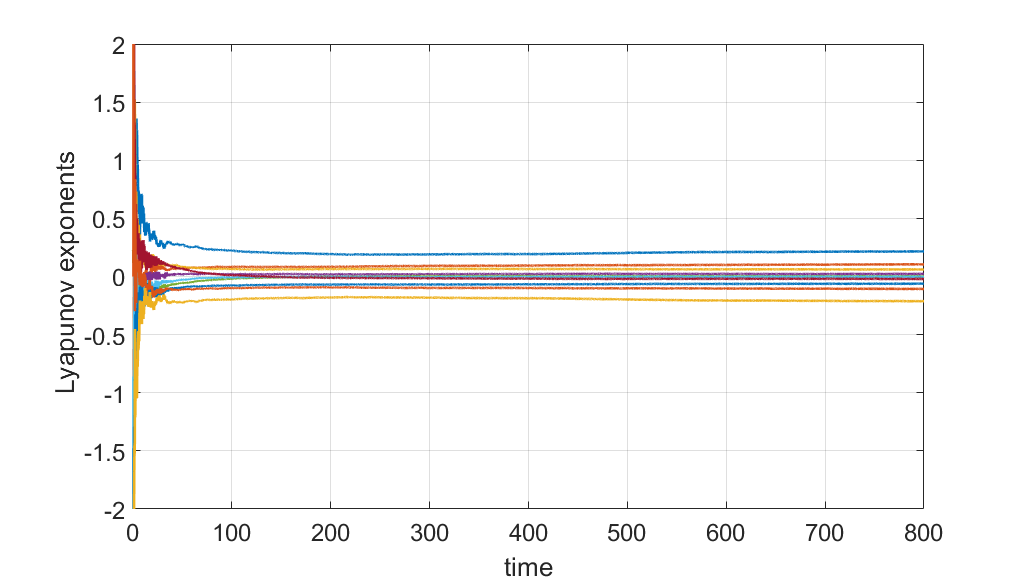}
			\caption{ $n=3$, $E=25$ ,$LLE=0.2159$}
			\label{fig:n3E25}
		\end{minipage}%
	\end{figure}
	\newpage
	\begin{figure}[!htb]
		\centering
		\begin{minipage}[t]{0.5\textwidth}	
			\centering
			\includegraphics[width=\linewidth,height=0.25\textheight]{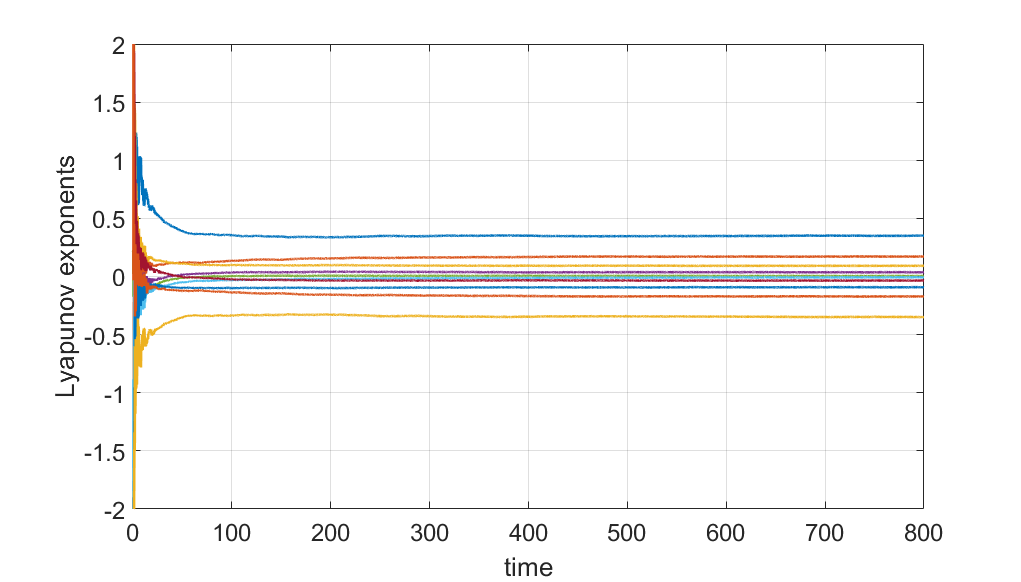}
			\caption{ $n=3$, $E=30$ ,$LLE=0.3515$}
			\label{fig:n3E30}
		\end{minipage}
	\end{figure}
	\begin{figure}[!htb]
		\centering
		\begin{minipage}[t]{0.5\textwidth}	
			\centering
			\includegraphics[width=\linewidth,height=0.25\textheight]{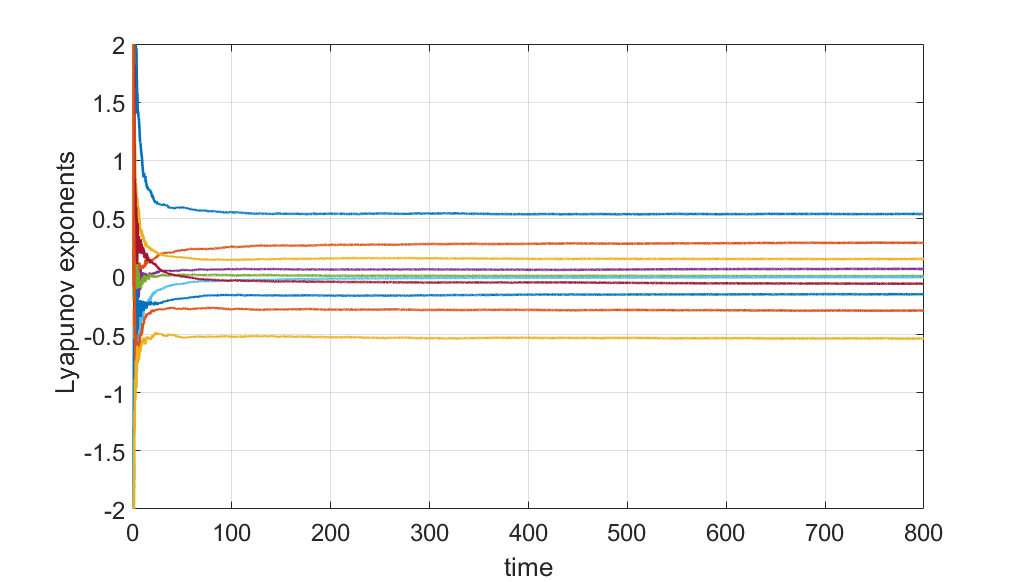}
			\caption{ $n=3$,$E=40$ ,$LLE=0.5371$}
			\label{fig:n3E40}
		\end{minipage}%
		\begin{minipage}[t]{0.5\textwidth}	
			\centering
			\includegraphics[width=\linewidth,height=0.25\textheight]{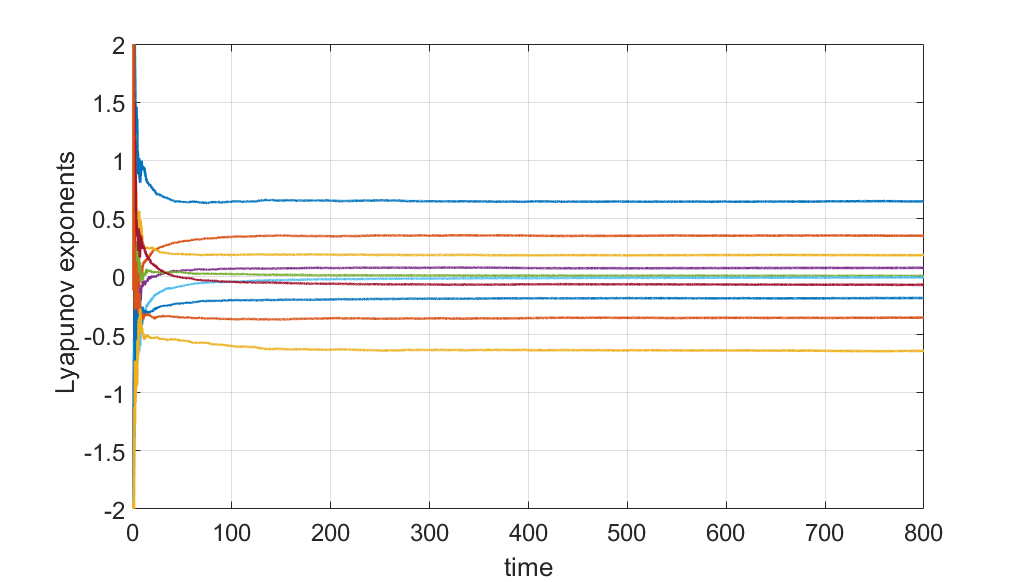}
			\caption{ $n=3$, $E=50$ ,$LLE=0.6450$}
			\label{fig:n3E50}
		\end{minipage}
	\end{figure}
	\begin{figure}[!htb]
		\centering
		\includegraphics[width=0.5\linewidth,height=0.25\textheight]{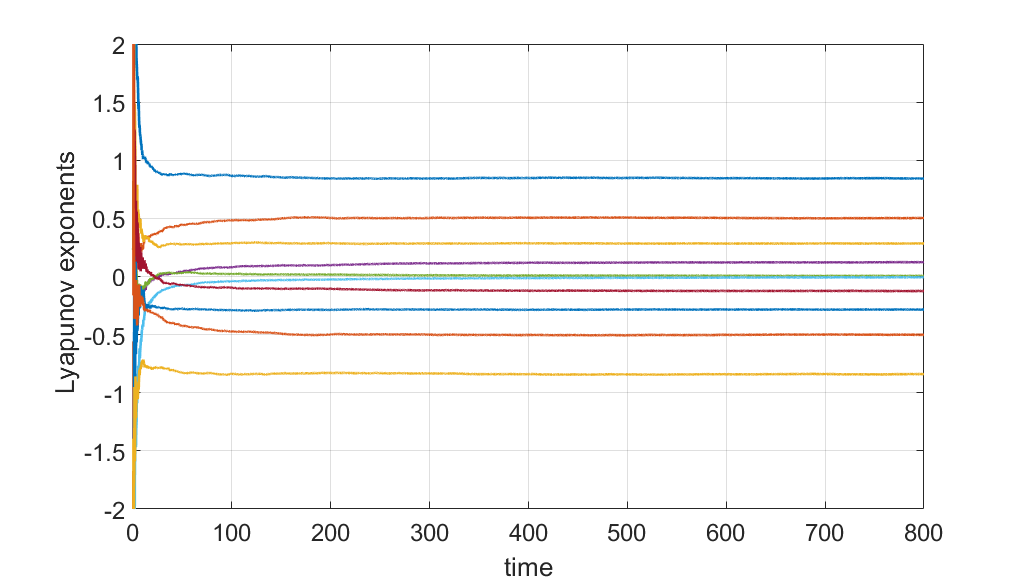}
		\caption{ $n=3$, $E=100$ ,$LLE=0.8430$}
		\label{fig:n3E100}
	\end{figure}
	\begin{figure}[!htb]
		\centering
		\begin{minipage}[t]{0.5\textwidth}	
			\centering
			\includegraphics[width=\linewidth,height=0.25\textheight]{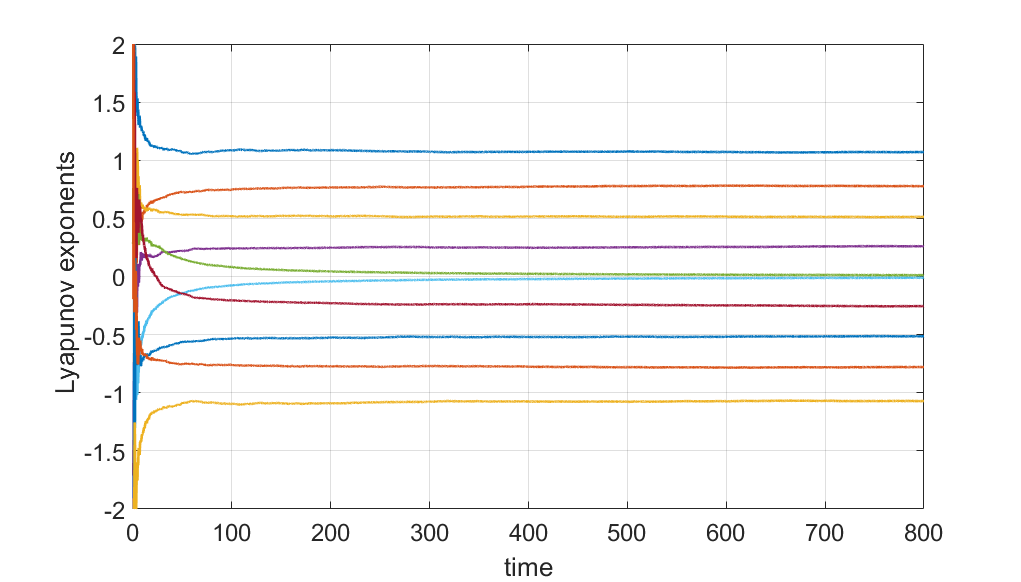}
			\caption{ $n=3$, $E=250$ ,$LLE=1.0699$}
			\label{fig:n3E250}
		\end{minipage}%
		\begin{minipage}[t]{0.5\textwidth}	
			\centering
			\includegraphics[width=\linewidth,height=0.25\textheight]{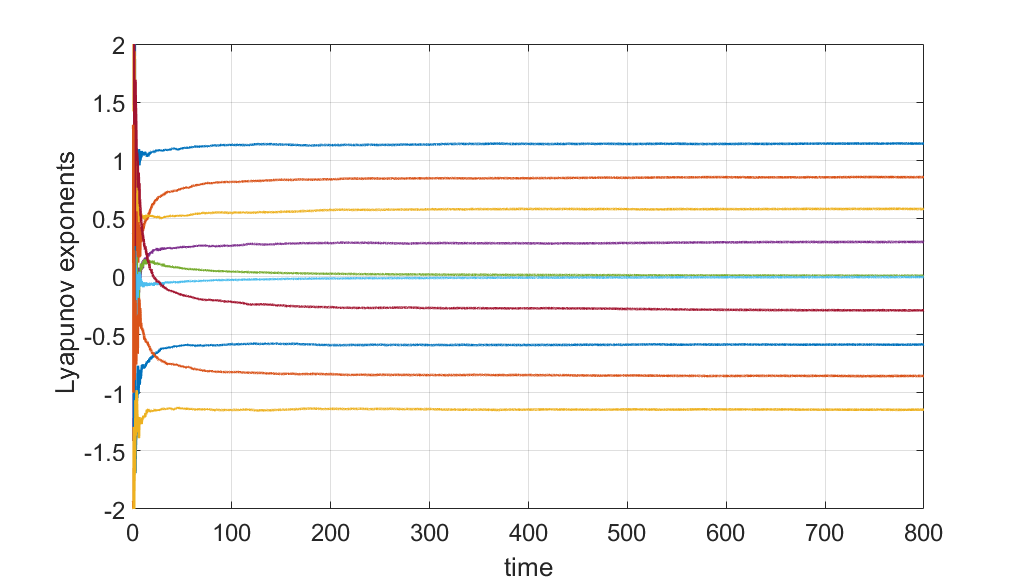}
			\caption{ $n=3$, $E=500$ ,$LLE=1.1439$}
			\label{fig:n3E500}
		\end{minipage}
	\end{figure}
	\newpage
	\begin{figure}[!htb]
		\centering
		\begin{minipage}[t]{0.5\textwidth}	
			\centering
			\includegraphics[width=\linewidth,height=0.25\textheight]{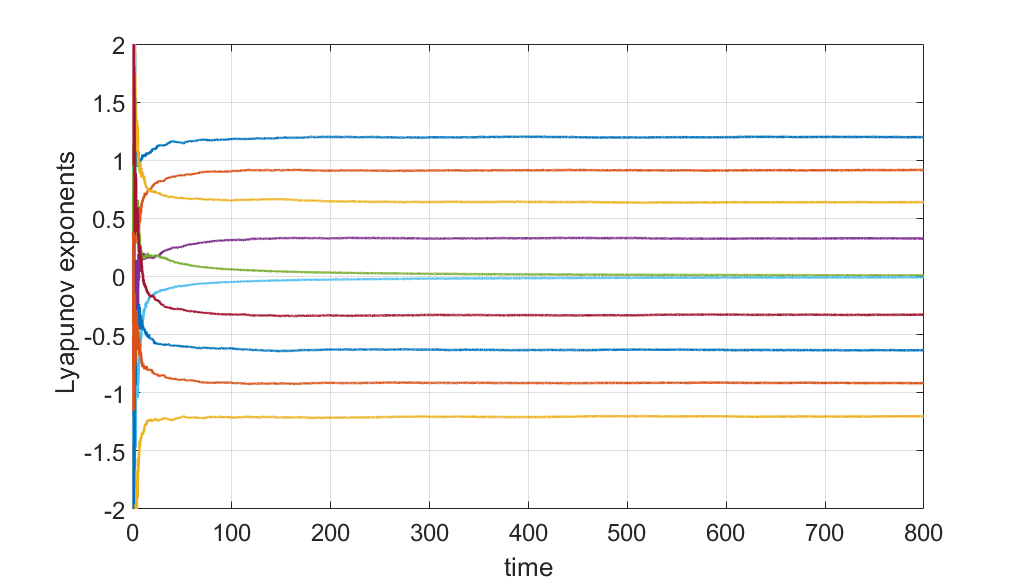}
			\caption{ $n=3$, $E=1000$ ,$LLE=1.1983$}
			\label{fig:n3E1000}
		\end{minipage}%
		\begin{minipage}[t]{0.5\textwidth}	
			\centering
			\includegraphics[width=\linewidth,height=0.25\textheight]{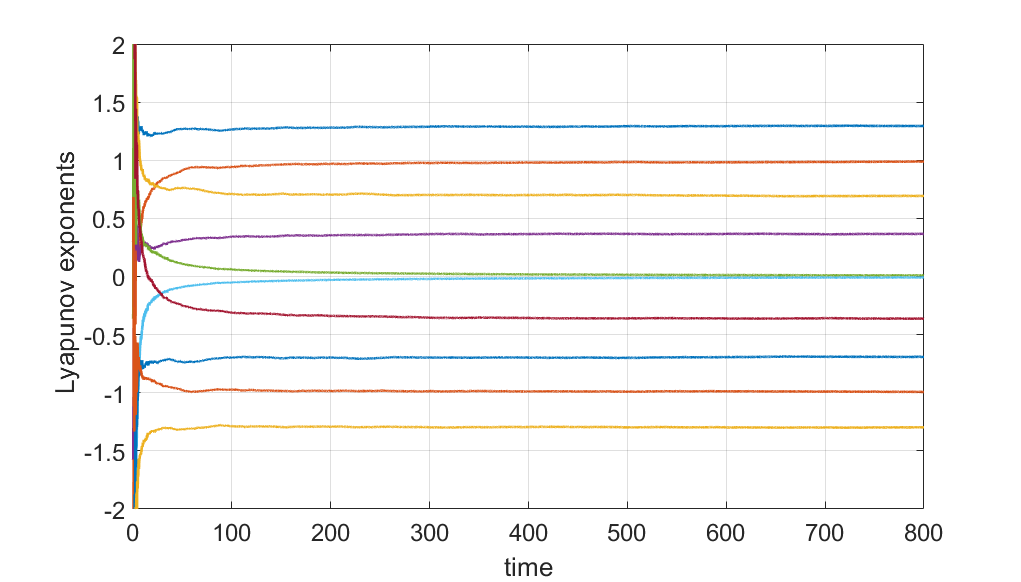}
			\caption{ $n=3$, $E=2000$ ,$LLE=1.2949$}
			\label{fig:n3E2000}
		\end{minipage}
	\end{figure}
	
	\begin{figure}[!htb]
		\centering
		\begin{minipage}[t]{0.5\textwidth}	
			\centering
			\includegraphics[width=\linewidth,height=0.25\textheight]{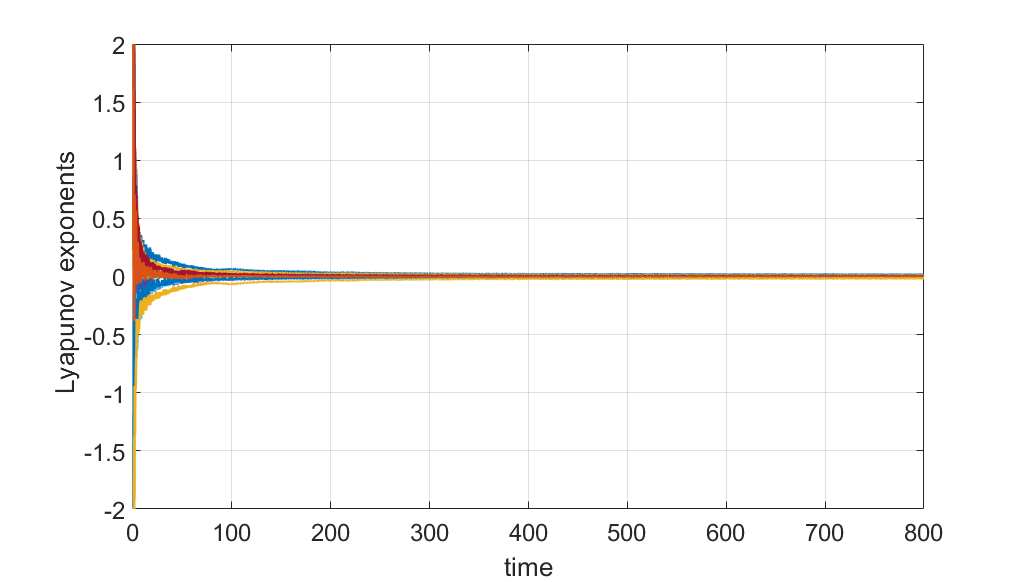}
			\caption{ $n=4$, $E=15$,$LLE=0.016$}
			\label{fig:n4E15}
		\end{minipage}%
	\end{figure}
	\newpage
	\begin{figure}[!htb]
		\centering
		\begin{minipage}[t]{0.5\textwidth}	
			\centering
			\includegraphics[width=\linewidth,height=0.25\textheight]{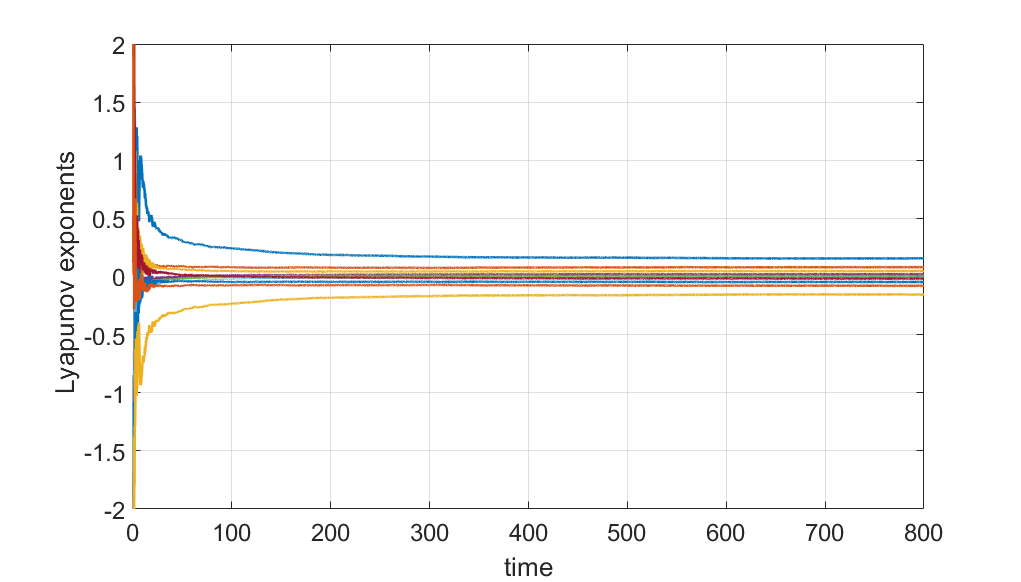}
			\caption{ $n=4$, $E=25$,$LLE=0.1563$}
			\label{fig:n4E25}
		\end{minipage}%
		\begin{minipage}[t]{0.5\textwidth}	
			\centering
			\includegraphics[width=\linewidth,height=0.25\textheight]{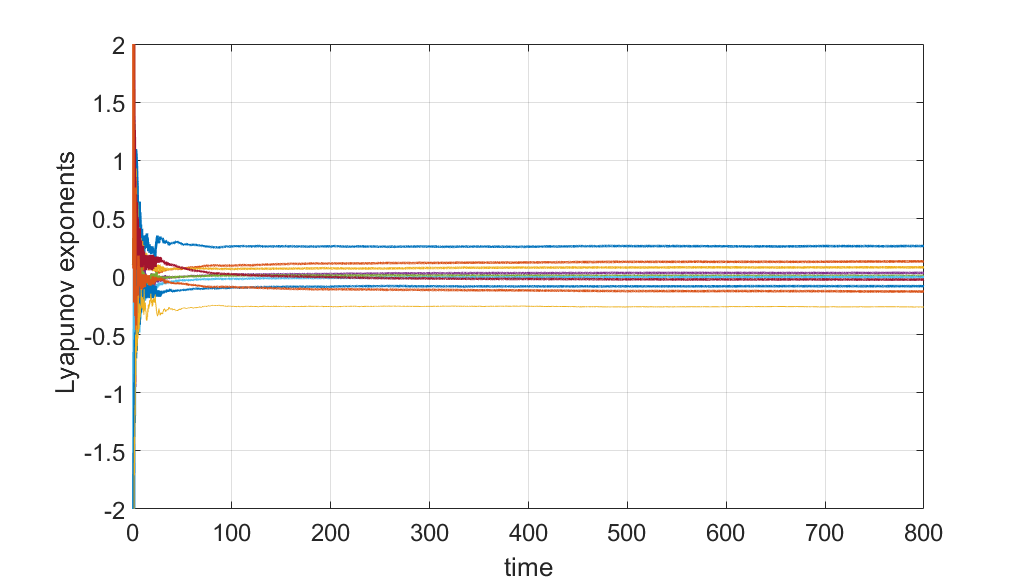}
			\caption{ $n=4$, $E=30$,$LLE=0.2623$}
			\label{fig:n4E30}
		\end{minipage}
	\end{figure}
	\begin{figure}[!htb]
		\centering
		\begin{minipage}[t]{0.5\textwidth}	
			\centering
			\includegraphics[width=\linewidth,height=0.25\textheight]{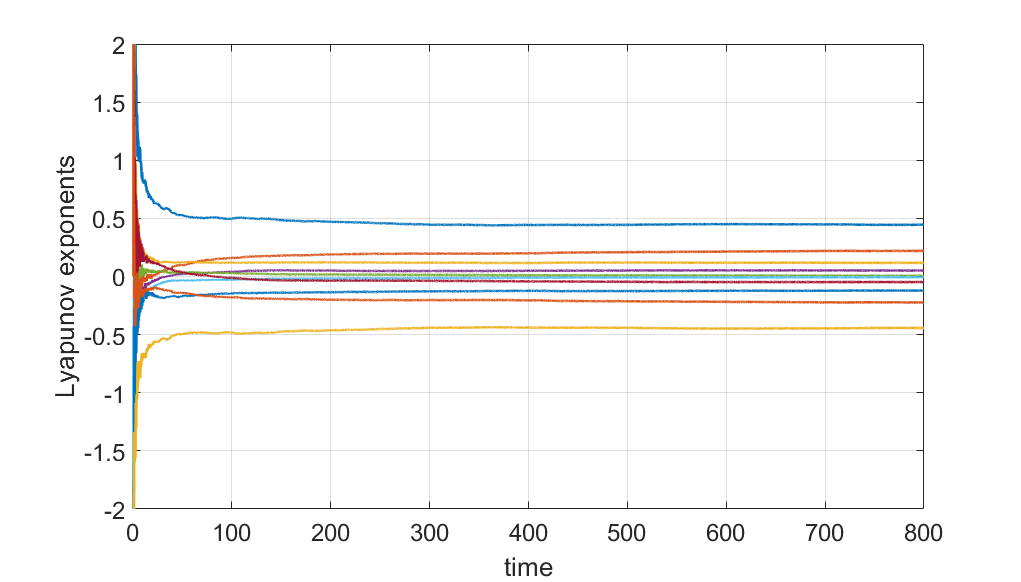}
			\caption{ $n=4$,$E=40$,$LLE=0.4453$}
			\label{fig:n4E40}
		\end{minipage}%
		\begin{minipage}[t]{0.5\textwidth}	
			\centering
			\includegraphics[width=\linewidth,height=0.25\textheight]{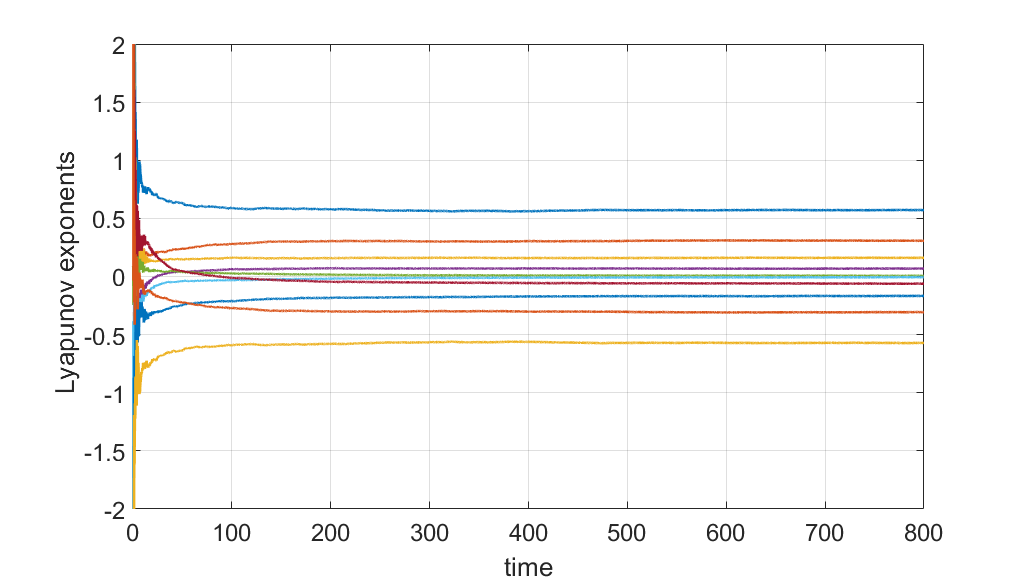}
			\caption{ $n=4$, $E=50$,$LLE=0.5710$}
			\label{fig:n4E50}
		\end{minipage}
	\end{figure}
	\begin{figure}[!htb]
		\centering
		\includegraphics[width=0.5\linewidth,height=0.25\textheight]{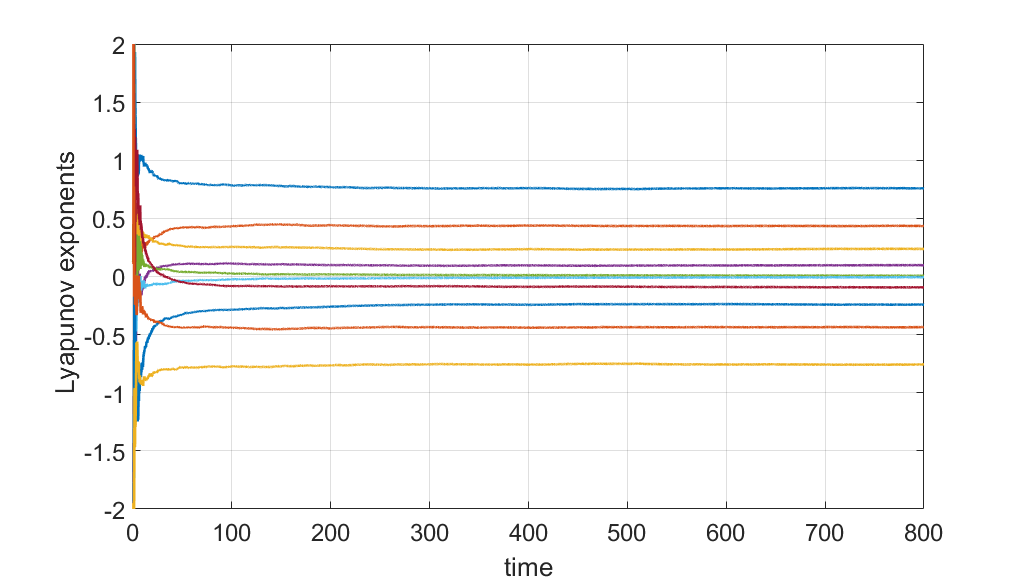}
		\caption{ $n=4$, $E=100$,$LLE=0.7578$}
		\label{fig:n4E100}
	\end{figure}
	\begin{figure}[!htb]
		\centering
		\begin{minipage}[t]{0.5\textwidth}	
			\centering
			\includegraphics[width=\linewidth,height=0.25\textheight]{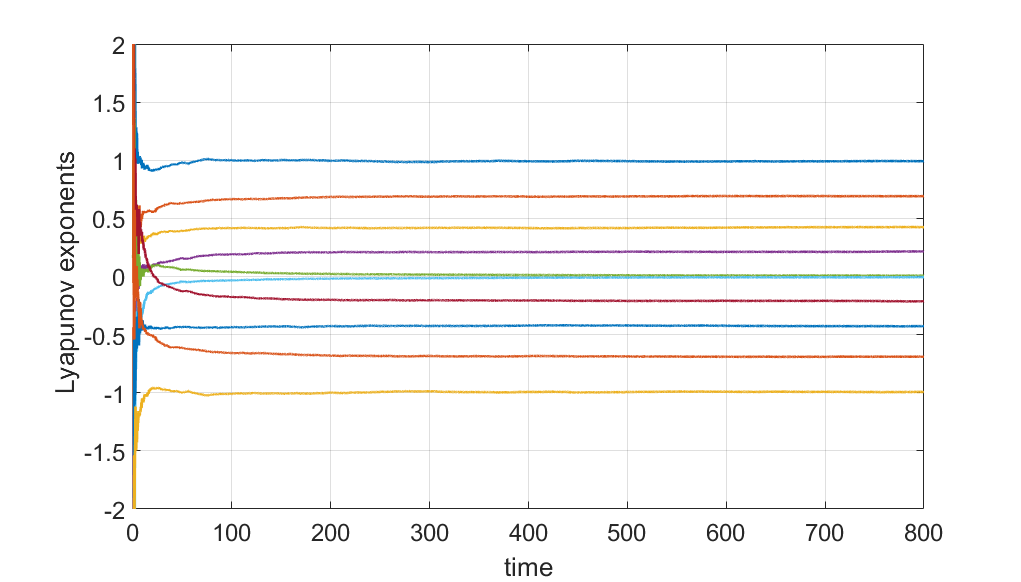}
			\caption{ $n=4$, $E=250$,$LLE=0.9922$}
			\label{fig:n4E250}
		\end{minipage}%
		\begin{minipage}[t]{0.5\textwidth}	
			\centering
			\includegraphics[width=\linewidth,height=0.25\textheight]{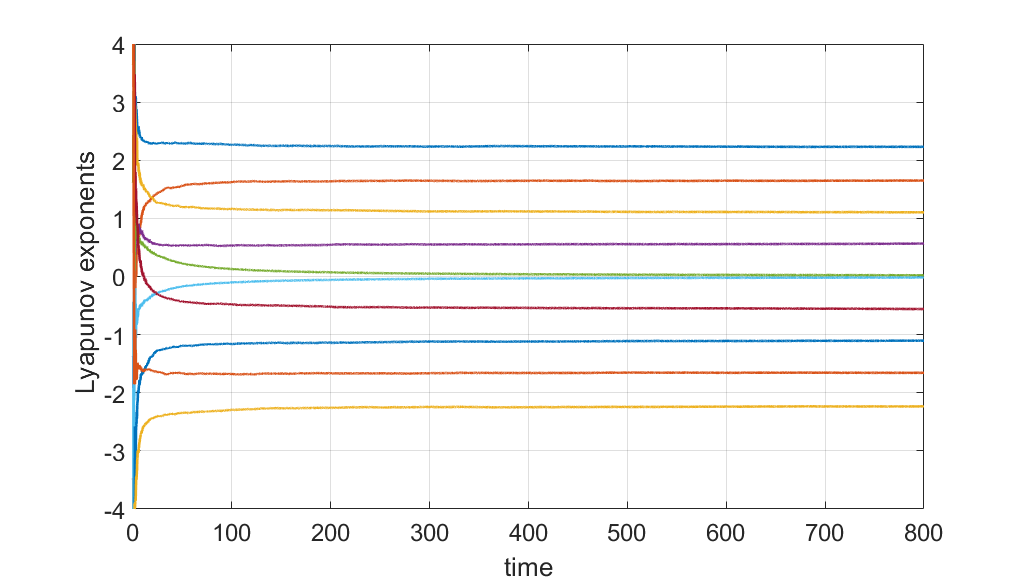}
			\caption{ $n=4$, $E=500$,$LLE=1.1064$}
			\label{fig:n4E500}
		\end{minipage}
	\end{figure}
	\newpage
	\begin{figure}[!htb]
		\centering
		\begin{minipage}[t]{0.5\textwidth}	
			\centering
			\includegraphics[width=\linewidth,height=0.25\textheight]{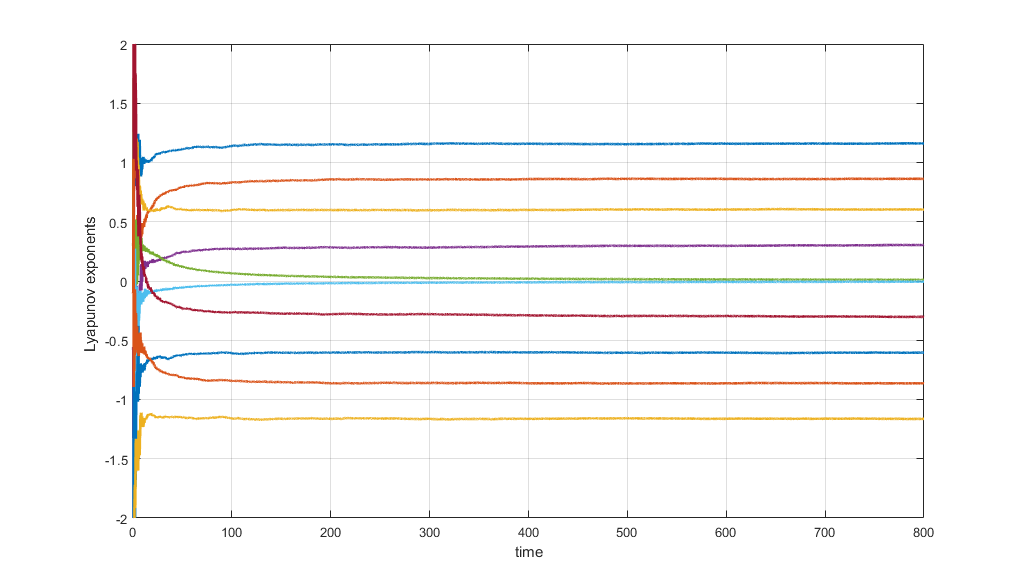}
			\caption{ $n=4$, $E=1000$,$LLE=1.161$}
			\label{fig:n4E1000}
		\end{minipage}%
		\begin{minipage}[t]{0.5\textwidth}	
			\centering
			\includegraphics[width=\linewidth,height=0.25\textheight]{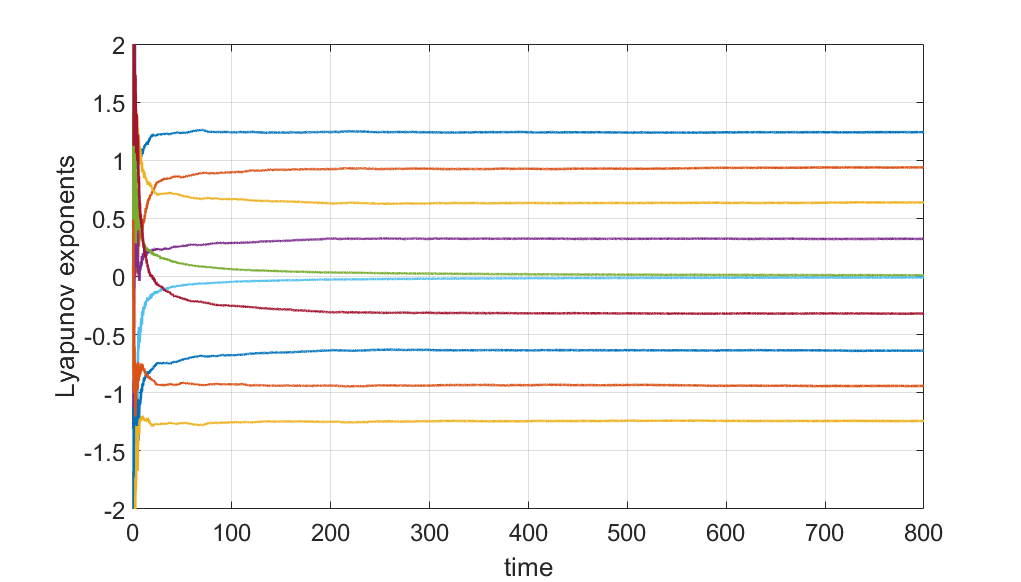}
			\caption{ $n=4$, $E=2000$,$LLE=1.2412$}
			\label{fig:n4E2000}
		\end{minipage}
	\end{figure}
	
	\begin{figure}[!htb]
		\centering
		\begin{minipage}[t]{0.5\textwidth}	
			\centering
			\includegraphics[width=\linewidth,height=0.25\textheight]{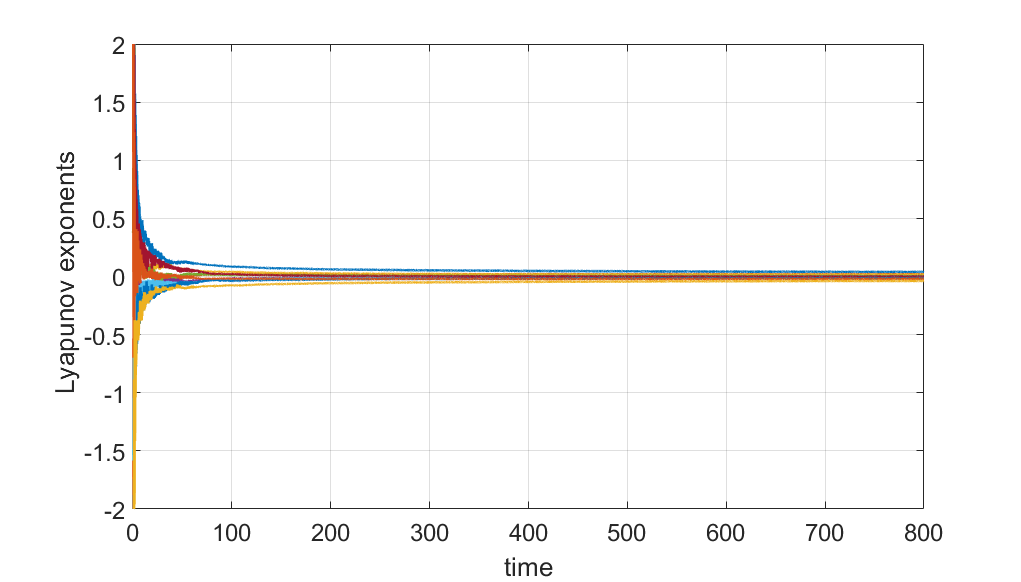}
			\caption{ $n=5$, $E=15$,$LLE=0.039$}
			\label{fig:n5E15}
		\end{minipage}%
		\begin{minipage}[t]{0.5\textwidth}	
			\centering
			\includegraphics[width=\linewidth,height=0.25\textheight]{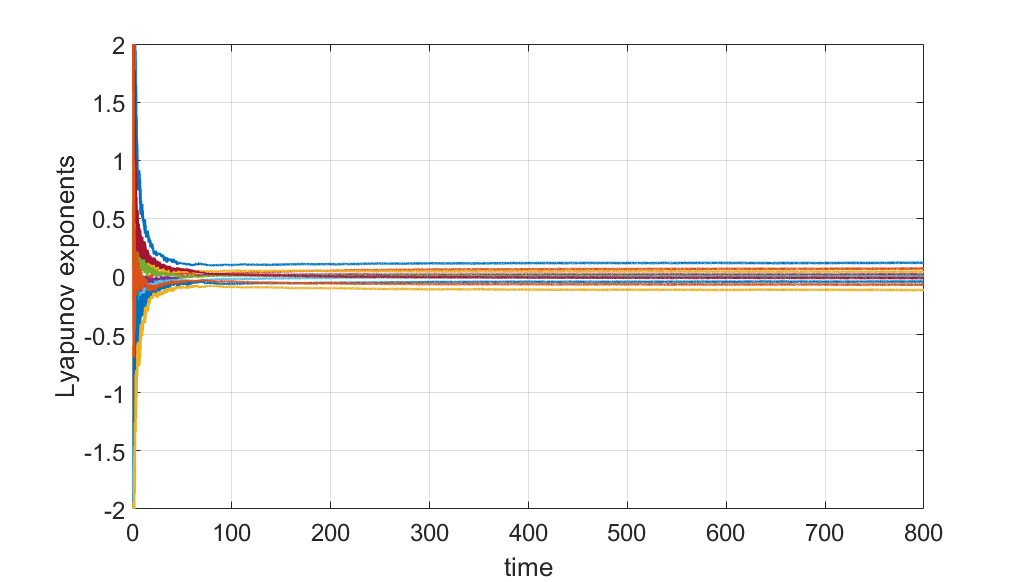}
			\caption{ $n=5$, $E=25$,$LLE=0.118$}
			\label{fig:n5E25}
		\end{minipage}%
	\end{figure}
	\newpage
	\begin{figure}[!htb]
		\centering
		\begin{minipage}[t]{0.5\textwidth}	
			\centering
			\includegraphics[width=\linewidth,height=0.25\textheight]{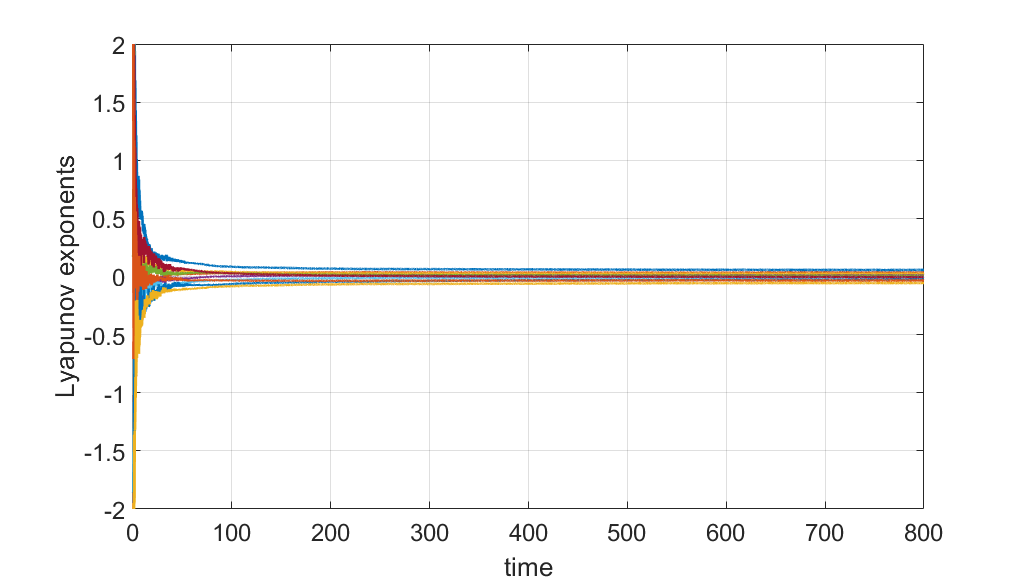}
			\caption{ $n=5$, $E=30$ ,$LLE=0.23616$}
			\label{fig:n=5E=30}
		\end{minipage}
	\end{figure}
	\begin{figure}[!htb]
		\centering
		\begin{minipage}[t]{0.5\textwidth}	
			\centering
			\includegraphics[width=\linewidth,height=0.25\textheight]{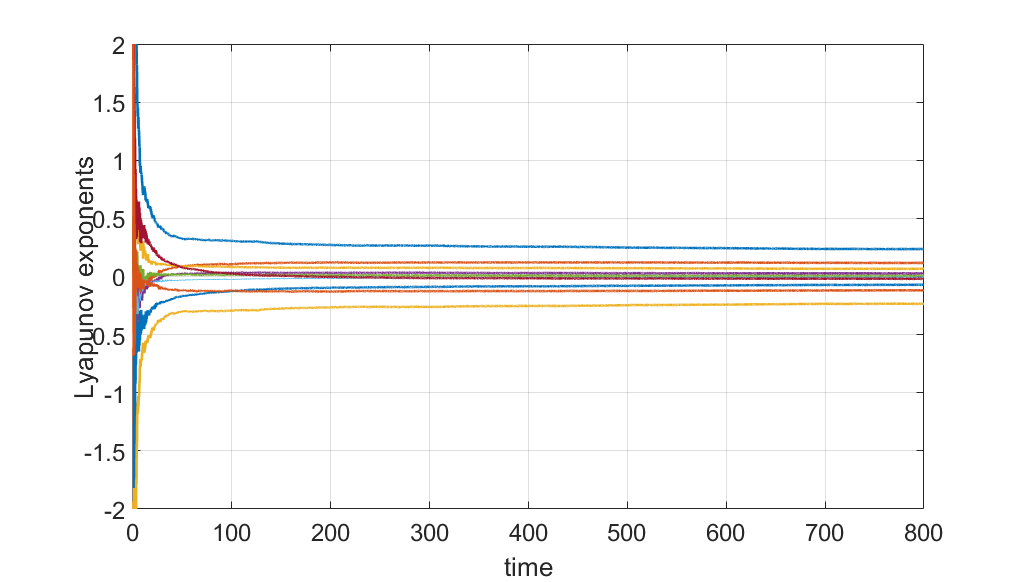}
			\caption{ $n=5$,$E=40$,$LLE=0.4393$}
			\label{fig:n5E40}
		\end{minipage}%
		\begin{minipage}[t]{0.5\textwidth}	
			\centering
			\includegraphics[width=\linewidth,height=0.25\textheight]{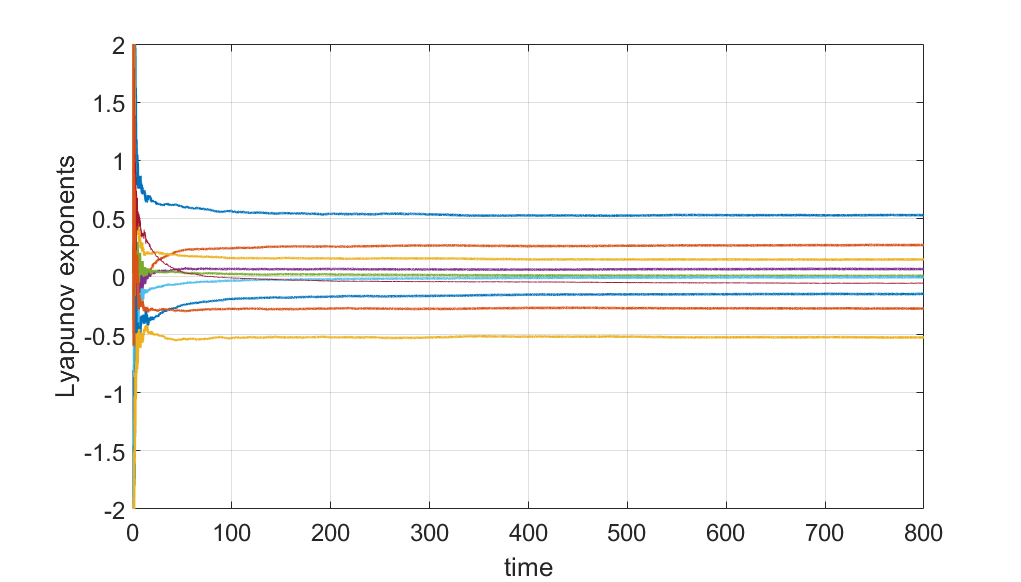}
			\caption{ $n=5$, $E=50$,$LLE=0.5276$}
			\label{fig:n5E50}
		\end{minipage}
	\end{figure}
	\begin{figure}[!htb]
		\centering
		\includegraphics[width=0.5\linewidth,height=0.25\textheight]{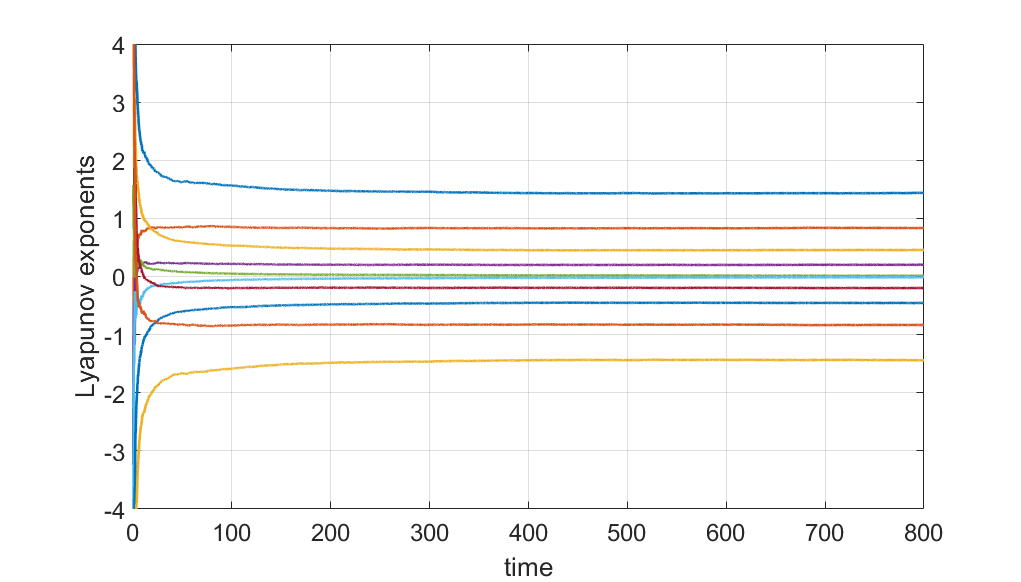}
		\caption{ $n=5$, $E=100$,$LLE=0.6972$}
		\label{fig:n5E100}
	\end{figure}
	\begin{figure}[!htb]
		\centering
		\begin{minipage}[t]{0.5\textwidth}	
			\centering
			\includegraphics[width=\linewidth,height=0.25\textheight]{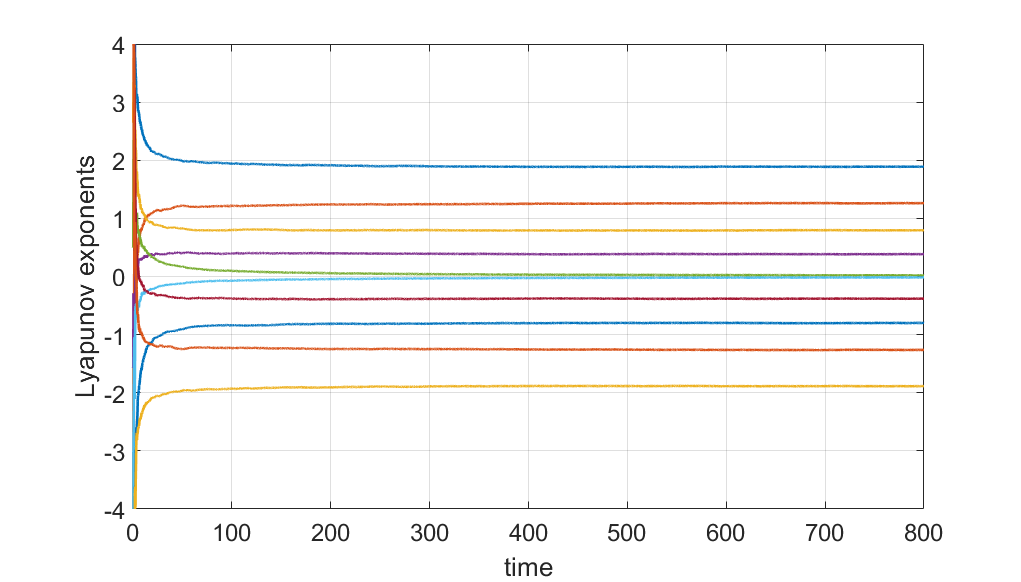}
			\caption{ $n=5$, $E=250$,$LLE= 0.9054$}
			\label{fig:n5E250}
		\end{minipage}%
		\begin{minipage}[t]{0.5\textwidth}	
			\centering
			\includegraphics[width=\linewidth,height=0.25\textheight]{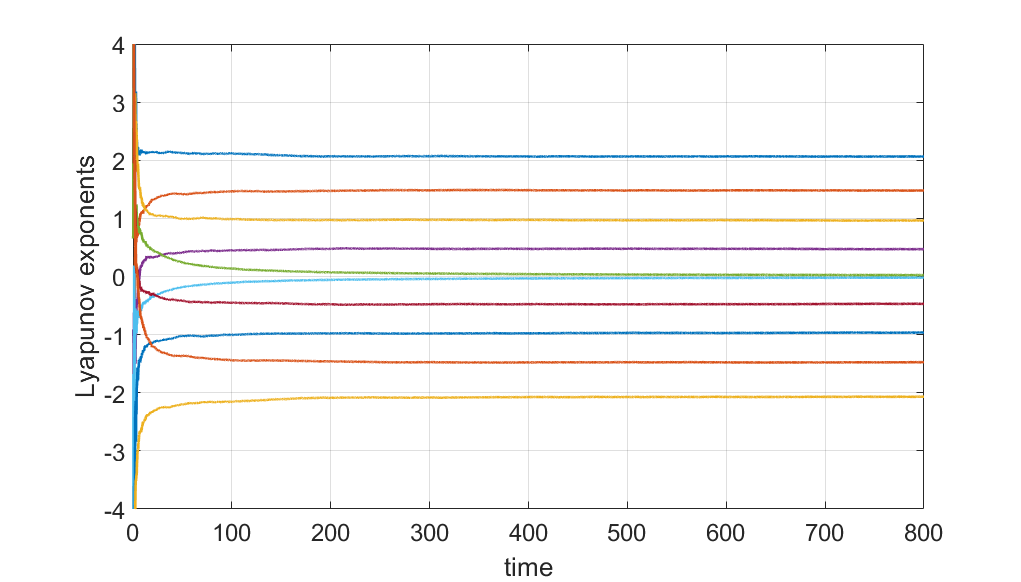}
			\caption{ $n=5$, $E=500$,$LLE=1.0405$}
			\label{fig:n5E500}
		\end{minipage}
	\end{figure}
	\newpage
	\begin{figure}[!htb]
		\centering
		\begin{minipage}[t]{0.5\textwidth}	
			\centering
			\includegraphics[width=\linewidth,height=0.25\textheight]{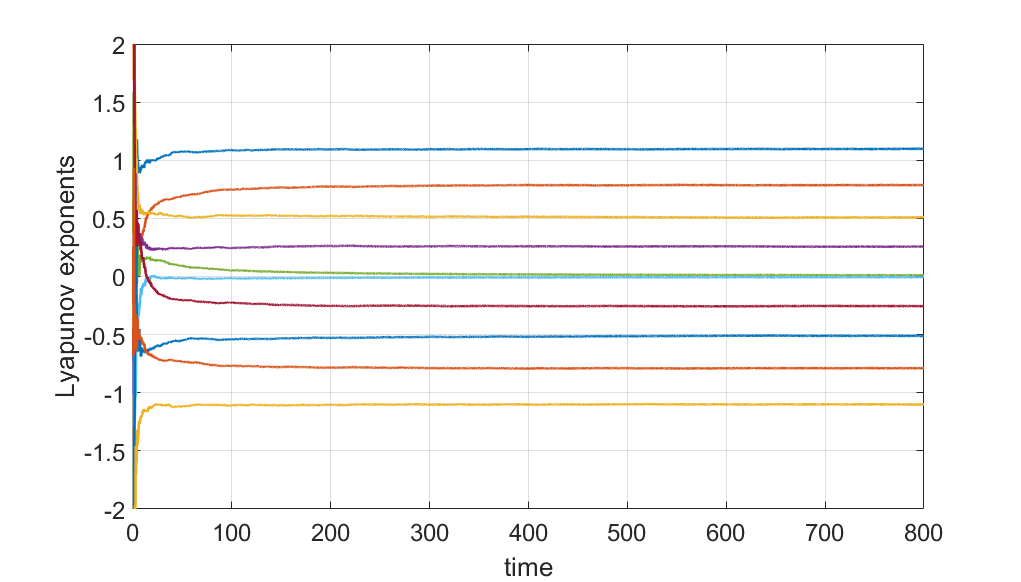}
			\caption{ $n=5$, $E=1000$,$LLE=1.0982$}
			\label{fig:n5E1000}
		\end{minipage}%
		\begin{minipage}[t]{0.5\textwidth}	
			\centering
			\includegraphics[width=\linewidth,height=0.25\textheight]{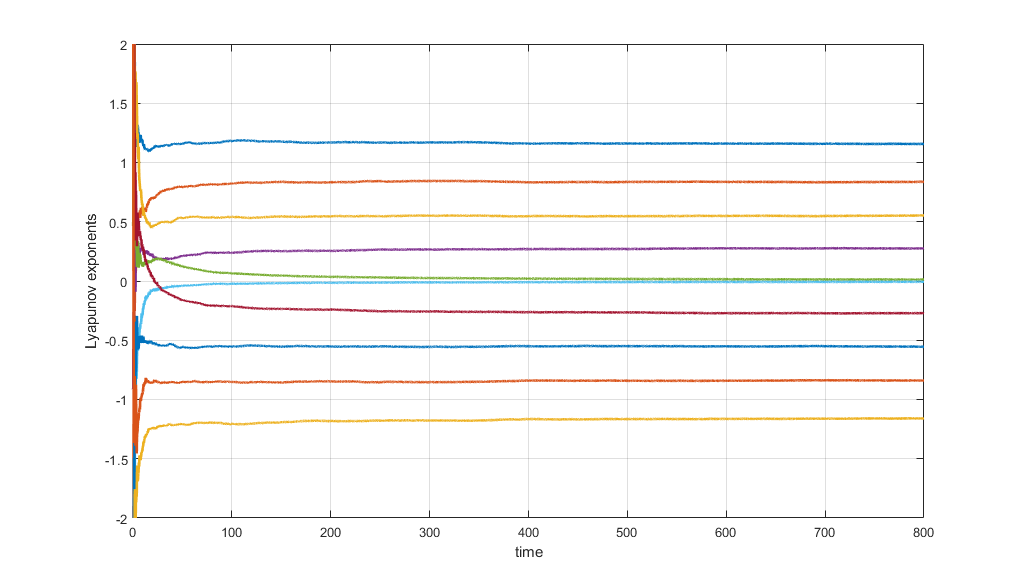}
			\caption{ $n=5$, $E=2000$,$LLE=1.1566$}
			\label{fig:n5E2000}
		\end{minipage}
	\end{figure}
	
	\subsection{Absolute minima of the potentials $V_{(n)}$}
	
	The absolute minima of the potentials $V_{(n)}$ associated to $L_{(n)}$ are given below.
	
	\vskip 1em
	
	For $n=2$
	
	\be
	\begin{split}
		&\left\{\phi _1\to 0,\phi _3\to 2.05,\phi _4\to 1.95,\chi _1\to \pm2.,\chi _3\to -6.47\right\},\\
		&\left\{\phi _1\to 0,\phi _3\to 2.09,\phi _4\to -0.16,\chi _1\to \pm2.,\chi _3\to -2.76\right\},\\
		&\left\{\phi _1\to 0,\phi _3\to -0.41,\phi _4\to 2.50,\chi _1\to \pm 2.,\chi _3\to -2.66\right\},\\
		&\left\{\phi _1\to 0,\phi _3\to -0.37,\phi _4\to 0.39,\chi _1\to \pm2.,\chi _3\to 1.05\right\},\\
		&\left\{\phi _1\to \pm2.,\phi _3\to 1.30,\phi _4\to 2.74,\chi _1\to 0,\chi _3\to -4.38\right\},\\
		&\left\{\phi _1\to \pm2.,\phi _3\to 1.44,\phi _4\to 2.60,\chi _1\to 0,\chi _3\to -1.28\right\},\\
		&\left\{\phi _1\to \pm2.,\phi _3\to 0.24,\phi _4\to -0.25,\chi _1\to 0,\chi _3\to -4.14\right\},\\
		&\left\{\phi _1\to \pm 2.,\phi _3\to 0.37,\phi _4\to -0.40,\chi _1\to 0,\chi _3\to -1.05\right\},\\
	\end{split}
	\label{n2vacua}
	\ee	
	
	For $n=4$
	
	\be
	\begin{split}
		&\left\{\phi _1\to 0,\phi _3\to 2.46\phi _4\to 2.90,\chi _1\to \pm1.,\chi _3\to -12.4\right\},\\
		&\left\{\phi _1\to 0,\phi _3\to 2.27,\phi _4\to -0.17,\chi _1\to \pm1.,\chi _3\to -7.91\right\},\\
		&\left\{\phi _1\to 0,\phi _3\to 0.87,\phi _4\to 4.67,\chi _1\to \pm1.,\chi _3\to -7.43\right\},\\
		&\left\{\phi _1\to 0,\phi _3\to 0.68,\phi _4\to 1.60,\chi _1\to \pm1.,\chi _3\to -2.94\right\},\\
		&\left\{\phi _1\to \pm1.,\phi _3\to 0.89,\phi _4\to 0.64,\chi _1\to 0,\chi _3\to -4.73\right\},\\
		&\left\{\phi _1\to \pm 1.,\phi _3\to 2.25,\phi _4\to 3.86,\chi _1\to 0,\chi _3\to -10.6\right\},\\
		&\left\{\phi _1\to \pm 1.,\phi _3\to 2.42,\phi _4\to 3.85,\chi _1\to 0,\chi _3\to -6.38\right\},\\
		&\left\{\phi _1\to \pm1.,\phi _3\to 0.72,\phi _4\to 0.66,\chi _1\to 0,\chi _3\to -8.97\right\},\\
	\end{split}
	\label{minima4}
	\ee
	
	For $n=5$
	
	\be
	\begin{split}
		&\left\{\phi _1\to 0,\phi _3\to 3.04,\phi _4\to 2.64,\chi _1\to \pm1.,\chi _3\to -16.9\right\},\\
		&\left\{\phi _1\to 0,\phi _3\to 0.92,\phi _4\to 2.08,\chi _1\to \pm1.,\chi _3\to -4.27\right\},\\
		&\left\{\phi _1\to 0,\phi _3\to 2.60,\phi _4\to -2.67,\chi _1\to \pm1.,\chi _3\to -11.0\right\},\\
		&\left\{\phi _1\to 0,\phi _3\to 1.36,\phi _4\to 7.38,\chi _1\to \pm1.,\chi _3\to -10.2\right\},\\
		&\left\{\phi _1\to \pm1.,\phi _3\to 1.06,\phi _4\to 0.28,\chi _1\to 0,\chi _3\to -6.35\right\},\\
		&\left\{\phi _1\to \pm1.,\phi _3\to 2.90,\phi _4\to 4.43,\chi _1\to 0,\chi _3\to -14.9\right\},\\
		&\left\{\phi _1\to \pm1.,\phi _3\to 3.11,\phi _4\to 4.80,\chi _1\to 0,\chi _3\to -9.04\right\},\\
		&\left\{\phi _1\to \pm1.,\phi _3\to 0.85,\phi _4\to -0.09,\chi _1\to 0,\chi _3\to -12.2\right\},\\
		\label{minima5}
	\end{split}
	\ee	
	
	\subsection{Asymptotic Profiles of the Kink Solution for $L_{(n=3)}$}
	
	Solutions of (\ref{linear}), which are regular as $\tau \rightarrow \infty$ are given below
	\begin{multline}
	s_1(\tau) = \left(3.1 c_1+0.49 c_2-6.52 c_3-1.52 c_4\right) e^{-3.45 \tau}+\left(0.07 c_1-0.005 c_2+0.25 c_3+0.16 c_4\right) e^{-4.56
		\tau} \\
	+\left(-1.17 c_1-0.3 c_2+1.29 c_3+0.26 c_4\right) e^{-7.36 \tau}+\left(-0.98 c_1-0.18 c_2+4.98 c_3+1.11 c_4\right) e^{-2.38 \tau} \,,
	\end{multline}
	
	\begin{multline}
	s_3(\tau)  = \left(0.92 c_1+ 0.15 c_2-1.95c_3-0.46 c_4\right) e^{-3.44 \tau} + \left(-0.139 c_1+0.01 c_2-0.48
	c_3-0.30c_4\right) e^{-4.56 \tau}  \\
	+\left(-0.14 c_1-0.035 c_2+0.15 c_3+0.03 c_4\right) e^{-7.36 \tau}+\left(-0.65 c_1-0.12 c_2+3.28 c_3+0.73 c_4\right) e^{-2.38 \tau}
	\end{multline}
	
	\begin{multline}
	s_4(\tau) = \left(7.74 c_1+1.45 c_2-39.25 c_3-8.73 c_4\right) e^{-2.38 \tau}+\left(5.79 c_1+0.93 c_2-12.24 c_3-2.86
	c_4\right) e^{-3.45 \tau}  \\
	+\left(0.25 c_1+0.065 c_2-0.28 c_3-0.055 c_4\right) e^{-7.36 \tau}+\left(0.016
	c_1-0.001 c_2+0.055 c_3+0.035 c_4\right) e^{-4.56 \tau}     
	\end{multline}
	
	\begin{multline}
	s_5(\tau) = \left(3.09 c_1+0.49 c_2-6.26 c_3-1.47 c_4\right) e^{-3.45 \tau}+\left(0.20 c_1 + 0.051 c_2 - 0.24
	c_3-0.048 c_4\right) e^{-7.36 \tau}  \\
	+\left(0.025 c_1+0.00025 c_2+0.03 c_3+0.03 c_4\right) e^{-4.56
		\tau}+\left(-73.40 c_1-13.99 c_2+404.89 c_3+89.86 c_4\right) e^{-2.38 \tau}
	\end{multline}


\end{document}